# NATO'S MISSION-CRITICAL SPACE CAPABILITIES UNDER THREAT:

## Cybersecurity Gaps in the Military Space Asset Supply Chain

BERENIKE KATHARINA VOLLMER

Non-Resident Visiting Scholar
Affiliated to NATO CCDCOE Tallinn

berenike.vollmer@sciencespo.fr

FEBRUARY 2021

Academic Supervision: Professor Dr. Ronald HATTO

# ABSTRACT

The North Atlantic Treaty Organization's (NATO) public-private Space Asset Supply Chain (SASC) currently exhibits significant cybersecurity gaps. It is well-established that data obtained from space assets is fundamental to NATO, as they allow for the facilitation of its missions, self-defence and effective deterrence of its adversaries. However, it was only in 2019 that NATO issued its first Space Policy and subsequently recognized space as an operational domain.

Any hostile cyber operation, suspending control over a space asset, severely impacts both NATO missions and allied Member States' national security. This threat is exacerbated by NATO's mostly unregulated cyber SASC. Hence, this thesis answers a twofold research question: *a) What are current cybersecurity gaps along NATO's global SASC; and b) How can NATO and its allied Member States gain greater control over such gaps to safeguard the supply of NATO mission-critical information?* An ontological field study is carried out by conducting nineteen semi-structured interviews with high-level representatives from relevant public, private, and academic organizations. This research was undertaken in collaboration with the NATO Cooperative Cyber Defence Centre of Excellence (CCDCOE) in Tallinn, Estonia.

This thesis concludes that current cybersecurity gaps along NATO's SASC are caused by cyber vulnerabilities such as legacy systems or the use of Commercial-Off-the-Shelf (COTS) technology. Inadequate cyber SASC management is caused by hindrances such as misaligned classification levels and significant understaffing. On this basis, NATO should consider two major collaboration initiatives: a) Raising awareness throughout the whole of the NATO system, and b) Pushing forward the creation of regulation through a standardized security framework on SASC cybersecurity. Doing so would enable NATO and its Member States to recognise cyberthreats to mission-critical data early on along its cyber SASC, and thus increase transparency, responsibility, and liability.

**Keywords**: *NATO, military, cybersecurity, space, SATCOM, supply chain, intelligence*

# TABLE OF CONTENTS

# LIST OF FIGURES

# LIST OF TABLES

# ABBREVIATIONS

| | |
|---|---|
| ACO | NATO Allied Command Operations |
| AGS | NATO Alliance Ground Surveillance |
| AJP-3.20 | NATO Allied Joint Publication for Cyberspace Operations |
| AJP-3.3 | NATO Allied Joint Doctrine for Air and Space Operations |
| AOM | NATO Alliance Operations and Missions |
| ASAT | Anti-Satellite Weapons |
| BDS | Chinese BeiDou Navigation Satellite System |
| C2 | Military Command and Control |
| C&C | IT Command-and-Control [C&C] |
| CCDCOE | Cooperative Cyber Defence Centre of Excellence |
| CDP | EU Capability Development Plan |
| CERT | Computer Emergency Response Team |
| CIA | Confidentiality, Integrity and Availability |
| CIK | Contribution in Kind |
| CISA | Cybersecurity and Infrastructure Security Agency |
| CMMC | Cybersecurity Maturity Model Certification |
| CNAD | Conference of National Armament Directors |
| CNES | Centre National d'Études Spatiales |
| CO | Cyberspace Operations |
| COE | Center of Excellence |
| COC | Code of Conduct |
| COTS | Commercial-Off-The-Shelf |
| COVID | Coronavirus Disease |
| CP5A0030 | NATO SATCOM Capability Package |
| CSC | US Cyberspace Solarium Commission |
| CSIRT | Computer Security Incident Response Teams |
| CSCRM | Cyber Supply Chain Risk Management |
| C4ISR | Command, Control, Communications, and Computers, Intelligence, Surveillance, and Reconnaissance |
| CyOC | Cyberspace Operations Centre |
| DARPA | US DOD Advanced Research Projects Agency |
| DHS | US Department of Homeland Security |
| DIA | US Defense Intelligence Agency |
| DIRLAUTH | Direct Liaison Authority |
| DISA | US DOD Defense Information Systems Agency |
| DOD | US Department of Defense |
| DoS | Denial of Service |
| PA DSC | NATO PA Defence and Security Committee |
| DSC | Defensive Space Control |
| EDA | European Defense Agency |
| EEAS | EU External Action Service |
| EGNOS | European Geostationary Navigation Overlay Service |
| ENISA | EU Agency for Network and Information Security |
| EU | European Union |
| EW | Electronic Warfare |
| GDPR | EU General Data Protection Regulation |
| GEO | Geosynchronous Orbit |
| GHz | Gigahertz |
| GLONASS | Russian Global Navigation Satellite System |
| GNSS | Global Navigation Satellite Systems |
| GPS | Global Positioning System |
| GSA | European Global Navigation Satellite Systems Agency |
| HQ | Headquarter |
| ICT | Information and Communication Technology |
| IEC | International Electrotechnical Commission |
| ISAC | Information Sharing and Analysis Center |
| ISO | International Standards Organization |
| ISR | Intelligence, Surveillance, And Reconnaissance |
| IT | Information Technology |
| ITU | International Telecommunication Union |
| JISR | NATO Joint Intelligence, Surveillance and Reconnaissance |
| JOA | Joint Operations Area |
| JPMO | NATO Joint Programme Management Office |
| LEO | Low-Earth Orbit |
| MA | Mission Assurance |
| MHz | Megahertz |
| MILAMOS | Manual on International Law Applicable to Military Uses of Outer Space |
| MILSATCOM | Military Satellite Communications |
| MISP | Malware Information Sharing Platform |
| MDA | Maritime Domain Awareness |
| MOU | Memorandum of Understanding |
| MS | Member States |
| NATO | North Atlantic Treaty Organization |
| NCIA | NATO Communications and Information Agency |
| NDPP | NATO Defence Planning Process |
| NGA | US National Geospatial-Intelligence Agency |
| NIST | US National Institute of Standards and Technology |
| NMAC | NATO Mission Access Centre |
| NSO | NATO Standardization Office |
| NSP2K | NATO SATCOM Post-2000 |
| ODNI | US Office of the Director of National Intelligence |
| OPCON | Operational Control |
| OSA | Orbital Security Alliance |
| PA | NATO Parliamentary Assembly |
| PARP | Partnership for Peace Planning and Review Process |
| PNT | Positioning, Navigation and Timing |
| POC | Point of Contact |
| PPP | Public-Private Partnership |
| RF | Radio Frequencies |
| ROI | Return on Investment |
| SASC | Space Asset Supply Chain |
| SATCOM | Satellite Communications |
| SC | Supply Chain |
| SCRM | Supply Chain Risk Management |
| SDA | US Space Development Agency |
| SG | Secretary General |
| SHAPE | NATO Supreme Headquarters Allied Powers Europe |
| SHF | Super High Frequency (X-Band) |
| SICRAL | Sistema Italiano per Comunicazioni Riservate ed Allarmi |
| SME | Subject Matter Expert |
| SPD5 | US Space Policy Directive - 5 |
| SpSC | NATO Space Support Coordination |
| SpSCE | NATO Space Support Coordination Elements |
| SSA | Space Situational Awareness |
| SSR | Space Support Request |
| SYRACUSE | Système de Radiocommunication Utilisant un Satellite |
| TCBM | Transparency and Confidence-Building Measures |
| TRJE18 | NATO Exercise Trident Juncture 2018 |
| TRJU19 | NATO Exercise Trident Jupiter 2019 |
| UHF | Ultra-High Frequency |
| UN | United Nations |
| UNOOSA | UN Office for Outer Space Affairs |

# 1 INTRODUCTION

When Sputnik I, the first artificial *satellite[1],* was launched in 1957, *space[2]* became critical for the security of international and national infrastructures (Unal, 2019, p. 3). Thirteen years later, NATO launched its own first satellite into space on 20 March 1970, to provide quick and secure strategic communication to its allied Member States (MS; NATO, 2020a). In total, eight satellites were launched, with the first two satellites PR/CP(70)2 and PR/CP(71)1 being launched in 1970 and 1972 respectively, and NATO IVA and NATO IVB, the final two satellites, being launched in 1991 and 1993. Subsequently, basing on cooperation with the US, NATO began installing over 20 ground communication terminals in Belgium, Canada, Germany, Italy, the Netherlands, the UK, and the US, as well as Denmark, Greece and, Norway, Portugal, and Turkey. However, as most NATO MS nowadays own their own satellites or invest in other space-based assets, NATO decided in 2005 to replace its own satellites with the *NATO Satellite Communications[3] (SATCOM) Post-2000 (NSP2K)* programme. This allowed direct access to cooperating MS's space assets and related services (NATO, 2020d; JAPCC, 2020). *Space assets* are defined as space-related systems collecting *intelligence*[4] and communication, which provide *internet[5]* and data for military manoeuvres, ships, aircraft, telecommunication and finances (Unal, 2019, p. 3). *Information[6]* retrieved from such MS space assets is fundamental for conducting NATO's missions. It allows NATO commanders and decision-making staff to access critical and highly sensitive data, facilitate the protection of NATO MS national security as well as properly implement NATO MS foreign policies and operations (Bimfort, 1995, p. 1; Unal, 2019, p. 3).

Following NATO's first *Space Policy* and *Recognition of Space as an Operational Domain* in 2019, NSP2K was replaced in 2020 by a Memorandum of Understanding (MOU) between NATO, France, Italy, the UK, and the US, to ensure fifteen more years of critical MS space capacity to NATO. This includes a wide range of services under supervision of NATO's *Communications and Information Agency* (NCIA), from communication, navigation and intelligence gathering to detecting missile launches and tracking forces worldwide. Additionally, the 2020 MOU aims at strengthening NATO's defence and *deterrence*[7] through heightened redundancy, flexibility, and resiliency, as well as national expertise.

---

[1] *Satellite:* Defined as "a device sent up into space to travel around the earth, used for collecting information or communicating by radio, television, etc." (Cambridge Dictionary, 2020h); computer-enabled objects in orbit "with peripherals and network connections" (Caudill, 2019, p. 4). Thereby, *orbit* is defined as "the curved path through which objects in space move around a planet or star" (Cambridge Dictionary, 2020f).

[2] *Space:* Defined as "the region beyond the earth's atmosphere" (Merriam-Webster, 2020e), involving **outer space**, thus "the physical universe beyond the earth's atmosphere" (Oxford Lexico, 2020), and **aerospace**, thus the "space comprising the earth's atmosphere and the space beyond" (Merriam-Webster, 2020a). *Throughout interviews and literature, those terms are used interchangeably.*

[3] *Satellite Communications (SATCOM):* see Chapter 3.2.1.

[4] *Intelligence*: The product resulting from "the directed collection and processing of information regarding the environment and the capabilities and intentions of actors […] to identify threats and offer opportunities for exploitation by decision-makers" (NSO, 2020, p. 9).

[5] *Internet:* Defined as "the large system of connected computers around the world that allows people to share information and communicate with each other" (Cambridge Dictionary, 2020e).

[6] *Information:* Harnessed from cyber operations support "commanders […] while integrating with other functions, to influence relevant actor perceptions, behaviour, action or inaction and decision making" (NSO, 2020, p. 9).

Space assets consist of three segments: a *space* (uplink, satellites), *ground* (downlink, operational ground stations), and *data* (crosslink) segment (Livingstone & Lewis, 2016, p. 16). This architecture is enabled by *commercially*[8] developed, market-led *cyber*[9] technology, consisting of hardware, software and further digital components. Thereby, cyber and space security are indivisibly linked (Livingstone & Lewis, 2016, p. 3). As stated by Paulauskas "space is unique in that it depends on other domains less than they depend on space with the exception of cyber, which is critical for space for data links" (Paulauskas, 2020, p. 8). NATO recognized *cyberspace*[10] as an operational domain in 2016, acknowledging its inherent similarities with outer space, specifically its lack of geographic boundaries. This marked an operational shift to focus on *mission*[11], and specifically *information assurance*[12], and thus the strategic deconfliction of cyberspace. However, NATO's Standardization Office acknowledges that

> "the assignment of classical operational boundaries in cyberspace is particularly difficult [since] is not only in constant flux but even more importantly, it may be used by anyone for almost any purpose [it] is also distinct in that its underlying physical elements are entirely man-made, which is different from land, air and space, and sea [and] may be managed through manipulation of the domain itself" (NSO, 2020, p. 3).

Due to cyberspace's ubiquitous nature, low-cost *cyberattack*[13] methods and techniques can cause *asymmetric*[14] effects against the technology-dependent Alliance or allied MS. This complicates attribution of *cyberthreat*[15] actors behind attacks, and disguises activities through or in cyberspace. To mitigate any cyber *vulnerability*[16] that may impact control over space assets, reliability of information or *availability*[17] of satellite *bandwidth*[18] is critical to protect the security of NATO missions, and national reliance on space asset-gathered information. If *hackers*[19] were to take control over space assets, consequences would be dire and cause major international, national, and regional security concerns. This may include widespread disruptions or permanent shut-down of space assets, denial of access to their services or even attacks on national *Critical Infrastructure*[20] (CI).

---

[7] **Deterrence**: Defined as "the threat of force in order to discourage an opponent from taking an unwelcome action […] through the threat of retaliation (deterrence by punishment) or by denying the opponent's war aims (deterrence by denial)" (Rühle, 2015, p. 1).

[8] **Commercial:** Intending to make a profit, thus „the exchange and buying and selling of goods or services on a large scale involving transportation from place to place" (Merriam-Webster, 2020b). *Throughout literature, interviews and this thesis, the terms 'commercial' and 'private' sector are used interchangeably. Therefore,* the **private sector** is defined as "the part of an economy which is not controlled or owned by the government" (Merriam-Webster, 2020d).

[9] **Cyber:** Everything "relating to, or involving computers or computer networks" (Merriam-Webster, 2020).

[10] **Cyberspace:** As defined within NATO AJP-3.20, is a "global domain consisting of all interconnected communication, information technology and other electronic systems, networks and their data, including those which are separated or independent, which process, store or transmit data […] a computerized environment, artificially constructed and constantly under development […] largely globally interconnected" (NSO, 2020, p. 3).

[11] **Mission assurance**: In this context the "operational impact of activities in or through cyberspace" (NSO, 2020, p. 5).

[12] **Information assurance**: The "security and defensive posture related to the protection of information and systems" (NSO, 2020, p. 5)

[13] **Cyberattack:** Defined as "disrupting, disabling, destroying, or maliciously controlling a computing environment/infrastructure; or destroying the integrity of the data or stealing controlled information" (CSRC, 2020a); "unlike electronic attacks, which interfere with the transmission of RF signals, cyberattacks target the data itself and the systems that use this data [such as] antennas on satellites and GSs, the landlines that connect GSs to terrestrial networks, and the user terminals that connect to satellites are all potential intrusion points for cyberattack" (Harrison et al., 2019).

[14] **Asymmetric cyberattacks:** Require a "relatively small number or low levels of resources […] by an attacker to cause a significantly greater number or higher level of target resources to malfunction or fail" (Radware, 2020).

[15] **Cyberthreat:** Defined as "any circumstance or event with the potential to adversely impact organizational operations (including mission, functions, image, or reputation), organizational assets, individuals, other organizations […] through an information system via unauthorized access, destruction, disclosure, modification of information, and/or denial of service" (CSRC NIST, 2020).

[16] **Cyber vulnerability:** Defined as a "weakness which can be exploited by a cyberattack to gain unauthorized access to or perform unauthorized actions on a computer system. Vulnerabilities can allow attackers to run code, access a system's memory, install malware, and steal, destroy or modify sensitive data" (UpGuard, 2020).

This threat is aggravated by NATO's mostly unregulated cyber *Space Asset Supply Chain (SASC)*. A *supply chain* is defined as the flow and movements of goods and services and related information between *customers and suppliers*[21] in up- and downstream relationships, from the good's or service's origin to its operation, and final disposal (Pandey et al., 2020, p. 2). NATO's SASC is composed of a broad network of internationally distributed actors from both the public and the commercial sector. NATO and its MS commercially procure space asset *software*[22], equipment and services, which becomes part of NATO's overall toolkit on collaborative efforts. The integration of this commercial SC is not necessarily a risk, however its risen complexity, need for security updates via remote connections and a lack of responsibility for individual component development make space assets vulnerable to cyberattacks (Harrison et al., 2019, p. 40; Livingstone & Lewis, 2016, p. 4). Likely attack vectors would be, for example, the interference with unencrypted data transmission, or attacking Ground Station (GS) antennas, satellites or user software, to monitor or insert false or corrupted data (Livingstone & Lewis, 2016, p. 18).

Reliability on NATO's cyber SASC is additionally threatened by commercial organizations, which are "far less sensitive to security than they are to profit" (Caudill, 2019, p. 4), thus trading off *cybersecurity*[23] measures (Livingstone & Lewis, 2016, p. 25). As market competition on space asset availability and sophistication increases, time for development shrinks and less scrutiny and attention are paid to an individual component's SC security. Moreover, the recent national trend to launch low-cost space assets, including *Commercial Off-The-Shelf*[24] (COTS) technology, augments the complexity of space asset ownership, liability and management (Falco, 2018b, p. 5). Up to this point, there exists no relevant international law, regulation, agreed mechanism or international organization that conceivably constitutes a regulatory basis for the cyber SASC. Resulting cyber vulnerabilities undermine space asset capabilities, confidence in mission-critical data required to reliably analyse strategic stability, and thus NATO credibility and deterrence capability (Unal, 2019, p. 2).

---

[17] **Availability:** means "making sure your data is available [through] correct firewall settings, updating your system regularly, backups of your data, documenting changes" (NASA, 2019).

[18] **Bandwidth:** Describes the "maximum data transfer rate of a network or internet connection. It measures how much data can be sent over a specific connection in a given amount of time" (TechTerms, 2020).

[19] **Hacker**: "Unauthorized user who attempts to or gains access to an information system" (CSRC, 2020b)..

[20] **Critical Infrastructure:** Are "assets, systems, and networks, whether physical or virtual, are considered so vital […] that their incapacitation or destruction would have a debilitating effect on security, national economic security, national public health or safety, or any combination thereof" (CISA, 2020a).

[21] S*upplier:* Defined as "a company, person, etc. that provides things that people want or need, especially over a long period of time" (Cambridge Dictionary, 2020i). In contrast, a **customer,** also *acquirer/ purchaser*, is a company or person "who buys goods or a service" (Cambridge Dictionary, 2020g).

[22] **Software**: A set of "instructions, data or programs used to operate computers and execute specific tasks. Opposite of **hardware**, which describes the physical aspects of a computer" (Akoto, 2020).

[23] **Cybersecurity**: Defined as "the application of security measures for the protection of communication, information, and other electronic systems, and the information that is stored, processed or transmitted in these systems with respect to confidentiality, integrity, availability, authentication and non-repudiation" (NSO, 2020, p. 4).

[24] **Commercial Off-The-Shelf (COTS) technology:** "Hardware and software IT products that are ready-made and available for purchase by the general public" (NIST, 2020a).

In consequence, the *integrity*[25] and correctness of data retrieved from this cyber SASC must be mandated for NATO officials to assure NATO's ability to operate its missions. Whilst there has been much research on the correlation of cyberspace and outer space, none has focused specifically on such cybersecurity vulnerabilities being enabled by security flaws in the NATO SASC (Bimfort, 1995; Falco, 2018b, p. 11; Harrison et al., 2019, p. 40; Livingstone & Lewis, 2016, p. 5; NIST, 2020, p. 34; Paulauskas, 2020, p. 8; Unal, 2019, p. 8). Furthermore, both literature and NATO organizational publications leave unclear the patterns of security assurance and organizational management of this SC. To fully understand the challenges to integrity of mission-critical data retrieved from space assets, it is important to gain a more complete picture of NATO's cyber SASC.

## 1.1 Research Question and Outline

This thesis aims to identify cybersecurity gaps along NATO's SASC in order to help readers to understand if and how such gaps threaten the integrity and security of NATO's missions. Additionally, it aims to develop policy recommendations to legitimize and streamline this SC between the private and the public sector, and thus to reduce cyber vulnerabilities. The *research goal* is to support the rise of NATO's organizational resilience against cyberattacks, so that NATO can better protect its missions against adversaries in cyberspace. Derived from the above introduced research problem, the *research question* guiding this thesis is twofold:

>    a) *What are current cybersecurity gaps along NATO's global SASC; and*
>
>    b) *How can NATO and its allied MS gain greater control over such gaps to safeguard the supply of NATO mission-critical information?*

This thesis is outlined as follows: *Chapter 1* presents an introduction to the research question and research methodology; *Chapter 2* discusses relevant academic literature and the consequential research gap. Subsequently, *Chapter 3* highlights the most actual use cases of space assets for NATO, and identifies contract evolution whilst examining the division of responsibilities and according mechanisms through the entire NATO system. *Chapter 4* presents crucial space asset cybersecurity gaps, describes malicious actors and motivations, provides insights into organizational negligence and identifies existing regulation. *Chapter 5* introduces NATO's cyber SASC and details cyberthreats and vulnerabilities. *Chapter 6* suggests policy recommendations to NATO, MS and commercial SC partners. For convenience, *Chapter 3 to 6* are each concluded by a summary. Finally, *Chapter 7* outlines the main conclusions and identifies both limitations to the study and recommendations for further research.

---

[25] **Integrity**: "making sure no bits were lost, making sure no web address was changed, and even making sure that unauthorized people cannot change your data" (NASA, 2019).

## *1.2 Research Design and Methodology*

### *Methodological Approach*

The research design and methodology of this thesis seeks to assess the current state of cooperation patterns along NATO's SASC, and how they are susceptible to cyber vulnerabilities. With the research focus outlined above, it becomes clear that the thesis positions itself in an internationally connected field. Thus, it aspires not only to investigate changes and dynamics within NATO, but also to link these findings to international SC partners across the private and commercial sector. Thereby, the analytical framework for this analysis is based on the overlapping scientific fields of cybersecurity, space asset and SC research. These are interrelated both in theory as well as in practice, as will be shown throughout the analysis. To succeed in this pursuit, methodological concerns are central to organise and focus the analysis. Adopting a structural, qualitative data perspective, the analytical strategy is based on a field ontology, to identify and study current collaboration structures. The following section elaborates this thesis's methodological approach, applied to illustrate such interconnections whilst providing an explanation of the methods used for the collection of data as well as research analysis.

It is my ambition to show how and why NATO faces cybersecurity vulnerabilities along its cyber SASC. However, as the nexus between cybersecurity, space assets and SC is so far consistently under-researched, I chose *in-person interviews with high-level representatives from a balanced mixture of most relevant public, private, and academia backgrounds* to be able to collect and analyse primary data for the purpose of this study. I am confident that this method provides an appropriate approach to answering the problem statement and research questions, allowing to incorporate the expertise of current and multi-layered insights by well-respected and well-experienced practitioners and scholars. Additionally, I complemented this approach by precedingly and continuously analysing secondary data within a thorough, academic literature review (*see Chapter 2*).

### *Methods of Qualitative Data Collection*

As outlined above, I decided to conduct a *qualitative research study* by holding 19 interviews with representatives from the following organizations and companies listed below:

Table i: Complete List of Interviewees

| NAME | Occupation | Organization/ Company | Sector |
|---|---|---|---|
| *Ashley BANDER,*<br><br>*John GALER,*<br>*Jason TIMM* | Director of Space Systems<br>Assistant VP, National Security Space<br>Assistant VP, National Security Policy | Aerospace Industries Association | *Private* |
| *Carlo CONTI* | Senior SATCOM Technician | NATO Communications and Information Agency (NCIA) | *Public* |
| **Crystal LISTER** | Co-Founder, Senior Director, Cyber | Global Professional Services Group (GPSG) | *Private* |
| **David LIVINGSTONE** | Innovation Associate Fellow, Cyber and Space Security | Chatham House | *Academia* |
| **Emma PHILPOTT** | Chief Executive Officer | IASME Consortium Ltd | *Private* |
| **Eric YINGST** | Cybersecurity Advisor, Senior Critical Infrastructure Protection Advisor | US Department of Defence (DOD) Defense Information Systems Agency (DISA) | *Public* |
| **Erin MILLER** | Executive Director, Operations | US National Cybersecurity Center (NCC), Space Information Sharing and Analysis Center (ISAC) | *Public* |
| **Frederic URBAN** | Head of Business Development AMBER™ | Horizon Technologies | *Private* |
| **Harrison CAUDILL,** | Founder | Orbital Security Alliance (OSA) | *Private-Academia* |
| **Gregory FALCO** | Postdoctoral Scholar Assistant Research Professor (incoming Assistant Professor) | Stanford University Johns Hopkins University | *Academia* |
| **Henry HEREN, Tim VASEN** (Separately) | Space Subject Matter Expert (SME) Space SME | NATO Joint Air Power Competence Centre (JAPCC) | *Public* |
| **Gil BARAM** | Head of Research Team, | Tel Aviv University, Cyber and Space Center of Excellence | *Academia* |
| **Laetitia ZARKAN** | Consultant, Space Security, Weapons of Mass Destruction and Other Strategic Weapons | United Nations Institute for Disarmament Research (UNIDAD) | *Public-Academia* |
| **Michael WIDMANN** | Commander, Strategy Branch Lead | NATO Cooperative Cyber Defence Centre of Excellence (CCDCOE) | *Public* |
| **Paul WELLS** **Roy SIELAFF** | VP, Government Satellite Communications Director Governmental SATCOM | LuxGovSat S.A. | *Private-Public* |
| **Steve LEE** | Manager, Aerospace Cybersecurity Program | American Institute of Aeronautics and Astronautics (AIAA) | *Public* |
| **Steven HILL** | Former NATO Chief Legal Advisor and Director (Head of Office) | NATO Office of Legal Affairs | *Public* |

Similar to the selection of appropriate academic literature, *organizations and companies* were selected based on

> *a) their organizational proximity to NATO along NATO's SC,* such as LuxGovSat or Horizon Technologies,
>
> *b) their organizational or managerial functions within NATO itself such as NCIA or JAPCC*, or
>
> *c) their NATO-significant company working output,* for instance legislation considering the nexus of SC, cybersecurity and space assets, such as OSA or AIA.

*Interviewees* were selected based on the research-significant positions and functions held within the respective organization or company. This repeatedly included positions such as Director, Vice President, Advisor or Strategist amongst fields such as Government Satellite Communications (*see 3.1.1*), Cyber, Space and National Security or Aerospace Cybersecurity. Whilst 17 interviews were conducted with interviewees separately, AIA and LuxGovSat participants decided to collaboratively take part in the interviews to complement each other's knowledge.

Five major fields determined the interview guidelines and aligned with the research goal of this thesis *(find the complete interview guideline in the Annex)*:

> *a) General Introduction;*
>
> *b) NATO and military use of space;*
>
> *c) Space asset cybersecurity;*
>
> *d) The nexus between space assets' cybersecurity and supply chain;* and
>
> *e) Policy recommendations.*

However, as interviews were conducted in a *semi-structured* manner, specific questions based on the preceding literature research were posed in response to the general direction of the interviewee's answers and field-specific knowledge. Due to the ongoing pandemic and geographically widespread locations of participants, interviews were exclusively conducted via online video conferencing services and lasted typically from one to two hours. All interviews were, with the interviewees' consent, recorded to ease later analysis. Due to frequently sensitive information, direct transcripts remain confidential, whilst all direct citations presented throughout the subsequent analysis were verified and agreed upon by the respective interviewees.

**Qualitative Methods of Analysis**

After following the steps introduced above, the gathered interviews were transcribed. Each field was examined with the aim to gain an understanding of the interviewee's perception and specific knowledge on the cyber SASC. Data was then processed and analysed using content analysis, categorizing, and subsequently discussing the meaning of phrases, words, and sentences, complemented by thematic analysis. This approach was chosen to allow for subsequent close examination of gathered data, identify common themes and patterns across interviews, and thus deduce relevant cybersecurity gaps and policy recommendations.

## *1.3 Theoretical Classification*

Turning to International Relations (IR) theory to support the interpretation and broader generalization of this study's results, *Neoliberal Institutionalism* stands out as possible theoretical framework put forward by scholars like Joseph Grieco (1988), Robert Jervis (1999), Arthur Stein (2008), and Quddus Snyder (2013). By applying this theoretical framework, the importance of commercial factors for the formation of NATO and national MS foreign policy becomes more pronounced. As stated by Snyder, *classical Liberal IR Theory* assumes that actors seek to maximize their gain, however it ignores systemic competitive pressure at the commercial level, as well as the fear to fall behind in the struggle to obtain a relative power upper hand (Stein, 2008, pp. 201–221). In contrast, the neoliberal institutionalist approach follows the principal economic and political role that international institutions place on IR and among states (Grieco, 1988, pp. 485-507). It focuses on the lack of cooperation in international political economy, and considers international institutions' main purpose to be a mediator and promote cooperation to resolve such global, political, and economic issues.

NATO and its MS showcase how cyber SASC gaps originate from commercial competition, causing pressure to maintain relative power, credibility and thus the ability to deter potential adversaries. As it is my goal to highlight how NATO and its MS may gain greater control over such gaps to safeguard the supply of NATO mission-critical information, I aim to contextualize subsequent findings with the theoretical assumptions of neoliberal institutionalism. However, as neoliberal institutionalist theory has rarely been applied by academics to the research nexus of cybersecurity, military space assets and SC, there is a significant gap of relevant, up to date scholarship. Therefore, this study is conceptualized as an expressively empirical, inductive study, that maximizes its implications for practical usage, whilst at the same time deriving systematic, transferable findings from field-specific academic literature.

# 2 LITERATURE REVIEW

This chapter seeks to provide an overview of recent scholarly research on mission-critical data that is threatened by cybersecurity gaps in NATO's SASC. It does so by focusing on two main thematic topics becoming apparent by analysing the current academic debate:

a) the *nexus between cybersecurity and space assets*, and

b) the *nexus between cybersecurity and space asset supply chains* (SASC).

To this end, this chapter offers a first insight into current SASC security challenges, before subsequently identifying the research gap underlying the aspiration for this study.

## 2.1 *Nexus Cybersecurity – Space Assets*

The main concern of scholars studying the relationship between cybersecurity and space assets is that consequences of cyberattacks on space assets could became a vital threat to national security, whilst NATO MS become increasingly dependent on such assets (Moon, 2017, pp. 8-10; Paulauskas, 2020, pp. 4-7). Thereby, scholars generally agree that three basic conditions allow for a lack of cybersecurity and thus for cyberattacks to unfold:

*The use of COTS-Technology:* As pointed out above, there is considerable debate about increasing private and public sector use of COTS hardware and software components (*such as for CubeSats, see Chapter 4.2.2*) to allow for "high quality commercial components [and a] considerable reduction of [governmental] qualification time and costs" (European Council, 2018). These papers suggest that such COTS components could be hacked and send malicious commands (Akoto, 2020; Falco, 2018; Falco, 2018b; Koch & Golling, 2016; Ziolkowski et al., 2013). Scholars further agree that the wide availability of COTS technology allows malicious actors to analyse them for vulnerabilities, especially by exploiting built-in *open-source* [26] technology to insert *backdoors* [27] and other vulnerabilities into a satellite's software.

*Attributing and understanding interests of adversaries:* According to Tucker (2019), the vast volume of data received on a day-to-day basis from space assets makes it difficult to determine if and to what extent the assets have been compromised by an adversary. Livingstone & Lewis (2016) suggest that this dilemma is further complicated through the fundamental difference between the rapid speed of hacker attacks, contrasting the decelerated reaction speed of governments and military entities. This enables cyber criminals to corrupt the integrity and accuracy of data, whilst facing a low probability of being discovered. Finally, Waterman (2019) suggests that in order to understand the impact of such attacks and bring up governmental reaction spees, technical and political collaboration must be increased and cyber-specific terminology be better explained to the political community.

---

[26] ***Open-source technology*:** Software in which "the source code used to create the program is freely available for the public to view, edit, and redistribute" (ESRI, 2020).
[27] ***Backdoors:*** Any method "by which authorized and unauthorized users are able to get around normal security measures and gain high level user access on a computer system, network, or software application […] to steal data, install additional malware, and hijack devices" (Malwarebytes, 2020a)

*Decade-long lifespan of space assets and old software:* Several recent papers discuss the decade-long duration of governmental space missions, making space assets unique in their lack of physical access (Falco, 2018; Koch & Golling, 2016). Once being deployed, operators and builders are unlikely to regain subsequent, physical access to the asset (Caudill, 2019, p. 10). Therefore, there is a vital scholarly debate on how to bring commercially trusted computing software for defence and aerospace applications at the speed of space asset technology innovations (Bailey, 2019; Bratton, 2020). Scholars also suggest that security problems unfold through the hindered supply of updated and spare parts and through products obsolescence over time, thereby creating outdated, *unpatched[28] legacy[29]* systems. Furthermore, there is considerable debate about whether or not GS running often outdated or common operational computer systems facilitate infiltration by hackers and allow the denial and disruption of vital information (Tucker, 2019).

## 2.2 Nexus Cybersecurity – Space Asset Supply Chain

Findings in the academic literature on the relationship between cybersecurity and space asset CS suggests that the multiplied involvement of suppliers and manufacturers in delivering mission-critical data for NATO opens up numerous cyber intrusion threats (Falco, 2018b; Koch & Golling, 2016):

*Outsourcing of day-to-day operational management to private companies:* There is considerable debate amongst scholars about the practice of outsourcing day-to-day management to private company services, such as software relying on space assets (Akoto, 2020; Falco, 2018b). Due to the highly technical nature of satellites, the increased involvement of multiple component manufacturers opens up an uncontrollable number of weaknesses to infiltrate the SC system, and subsequently the data reaching NATO (Koch & Golling, 2016). In short, scholars generally agree that potential threats include the substitution of hardware or software components by counterfeit parts, software vendors introducing security backdoors, or the execution of invisible changes during the manufacturing of satellite components (Falco, 2018a; Unal, 2019). Scholars also express concern that space assets are not owned by the same organizations that manage their infrastructure, thus leaving open the scramble for liability if they are attacked (Bailey, 2019; Sachdeva, 2019). This includes little control over the specific technicians or software engineers assigned to the production of a component, leaving ownership of space assets unclear (Falco, 2018a; Falco, 2018b).

*Lack of internationally binding regulations:* A wide range of scholars agree on the fundamental lack of cybersecurity standards and regulations along the public-commercial satellite supply chain (Akoto, 2020; Falco, 2018; Sachdeva, 2019). To this date, there exists no coherent international organization regulating cybersecurity in space implementing and cybersecurity controls on the SASC. Scholars agree that the missing involvement of governments in the commercial space sector, and in the development and regulation of cybersecurity standards for space assets lead to a lack of clarity on who bears responsibility and liability for cyber breaches (Falco, 2018a; Falco, 2018b). Therefore, a major academic concern holds that this practice breeds complacency and hinders efforts to secure these systems along the SC. However, recent research shows that current private sector attempts to increase structure and regulation of the commercial SASC increase resistance against cyber intrusion on space asset components (Livingstone & Lewis, 2016). Scholars agree that, as companies are constantly seeking rapid market advantage through innovation and exploitation of business opportunities, overregulation and rigid control of the space market on a national level typically causes system developers to seek ways to bypass official regulation, such as moving to another jurisdiction or country.

---

[28] *Patch:* A patch is "a software update compris[ing] code inserted (or patched) into the code of an executable program. Typically, a patch is installed into an existing software program [to] Address new security vulnerabilities, Address software stability issues, Upgrade the software"

[29] *Legacy system*: Defined as "outdated computing software and/or hardware that is still in use" (Talend, 2020; *see 4.2.3.3*);

*Lack of appropriate cyber SC risk concerns:* As discussed by Akoto (2020), commercial market forces pressure companies along the SASC into cutting costs and speeding up production and development at the cost of suppressing cybersecurity concerns. Thus, a major concern amongst scholars is a SC security failure, due to a lack of reporting of incidents, uniform language, streamlined processes and responsiveness of suppliers (Akoto, 2020; Caudill, 2019; Falco, 2018; Falco, 2018b; Koch & Golling, 2016; Ziolkowski et al., 2013; Yuval, 2019). Furthermore, an implied risk includes that the awareness of the need for security is not present in commercial SC's, and that this needs to be overcome by education and the use of security experienced space expertise. This is aggravated by the ongoing Coronavirus Disease (COVID) pandemic: according to Lister (2020), further SC volatilities are exposed within the SASC, such as product shortages or lagging timelines because of COVID. Finally, Vasen (2020) cites significant understaffing in handling such SC concerns within NATO, slowing down the management of space asset retrieved data.

## *2.3  Research Gap and Hypotheses*

This thesis aims to identify current cybersecurity gaps along the global cyber SASC and highlight policies which enable NATO and its MS to gain greater control over such gaps and safeguard the supply of mission-critical information. As has been shown, contemporary academic debate considers both the relationship between space assets, as well as the relationship between cybersecurity and such asset SC, to broadly relate to the diminished security of NATO mission-critical data to these fields. Whilst there has been much research on the respective nexuses of cyberspace, outer space and SC, *there exists yet no specific academic literature and analysis on NATO's own cyber SASC.* Additionally, both NATO official publications and academic literature leave the management and responsibilities throughout NATO's cyber SASC vague. However, it is key for NATO and its MS to fully understand integrity challenges to the mission-critical data retrieved from its space assets in order to secure both the organizational system's cybersecurity and the physical security of personnel on the ground. Hence, this thesis aims to fill this research gap, by providing a broader understanding of cooperation challenges along the NATO cyber SASC, and thus lifting this veil of obscurity. Based on the above outlined research question and in accordance with the existing literature (*see Chapter 2*), the following two hypotheses have been developed:

(H1) *Cybersecurity gaps along NATO's SASC are likely to originate from a) outsourcing of space capabilities to private companies, b) the use of outdated software and c) an almost complete lack of international laws and regulations.*

(H2) *A safer supply of NATO mission-critical information is likely to be ensured through the establishment of a strict international legal and operational framework, establishing transparent development, design, management and ownership of space assets.*

In laying the foundation for verifying or disproving these hypotheses, the subsequent *Chapter 3* aims to analyse current use cases of space assets for NATO as well as surrounding international regulation and contract evolution.

# 3 NATO USE OF SPACE

## 3.1 Fields of Use

Nowadays, NATO's most advanced technological systems inevitably depend on space assets (Moon, 2017, p. 8). They grant military commanders a comprehensive picture of the situation on the ground, perform surveillance near real-time, and thus allow for a wider range of missions and operations. In 2016, NATO's Standardization Office (NSO) issued an *AJP-3.3: Allied Joint Doctrine for Air and Space Operations, Edition B Version 1,* acknowledging that

> "while NATO neither owns nor directly operates any *spacecraft*[30], NATO's combined command and force structure depends on space-based capabilities across the spectrum of operations. *Space situational awareness*[31] enables the efficient use and protection of those space-based systems" (NSO, 2016, p. 5-4).

Thereby, NATO uses space assets exclusively for Defensive Space Control (DSC), which mean "passive and active measures [are] precautionary and proactive in order to prevent the adversary from disrupting NATO operations in all domains" (NSO, 2016, p. 5-7). The following subchapters aim at specifying the current major use cases for space assets by NATO, as identified throughout the preceding field research.

### 3.1.1 Satellite Communications (SATCOM)

*Satellite Communications* (SATCOM) are the most important function of space assets for NATO. It provides "NATO commanders with the ability to establish or augment telecommunications in remote regions that may lack suitable terrestrial infrastructure" (NSO, 2016, p. 5-7). NATO defines SATCOM as

> "instant global connection to NATO communication infrastructure and to the nations, transmission of critical intelligence, the ability to tie sensors to shooters, and establish survivable communications in austere locations with limited or no infrastructure […] including governmental, military, civil and commercial SATCOM systems and applications" (NSO, 2016, p. 5-7).

Communication satellites used for SATCOM provide global broadband internet, mobile services, voice communications, television broadcasts, and data transfer services for military, civil, and commercial users (US DIA, 2019, p. 7). As stated by Yingst (2020), the military uses SATCOM to ensure data integrity and effective communication among warfighters and commanders deployed in the air, on the ground or on water. Additionally, Wells and Sielaff point to the growing importance of SATCOM for military *Beyond Line-of-Sight Communications* (Wells & Sielaff, 2020), and thus the ability to act beyond the boundaries of adversaries' infrastructure and borders, which restricts constant connectivity.

---

[30] **Spacecraft:** defined as "a general term for objects launched into space—e.g., Earth-orbiting satellites and space probes, experiment capsules, the orbiting modules of some launch vehicles (e.g., the US space shuttle or the Russian Soyuz), and space stations" (Encyclopedia Britannica, 2020). *Throughout literature, the terms spacecraft and satellites are used interchangeably.*

[31] **Space Situational Awareness (SSA):** defined by NSO as "the requisite current and predictive knowledge of the Space environment and the operational environment and their effect on NATO operations […] This includes knowledge about space systems capabilities, operational readiness, limitations, as well as environmental conditions, events, threats and activities (both current and planned) in, from, toward or through Space […] a complement of space support to NATO operations in order to support all levels of planning, decision making and operation execution across the full spectrum of NATO operations in all domains" (NSO, 2016, p. 5-6).

### 3.1.2  Intelligence, Surveillance, Reconnaissance (ISR)

As stated by NATO's Standardization Office (NSO), *Intelligence, Surveillance and Reconnaissance (ISR)* includes "monitoring areas of interest from space helps provide information on adversary location, disposition and intent; aids in tracking, targeting, and engaging the adversary; and provides a means to assess these actions through tactical battle" (NSO, 2016, p. 5-6). NSO precises that for NATO, "space-based sensors may support: the cueing of ISR systems operating in the JOA [*Joint Operations Area*], damage assessment, and operational combat assessment. It also provides situational awareness, warning of attack, and feedback" (NSO, 2016, p. 5-6). ISR satellites are key enablers for evidence-based decision-making, countering *false information* [32], recognition and attribution (Vasen, 2020b).

Furthermore, US DIA states that "ISR satellites provide remote sensing data, which include data on the Earth's land, sea, and atmosphere [and] support a variety of military activities by providing signals intelligence [...] battle damage assessments, and military force disposition" (US DIA, 2019, p. 7). ISR satellites allow for *Terrestrial and Space Environmental Monitoring* to contribute "data on meteorological, oceanographic and space environmental factors that might affect military operations [through] forecasts, alerts, and warnings" (NSO, 2016, p. 5-6). This is done so to "provide joint *force planners* [33] with current, multi-spectral information on subsurface, surface, and air conditions (e.g., traffic capability, beach conditions, vegetation, and land use) [… and] to avoid adverse environmental conditions while taking advantage of other conditions to enhance operations" (NSO, 2016, p. 5-6). Thus, ISR is critical for both the military and civilian side and forms the operational backbone of the military.

### 3.1.3  Positioning, Navigation & Timing (PNT)

*Positioning, Navigation and Timing (PNT)* is defined as "data that enable civilian, commercial, and military users to determine their precise location and local time" (US DIA, 2019, p. 7). For NATO, it is

> "vital to military operations and a key enabler for a host of mission types, Command, Control, Information and weapon systems and platforms. The provision of accurate location and time of reference is a prerequisite for synchronized, precise, network enabled operations in all domains" (NSO, 2016, p. 5-7).

---

[32] **Misinformation:** defined as "false or inaccurate information, especially that which is deliberately intended to deceive" (Cambridge Dictionary, 2020c).

[33] **Space Force Enhancement:** defined by NSO as "the exploitation of space based products and services that contribute to maximizing the effectiveness of military operations in all domains" (NSO, 2016, p. 5-6).

This includes enabling the movement of heavy equipment across land, air or sea using the US space asset-enabled *Global Positioning System*[34] (GPS) or other navigation systems such as the European Union's (EU) *Galileo* [35] or *European Geostationary Navigation Overlay Service (EGNOS)* [36] navigational systems, at any time. Thus, navigation satellites used by the military are vital for NATO communications and transportation networks, especially when navigating in maritime areas without terrestrial orientation (Yingst, 2020).

### *3.1.4 Deterrence*

Following the field research conducted for this thesis, one further reason for NATO activity in space is the signalling of capability to *deter*. Paulauskas points out that NATO aims at a 360-degree approach to deterrence "to respond to any threats from wherever they arise […] *And space is as 360-degree as it gets*" (Paulauskas, 2020, p.7).

On the whole, NATO's MS currently control over fifty percent of active space assets globally, giving it an unprecedented advantage in past and present operations. However, therefore, NATO relies heavily upon those capacities, making them a significant vulnerability to the Alliance. This threatens NATO's ability to counter or deter aggression (Paulauskas, 2020, p. 9): if an adversary conducts a major surprise attack on NATO space assets, reliability and credibility of NATO operations will be undermined and hence may be precursor to a major global conflict. As stated by the United Nations Office for Outer Space Affairs (UNOOSA), the

> "growing dependence on space-based platforms and the increasing strategic value of outer space raises the likelihood that a terrestrial conflict could spill over into an already fragile space environment, with potentially devastating consequences" (UNOOSA, 2017, pp. 2-3).

Furthermore, NATO Secretary General (SG) Jens Stoltenberg warns that "some nations – including Russia and China – are developing anti-satellite systems which could blind, disable or shoot down satellites" (Posaner, 2020). Thereby, US DIA states that

> "while China and Russia are developing counterspace weapons systems, they are promoting agreements at the United Nations that limit weaponization of space. Their proposals do not address many space warfare capabilities, and they lack verification mechanisms, which provides room for China and Russia to continue to develop counterspace weapons" (US DIA, 2019, p. 7).

The space-related military programmes of these countries led to a worrisome trend of an ongoing arms-race and weaponization of space (Baram & Wechsler, 2020, p. 1). Russia has already reportedly jammed GPS signals during NATO exercises, where troops majorly rely on GPS as a navigational tool, whilst both North Korea and Iran developed indigenous space asset and launch capabilities. Meanwhile, India tested its first anti-satellite (ASAT) weapons in March 2019, showing the magnitude of this arms race. In July 2020, the US blamed Russia for getting too close to its satellites. In response, France and the US established specific space commands in 2019 through the *Armée de l'Air et de l'Espace Française* and the *US Space Force (USSF),* respectively.

---

[34] *Global Positioning System (GPS):* GPS is a "US-owned utility that provides users with positioning, navigation, and timing (PNT) services […] The US Air Force develops, maintains, and operates the space and control segments" (GPS, 2020).
[35] *Galileo*: Europe's *Global Navigation Satellite System (GNSS)*, thus providing improved "positioning and timing information with significant positive implications for many European services and users" (GSA, 2011a).
[36] *European Geostationary Navigation Overlay Service (EGNOS):* Europe's "regional satellite-based augmentation system (SBAS) that is used to improve the performance of global GNSSs, such as GPS and Galileo. It has been deployed to provide safety of life navigation services to aviation, maritime and land-based users over most of Europe" (GSA, 2011b).

## 3.2  Contract Evolution

To ensure access to services of allied MS's space assets, NATO developed a series of key contracts over the past decades. The following subsections will outline those specific contracts, with acquired space asset capabilities being concluded in *3.2.6: Overview*.

### 3.2.1  1966: SATCOM 1

In 1966, aligned with the development of the UN Outer Space Treaty, NATO began to investigate a potential cooperation with the US for the development of a SATCOM-cooperation programme (NATO, 2020a, p. 1). For missions, NATO stated that it required "availability and access assurance [as well as] anti-jamming features and [cybersecurity] hardening [which] lead to the use of Military-specific SATCOM (MILSATCOM)" (Plachecki, 2015, p. 4). In 1967, the SATCOM project was born, allowing NATO to install and operate two space asset *Ground Segments* (GS) for communication with a US satellite. A GS consists of

> "all the ground-based elements of a spacecraft system used by operators and support personnel, as opposed to the space segment and user segment [including] Ground (or Earth) stations, which provide radio interfaces with spacecraft; Mission control (or operations) centres, from which spacecraft are managed; Ground networks, which connect the other ground elements to one another; Remote terminals, used by support personnel; Spacecraft integration and test facilities; [and] Launch facilities" (Elbert, 2014, p. 141).

Initially seven countries signed a MOU which sponsored this programme: Belgium, Canada, Germany, Italy, the Netherlands, the UK and the US. Furthermore, Denmark, Greece, Norway, Portugal and Turkey participated in installing these GS. In total, eight SATCOM satellites were launched, with the first two satellites *PR/CP(70)2* and *PR/CP(71)1* respectively launched in 1970 and 1972, and *NATO IVA* and *NATO IVB*, the final two satellites, launched in 1991 and 1993. Those NATO-owned satellites were provided until 2004 (Plachecki, 2015, p. 4).

### 3.2.2  2005: NSP2K

Between 2005 and 2019, a new programme - *NATO SATCOM Post-2000 (NSP2K)* - replaced SATCOM1, expressing the paradigm shift that "space segment capabilities [are] no longer NATO-owned and operated but provided through national MILSATCOM capabilities" (Plachecki, 2015, p. 5). Whilst replacing "the two NATO-owned and -operated NATO IV communications satellites, which stopped their operational services in 2007 and 2010" (NATO, 2011), NSP2K enables additional and updated capability access to MS SATCOM assets "which is important as NATO forces take on expeditionary missions far beyond the Alliance's traditional area of operations" (NATO, 2011). The *SATCOM Capability Package CP5A0030* underlying NSP2K was approved as a MOU in 2001 between

a) France, providing its SATCOM system *SYRACUSE: Système de radiocommunication utilisant un satellite*;

b) the UK, providing *Skynet*; and

c) Italy, providing *SICRAL: Sistema Italiano per Comunicazioni Riservate ed Allarmi.*

NATO itself maintains seven terrestrial SATCOM GS (SGS), being currently reduced to four SGS (Conti, 2020). Furthermore, several projects for transportable, deployable SGS's have been implemented, working as an interface between the SGS's left. Those "terrestrial SATCOM anchor stations, transportable [mobile] space asset ground terminals and equipment" (NSO, 2016, p. 5-4) are managed by two single NATO *signal* [37] battalions (Vasen, 2020b). However, CP5A0030 allows NATO commanders to add complementary, crucial access to military *Ultra-High Frequency* [38] (UHF) band and *Super High Frequency* [39] (SHF; X-Band) band (NATO, 2011). The overall cooperation between emerging NATO and MS requirements is managed by NATO's *Joint Program Management Office* (JPMO) in Paris, France.

### 3.2.3  2016: NATO - Luxembourg Cooperation Agreement

In April 2016, NATO began cooperation with the Luxembourg government on its provision of SHF and commercial *Ku-Band* [40] as a *Contribution-in-Kind* [41] (CIK). This service allows for enhanced *Alliance Ground Surveillance* (AGS; Plachecki, 2015, p. 4), which is the

"protection of ground troops and civilian populations, border control and maritime safety, the fight against terrorism, crisis management and humanitarian assistance in natural disasters [by] giv[ing] commanders a comprehensive picture of the situation on the ground" (GOVSAT, 2016).

Therefore, NATO's *Communications and Information Agency* (NCIA) cooperates with the company LuxGovSat in "a public-private joint venture between the Luxembourg government and SES, the world-leading satellite operator […] dedicated entirely to governmental and institutional users" (GOVSAT, 2019). In addition, the Luxembourg government provides secured management of such commercial MILSATCOM Ku-band capacities and SHF (X-band) through LuxGovSat.

---

[37] ***NATO Signal Battalions:*** Provide „communications and information systems facilities and services to all NATO deployed Headquarters […] sets up transmission media by establishing, operating, and maintaining two independent Command and Control and Technical Control Centers (C2TCC) […] to administer the deployed LAN and provides support and maintenance services up to field level; provides support to testing of new Deployable communication equipment in support of any operation or exercise within NATO's area of responsibility" (NATO NCISG, 2020).

[38] *Ultra-High Frequency Band (UHF):* Designation by the International Telecommunication Union (ITU) for radio frequencies (RF) in the range between 300 megahertz (MHz) and 3 gigahertz (GHz); "besides their use in television broadcasting, UHF waves are utilized in ship and aircraft navigation systems and for certain types of police communications [as well as] radio communications between spacecraft and Earth-based tracking stations" (Britannica, 2013). *Frequency* is "in physics, the number of waves that pass a fixed point in unit time; also, the number of cycles or vibrations undergone during one unit of time by a body in periodic motion (in physics, motion repeated in equal intervals of time)" (Britannica, 2020), whilst *wavelength* describes the "distance between corresponding points of two consecutive waves. "Corresponding points" refers to two points or particles in the same phase—i.e., points that have completed identical fractions of their periodic motion" (Britannica, 2020b).

[39] *Super High Frequency Band (SHF, X-Band):* ITU designation for RF band from 3 and 30 GHz; "given their ability to carry enormous amounts of data, super high frequency waves are used for relaying broadcast programs, radar (weather, etc.), SATCOM and satellite broadcasting, and other applications" (DKK, 2001).

[40] *Ku-Band:* ITU designation for microwave frequencies from 12 to 18 GHz. Thereby, "radio waves have wavelengths of 1 m [meter] up. [ [in contrast,] microwaves have wavelengths of 1 mm (millimeter) to 1 m" (Mathsisfun, 2020).

[41] *Contribution-in-Kind (CIK):* Divided between direct and indirect: "Indirect – or national – contributions are the largest and come, for instance, when a member volunteers equipment or troops to a military operation and bears the costs of the decision to do so. Direct contributions are made to finance requirements of the Alliance that serve the interests of all 30 members - and are not the responsibility of any single member - such as NATO-wide air defence or command and control systems" (NATO, 2020b).

### 3.2.4  2019: NATO Space Policy

NATO did not have any specific space policy, mandate, or concept until 2019 (Paulauskas, 2020, pp. 5-7). The need for a NATO Policy on Space, was firstly officially recognized at the 2018 Brussels Summit, being acknowledged as "a highly dynamic and rapidly evolving area, which is essential for the Alliance's security" (NATO, 2020d). Driving reasons for the establishment of the policy included (Hill, 2020):

- The upcoming termination of NSP2K, hence the need for a policy lead to provide legal support to the use of space asset capabilities;

- Increasing adversarial activity in space, with NATO lagging behind due to legal framework challenges in aligning MS positions; [42]

- Increasing dependence on NATO's cyber and space infrastructure for operations;

- The rise of sensible ISR exchange within NATO for better imagery and data analysis through a newly set up *NATO Joint Intelligence, Surveillance and Reconnaissance (JISR)* division.

Subsequently, at NATO's June 2019 meeting of Defence Ministers, NATO's first *Space Policy* was adopted, serving to provide an overarching approach to Space for NATO. Subsequently, in December 2019, NATO separately *recognized Space as an Operational Domain, alongside air, land, sea and cyberspace.* Heren points out that these new policies are a pathway to reduce inutile replication through MS individually providing data, underlines the need for international collaboration on space security, facilitates administrative processes, and thus more rapidly the provision of operational data (Heren, 2020).

### 3.2.5  2020: MOU NATO, UK, US, France, and Italy

Currently, most of the space asset support for NATO missions is provided by the US, France, UK, Italy, and Germany (Vasen, 2020b). Thereby, the national or commercial company owning a space asset voluntarily determines the extent of its support to NATO. A more binding MOU was set up in 2019 between NATO, the US, UK, France, and Italy, affirming *Capability Package CP9A0130* (NCIA, 2020, p. B-1). CP9A0130 will provide NATO with fifteen more years of SHF, UHF and *Extremely High Frequency* [43] (EHF) band (Plachecki, 2015, p. 4; Space Intel Report, 2018). CP9A0130 considers

- *Future SATCOM requirements*, including protected core capabilities such as higher bandwidth and smaller GS, EHF for highly protected communication and increased coverage flexibility, increased tactical use of UHF SATCOM, enhanced flexibility of capacity extension through pre-arranged contracts and nationally leased capabilities and services (Plachecki, 2015, p. 9); as well as

---

[42] Key question: '*Where does outer space begin and why nations have to go to space?'* (Hill, 2020).
[43] *Extremely High Frequency Band (EHF):* ITU designation for RF band from 30 to 300 GHz; used for more reliable communications: higher survivability under physical attack, minimized susceptibility to adversary eavesdropping and *jamming* (see 4.2.3.1), "intersatellite communication and satellite radio navigation" (Britannica, 2020a).

- *Future Ground Requirements,* including new deployable and transportable, multi-band capable GS, additional remote control and *modem*[44] capabilities, new *broadcast*[45] capability, higher data rate, enhanced control and access of UHF capability as well as easier acquisition from commercial suppliers (Plachecki, 2015, p. 10).

As stated by Widmann and Vasen, this MOU is intended for peacetime and to ensure support for ongoing NATO operations; if NATO should have to act as a defence alliance in conflict, each NATO MS is equally expected to provide space asset services, to allow for heightened redundancy and frequency bandwidth and thus better operational outcome and force protection (Vasen, 2020b; Widmann, 2020).

### 3.2.6 Overview

In conclusion, NATO currently has access to four different types of space asset-enabled frequency bands. *Table ii* summarizes these capabilities, as well as their most prominent use cases and those country-respective space assets from which services are obtained. All entries are based on currently available public information.

Table ii: Overview: NATO-Used Frequency Bands and Utilities (Global Security, 2011; Microwaves 101, 2020; NATO, 2011; NATO, 2020; Plachecki, 2015, p. 4)

| | UHF | | SHF (X-Band) | | Ku-band | | EHF (K/ Ka-Band) |
|---|---|---|---|---|---|---|---|
| *Frequency* | 300 MHz - 3 GHz | | 3 - 30 GHz | | 12 - 18 GHz | | 30-300 GHz |
| *Use case* | Television broadcasting, ship/ aircraft navigation, tactical SATCOM | | Relaying broadcast programs, radar (e.g. for weather, reconnaissance), enhanced SATCOM | | Enhanced AGS | | Higher survivability, minimized susceptibility, intersatellite communication |
| *Type of Space Asset/ Country* | SICRAL 1/ 1Bis | *Italy* | SYRACUSE 3 space asset | *France* | Transponder (Afghanistan); SkyWan network and GS; Commercial Ku-band capacity/ Europe (45 days per year) | *Luxembourg* | *US, France, Luxembourg, UK, and Italy* |
| | Skynet 4/5 | *UK* | SICRAL 1/ 1Bis | *Italy* | | | |
| | | | GovSat-1 | *Luxembourg* | | | |
| | | | Skynet 4/5 | *UK* | | | |

---

[44] ***Modem:*** "device that converts signals produced by one type of device (such as a computer) to a form compatible with another (such as a telephone) and that is used especially to transmit and receive information" (Merriam-Webster, 2020c).
[45] ***Broadcast:*** the spread of "information to a lot of people" (Cambridge Dictionary, 2020a).

## 3.3  Division of Responsibilities and Mechanisms

Globally, responsibilities for the use of MS space asset related services as well as NATO-owned GS are divided between NATO and NATO-affiliated authorities, as well as the MS themselves. The subsequent subsections aim to define and divide the mechanisms underlying those responsibilities.

### 3.3.1  NATO and NATO-Affiliated Divisions

This subchapter outlines the concrete responsibilities between NATO as well as NATO-affiliated divisions. A graphic overview is provided in *3.4: Overview.*

#### 3.3.1.1  SATCOM: NATO Communications and Information Agency (NCIA)

As stated by NSO, NATO exercises operational control (OPCON) over its organic SATCOM units and capabilities (see *3.2.2*). Thereby, NCIA manages NATO's SATCOM network monitoring, service allocation, and telecommunications. It is also responsible for providing *Command, Control[46], Communications, and Computers, Intelligence, Surveillance, and Reconnaissance* (C4ISR), working on both ground and wireless communication to ensure compatibility (NSO, 2016, p. 5-4; Heren, 2020). It is responsible for procurement, partial operation of equipment, maintenance of services throughout missions, or identification of such gaps through NATO's *Defence Planning Process[47]* (NDPP). NDPP is NATO's "primary means to identify and prioritise the capabilities required for full-spectrum operations, and to promote their development and delivery" (NATO, 2020, p. 2).

JPMO reports to NCIA (Plachecki, 2015, p. 7). It manages the overall SATCOM cooperation between NATO and MS. Requirements for NATO's Allied Command Operations (ACO), which are responsible for the determination of the minimum needed military and operation capability requirements, are planned together with NCIA, being then discussed with JPMO to ensure that space asset capacity is suitable and available. NCIA liaises with the co-located *NATO Mission Access Centre (NMAC),* which is manned by national contractors as Point of Contact (POC) between national space asset control centres and operators of NATO. Thus, NCIA ensures availability of commercial SATCOM services throughout different MS (Vasen, 2020b). As the security of space assets and their architecture is responsibility of the MS, NCIA cares mostly for data *encryption[48],* verification and secure transmission.

---

[46] ***Command and Control (C2):*** In military terms describes "the exercise of authority by a properly designated commander over assigned and attached forces, performed through an arrangement of personnel, equipment, communications, facilities and procedures" (NATO C2COE, 2020). In contrast, in IT terms, a ***Command-and-Control (C&C) server*** is a "computer controlled by an attacker or cybercriminal which is used to send commands to systems compromised by malware and receive stolen data from a target network" (Trend Micro, 2020).

[47] ***NATO's Defence Planning Process (NDPP):*** see *Chapter 5.1.2.*

[48] ***Encryption***: The method by which "information is converted into secret code that hides the information's true meaning. The science of encrypting and decrypting information is called cryptography" (Searchsecurity, 2020a). In contrast, ***decryption*** describes the process of "transforming data that has been rendered unreadable through encryption back to its unencrypted form" (Techopedia, 2020).

### 3.3.1.2 ISR: NATO Joint Intelligence, Surveillance and Reconnaissance Division (JISR)

JISR gathers data and information for NATO AGS through a wide variety of national ISR assets from the space, air, land, maritime and cyber domains (NATO, 2018). Intelligence satellite-related information is handled by NATO's intelligence community at the NATO Headquarters in Brussels, Belgium and at the NATO Supreme Headquarters Allied Powers Europe (SHAPE) near Mons, Belgium; processes and agreements with MS are here created to secure access to nationally gathered data (Heren, 2020). Most agreements are established with the US, UK, France, Germany, and Italy. However, as commercial ISR has significantly increased over the past two decades, NATO also acquires commercial data for these purposes.

### 3.3.1.3 Awareness: NATO Space Centre

In October 2020, NATO SG Stoltenberg announced the establishment of a new *NATO Space Centre* in Ramstein, Germany, which

> "will be a focal point to support NATO missions with communications and satellite imagery, share information about potential threats to satellites and coordinate our activities in this crucial domain […] to increase NATO's awareness of challenges in space, and the Alliance's ability to deal with them" (NATO, 2020).

Furthermore, NATO states that the NATO Space Centre "will be housed within existing facilities at Allied Air Command and will initially be staffed with a small team of officers and experts already working at the command" (NATO, 2020). The Space Centre is expected to start working and partnering with MS at the beginning of 2021, to provide space-related services and products to support NATO missions, operations, and activities.

### 3.3.1.4 Coordination: NATO Space Support Coordination (SpSC)

*Space Support Coordination (SpSC)* within NATO operations is strategically hatted by ACO at SHAPE. Its role is defined in *AJP-3.3*:

> "Commanders at every level have responsibility […] to include the use of space-based capabilities […] The space coordination function will have responsibility for planning the integration of space force enhancement tools and capabilities; the integration and coordination of space control and space situational awareness activities; and provide space analysis expertise and space related products" (NSO, 2016, p. 5-8).

AJP 3.3 (B) further details that SpSC is responsible to "provide timely recommendations to courses of action […] collect, understand and process requests for space support [and] support the identification and recommendation for prioritization of delivery of space related products and services" (NSO, 2016, p. 5-8). Tactically, SpSC is *ad hoc*-created through an *SpSC Element* (SpSCE) *per mission,* being assigned to NATO corps air, naval and land operations, and are permanently staffed with one or two Subject Matter Experts (SME; Vasen, 2020, pp. 23-25).

### 3.3.1.5   Counselling: Joint Air Power Competence Centre (JAPCC)

The *Joint Air Power Competence Centre* (JAPCC, 2020), located in Kalkar, Germany, is NATO-accredited COE, consisting of independent, multinational experts (Heren, 2020). Existing since 2005, it provides

> "subject matter expertise across a broad range of Joint Air and Space Power mission areas and leads NATO in the development of Concepts and Doctrine, Capability Development, Education and Training, Exercise Development and Execution, and Lessons Learned" (NATO ACT, 2020).

As of June 2020, JAPCC maintains a separate branch focused exclusively on space that aims to gather cross-community experts such as from the fields of intelligence, communications or meteorology in order to bring in more perspectives and support educated operational decisions by NATO (JAPCC, 2020). Furthermore, the issue of standing-up a new NATO Space COE is currently being discussed between the existing, and potentially expanding, role of the JAPCC and a proposal for a new NATO-accredited organisation in France.

### 3.3.1.6   Counselling: Cooperative Cyber Defence Centre of Excellence (CCDCOE)

Like JAPCC, the Cooperative Cyber Defence Centre of Excellence (CCDCOE) is a NATO-accredited hub to support MS and NATO with expertise on *cyber defence*. Thereby, Cyber defence is defined as the protection of NATO's

> "own networks (including operations and missions) and enhanced resilience across the Alliance [including] awareness, education, training, and exercise activities, and […] further progress in various cooperation initiatives, including those with partner countries and international organisations. It also foresees boosting NATO's cooperation with industry, including on information-sharing and the exchange of best practices. Allies have also committed to enhancing information-sharing and mutual assistance in preventing, mitigating and recovering from cyberattacks" (NATO, 2020a).

Within CCDCOE, this includes a "diverse group of experts from 29 nations [which] brings together researchers, analysts and educators from the military, government, academia and *industry*[49]" (CCDCOE, 2020). There is also support for research and reporting on NATO outer space-related vulnerabilities (CCDCOE, 2019).

## 3.3.2   Allied Member States

As NATO does not own its own on-orbit space assets, it leaves it up to its MS "to determine whether they provide access to their satellite capabilities [through] memoranda of understanding with allies for possible use of space products and services" (Unal, 2019, p. 9). This cooperation allows NATO's military commanders and decision-making staff to have critical and highly sensitive data at their command to facilitate NATO missions (Bimfort, 1995, p. 1; Unal, 2019, p. 9; *see 3.2.2 and 3.2.5*). The NSO states that whilst MS space support is not mandatory,

"formal agreements should regulate NATO's degree of access, service level and coordination mechanisms. Nations sharing space services and products, or otherwise contributing space support to NATO operations, should consider providing planning information and designating a point of contact for coordination purposes" (NSO, 2016, p. 5-5).

## 3.4  Summary

Space assets are key enablers for NATO's most advanced technological systems, missions, and operations. Key functions to enable core tasks like SSA and DSC are:

Table iii: Space Asset Key Functions for NATO

| Field of Use | Key Functions |
|---|---|
| *SATCOM* | Provides NATO commanders with crucial telecommunications such as broadband internet, mobile services and voice communications for remote terrestrial areas and therefore allows for beyond line-of-sight communications; |
| *ISR* | Monitors terrestrial areas of interest, track and provide information on adversary intent and thus provides joint force planners with multi-spectral information on subsurface, surface, and air conditions; |
| *PNT* | Provides precise navigation services through GPS, Galileo or EGNOS, as well as precise coordinated time, making it vital to military operations; |
| *Deterrence effect* | Facing a rapid arms race in and through space. |

Whilst NATO launched in total 8 satellites between 1970 and 1993 through SATCOM1, this programme was replaced in 2005 by NSP2K. Even though NATO maintains four terrestrial GS, there has been a marked paradigm shift from owning and operating proper satellites to using national MILSATCOM services, which are managed by JPMO. In 2019, NATO published its first Space Policy and separate recognition of space as an operational domain alongside air, land, sea, and cyberspace. Hence, it responded to the need for aligning MS positions, operational legal space asset support, increasing the operational dependence on space assets, and assisting the rise of sensible ISR exchange within NATO for better imagery and data analysis through JISR. Finally, in 2020, a newly established MOU between NATO, US, France, UK, and Italy ensures space asset support for fifteen more years, with NCIA handling most of SATCOM, and JISR being mainly responsible for ISR retrieval and exchange. Awareness and counselling are provided by JAPCC, CCDCOE and the soon-to-be-established Space COE, whilst SpSC coordinates and integrates operational space functions.

However, this wide spread of responsibilities over the whole of NATO's internal and external systems, being visualized in *Figure 1 (see below),* might obscure, or leave open the quest for liability in case of an attack. This is aggravated by the lacking security of space assets themselves. Whilst space technology and components are indivisibly linked to cyberspace, cyberspace's ubiquitous and interconnected nature makes attribution of malicious actors (*see Chapter 4.1*) extremely difficult. Such cyber insecurities add additional pressure on the safeguarding and integrity of NATO missions. Subsequently, *Chapter 4* identifies and analyses such vulnerabilities.

---

[49] *Industry:* defined as "the companies and activities involved in the process of producing goods for sale, especially in a factory or special area" (Cambridge Dictionary, 2020d).

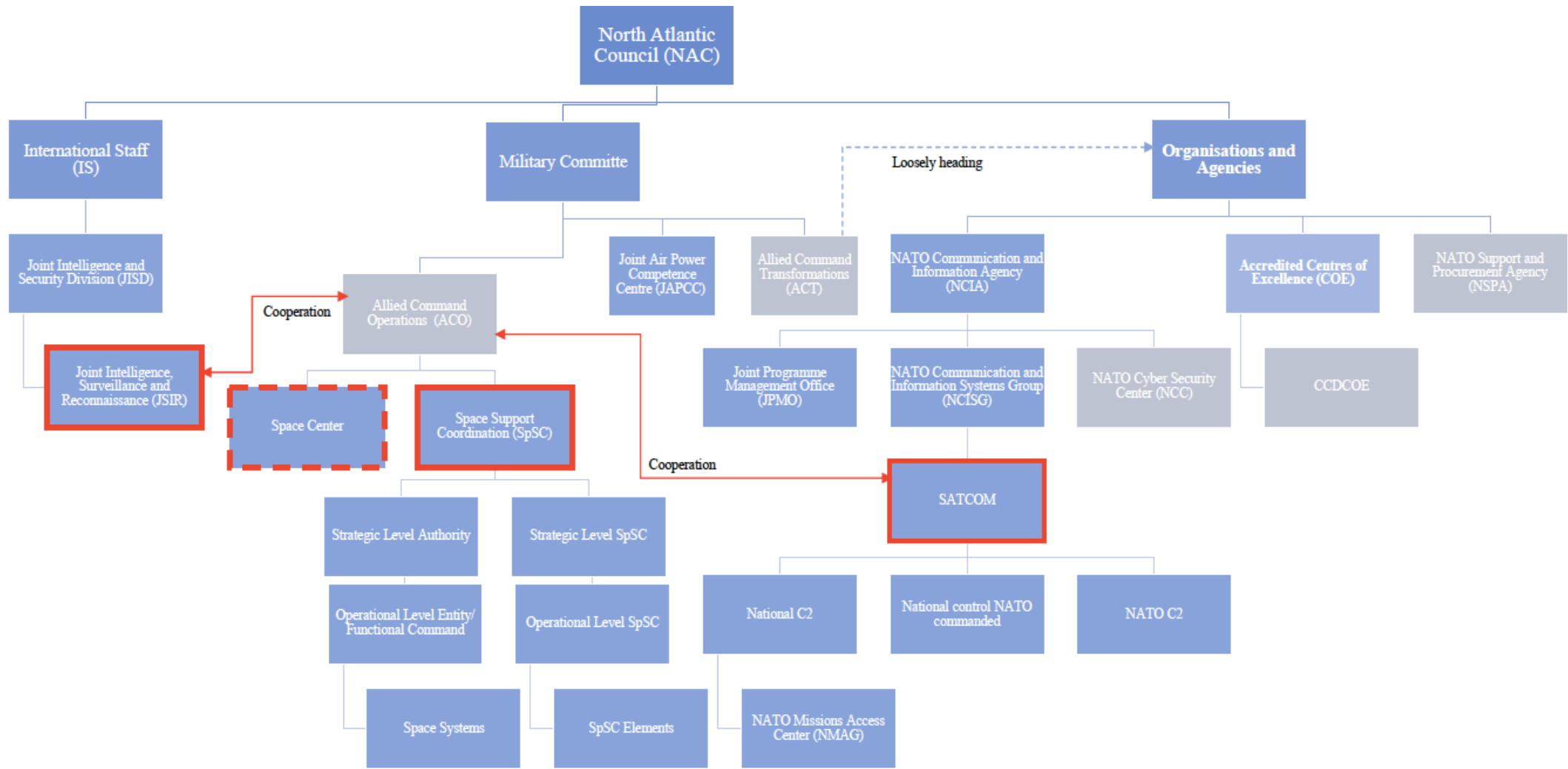

Figure 1: Overview of Most Relevant Space Functions in NATO (own organizational chart, basing on NATO, 2018; NCISG, 2020; NSO, 2016; Vasen, 2020)

# 4 SPACE ASSET CYBERSECURITY GAPS

Sophisticated cyberattacks on NATO or its MS's space assets have a distinct new impact on NATO operational decision-making as they "render NATO countries vulnerable to [adversarial] deception and information operations" (Unal, 2019, p. 18). Physically, space remains a remarkably transparent domain, unlike air or sea (Paulauskas, 2020, pp. 4-7). However, NATO's growing military dependence on commercial space-based assets increases the number of cyberattacks and vulnerabilities, denying and disrupting vital information flow (Moon, 2017, pp. 4-10). Any attack on SATCOM, ISR, or PNT through cyberspace jeopardises NATO's missions and services and diminishes and *confidentiality*[50]. Consequences could run "from a mild inconvenience to a vital security threat […] to the inability to coordinate and operate modern forces" (Moon, 2017, p. 12). This situation is aggravated by the absence of an international framework adapted to today's space environment realities such as privatisation and *democratisation* (*see Chapter 4.2.4*). For NATO, this means a significant threat to critical force manoeuvres, situational and strategic awareness, as well as precision-strike capabilities and efficient and quick reactions to a crisis. The subsequent chapter aims at identifying current space asset cybersecurity gaps, therefore diminishing NATO's credibility and hence capability to deter.

## 4.1 Malicious Actors and Motivations

Actors behind cyber-attacks on NATO space assets are grouped into either *nation-state actors*, *professional or amateur hackers*, *organised criminals*, or *insiders* (Moon, 2017, p. 7). Thereby,

- *Nation-state actors* seek military advantage through stealing intellectual property and pose the biggest threat to Alliance operations and missions (AOM), because they have the resources and capacities to conduct activities such as "terrorism, espionage, subversion, sabotage, organised crime [or] the relatively safe and global reach of espionage actions via cyberspace" (NSO, 2020, p. 6; Heren, 2020).

- *Professional or amateur hackers,* thus *non-state* actors, generally with the aim of proving their skills or motivated by personal interests such as *hacktivism*[51]. These actors can either operate individually or in groups. However, they may as well complement nation states with required cyber knowledge, skills, and resource means. Further motivation may include the monitoring of data and communication traffic patterns via space assets or inserting corrupted or false data to the system (Harrison et al., 2019, p. 5).

---

[50] **Confidentiality** being defined as "privacy. Making sure only the people who require access to data have access, while also making sure that everyone who needs the data is able to access it" (NASA, 2019).

[51] **Hacktivism**: the act of "misusing a computer system or network for a socially or politically motivated reason […it] is meant to call the public's attention to something the hacktivist believes is an important issue or cause, such as freedom of information or human rights. It can also be a way for the hacktivists to express their opposition to something by, for instance, displaying messages or images on the website of an organization they believe is doing something wrong" (Searchsecurity, 2020b).

- *Organised criminals* usually seek financial gain by exploiting cyberspace's anonymity and connectivity, possibly impacting military undertakings (NSO, 2020, p. 6). This might include *ransomware*[52] attacks on infrastructure or data.

- *Insiders* such as disgruntled NATO personnel may deliberately aim to exploit cyberspace to harm NATO. This takes place regardless of seniority or role and may give adversaries an opening to circumvent existing cybersecurity measures (NSO, 2020, p. 6). This threat is usually underestimated and requires specific guidelines to be dealt with.

Thereby, timely and accurate cyberattack attribution becomes extremely difficult, since attackers may use many identity-concealing methods, such as are *false flag campaigns*[53] (Harrison et al., 2019, p. 5), as well as due to both cyberspace's and outer space's ambiguous and completely borderless nature. Nonetheless, it remains an important piece of deterrence by combining several factors such as an adversary's intent, capability, technical *Indicators of Compromise*[54], geopolitical status and solid intelligence. Attribution is needed to ensure that adversaries realize that there will be consequences for actions taken in the cyber domain, hence adding to NATO's resilience posture.

## 4.2 Vulnerabilities

Space assets consist of three technological segments, forming their Satellite C2 Architecture, thus "how users control and communicate with satellites" (US DIA, 2019, p. 7). This includes the *physical GS*, the *space segment* as well as the *data link connection* between the two of them (Baram & Wechsler, 2020, p. 2; Unal, 2019, p. 3). By taking out even just one of these segments through a cyberattack, the whole of the space asset system is likely to render inoperable (Paulauskas, 2020, p. 7).

### 4.2.1 Ground Segment

As stated by Caudill (2020), every GS not deliberately secured "is to be assumed unsecure, starting from its system design". This negatively impacts the *Information Technology*[55] (IT) used within the space asset (Bander et al., 2020). Thereby, GSs "constitute a critical vulnerability, as [such] a terminal is an access point to a satellite and is usually not protected by authentication in order not to hinder operational actions" (Unal, 2019, p. 8). Thus, computer systems connecting GS internally and to their surroundings are likely to be attacked, as they might be commingled with unsecure enterprise computer applications (Lee, 2020). GS run often outdated or common operational computer systems controlling antennas for SATCOM data transfer, instead of specialized, secured systems (Tucker, 2019). This is aggravated by frequently *unsegmented*[56] networks especially in the commercial sector, which are critical satellite applications installed on the same computer as commercial, unsecured software. Therefore, transmission of sensible data may easily be intercepted through cyberspace, a common and consistent method across cyberattacks on GSs (Conti, 2020).

---

[52] **Ransomware:** Malicious software "that infects your computer and displays messages demanding a fee to be paid in order for your system to work again" (Kaspersky, 2020).

[53] **False Flag campaigns:** Refer to cyberattack tactics applied "to deceive or misguide attribution attempts including the attacker's origin, identity, movement, and exploitation" (Skopik & Pahi, 2020, p. 1), such as buying, leasing, or compromising infrastructure in the country being blamed, using internet addresses referring to a specific country, change language used in the malware, or use personal data of uninvolved individuals, often retrieved from open-sources like social media.

[54] **Indicators of Compromise:** IoC's are defined as "forensic evidence of potential intrusions on a host system or network. These artifacts enable information security (InfoSec) professionals and system administrators to detect intrusion attempts or other malicious activities. Security researchers use IOCs to better analyze a particular malware's techniques and behaviors. IOCs also provides actionable threat intelligence [and may include for example] Unusual traffic going in and out of the network; Unknown files, applications, and processes in the system; Suspicious activity in administrator or privileged accounts; Irregular activities such as traffic in countries an organization doesn't do business with; Anomalous spikes of requests and read volume in company files; Network traffic that traverses in unusually used ports; Tampered file, Domain Name Servers (DNS) and registry configurations" (Trendmicro, 2020).

[55] **Information Technology:** Physical entities and their tangible components such as "computers, servers, routers, hubs, switches, wiring and other equipment crucial to data storage, data processing and data transmission" (NSO, 2020, p. 3).

Additionally, the supporting CI around a GS poses a major vulnerability because it is vital to ensure GS power supply (Yingst, 2020). This includes energy resources such as pipelines or as well transportation networks. Corresponding attack vectors mainly include exploiting above stated system vulnerabilities or luring GS personnel to download *malwares*[57] to GS computers through *phishing*[58], *spearphishing*[59][60], *identity theft*[61], or *social engineering*[62] (Lee, 2020) attacks. In particular, social engineering is often underestimated and requires a thorough training of all involved staff members to be successfully thwarted. Further major GS-affiliated cyberattack types include *Denial of Service*[63] *(DoS)* or ransomware attacks, allowing attackers to control and maybe even damage the GS (*see Chapter 5*). Thereby, it is important to note that historically, the internet itself, connecting such GS and facilitating data links, was not built focusing on cybersecurity or *Confidentiality, Integrity, and Availability*[64] *(CIA)*, but "to allow multiple computers to communicate on a single network" (Andrews, 2019, p. 1).

### 4.2.2 Space Segment

*Space segment* describes those space assets being installed on-orbit, such as satellites themselves (Paulauskas, 2020, p. 4). It is the least vulnerable, since "once assets are launched, it becomes extremely difficult to physically attack the hardware, but allows mainly to attack only via cyber means" (Bander et al., 2020). Additionally, Sielaff states that few countries have the resources and capabilities to physically approach a space segment on-orbit for an attack. However, remaining vulnerabilities include:

- *Legacy systems* in case of aged satellites, whose systems having been built with no cybersecurity awareness and are entry points for more up-to-date cyberattack technology (Bander et al., 2020; *see 4.2.3.3*); and

- *The use of satellites reduced in size and weight* (*Small Satellites/ CubeSats*), produced fast and cheap by suppliers seeing investment in cybersecurity as a hurdle (*see below*).

---

[56] **Network Segmentation:** describes the process of dividing "a computer network into smaller parts […] to improve network performance and security […] improve cybersecurity by limiting how far an attack can spread. For example, segmentation keeps a malware outbreak in one section from affecting systems in another [, and] stop harmful traffic from reaching devices that are unable to protect themselves from attack" (Cisco, 2020).

[57] **Malware**: Also *malicious software;* "any malicious program or code that is harmful to systems" (Malwarebytes, 2020c).

[58] **Phishing:** is defined as "a form of fraud in which an attacker masquerades as a reputable entity or person in email or other forms of communication [usually through] phishing emails to distribute malicious links or attachments that can perform a variety of functions" (Searchsecurity, 2020c).

[59] **Spearphishing:** A targeted phishing attempt at "a specific individual, company, or industry will be targeted by the adversary […] Adversaries may send victims emails containing malicious attachments or links, typically to execute malicious code on victim systems or to gather credentials" (Mitre Att&ck, 2020).

[60] **Ransomware:** Malicious software "that infects your computer and displays messages demanding a fee to be paid in order for your system to work again" (Kaspersky, 2020).

[61] **Identity Theft:** Occurs „when a criminal obtains or uses the personal information [...] of someone else to assume their identity or access their accounts for the purpose of committing fraud, receiving benefits, or gaining financially in some way" (Malwarebytes, 2020b).

[62] **Social Engineering:** Attack type accomplished "through human interactions [which] uses psychological manipulation to trick users into making security mistakes or giving away sensitive information" (Imperva, 2020).

[63] **Denial-of-Service (DoS)**: "attack meant to shut down a machine or network, making it inaccessible to its intended users. DoS attacks accomplish this by flooding the target with traffic, or sending it information that triggers a crash" (Palo Alto, 2020).

[64] **Confidentiality, Integrity, and Availability (CIA) Triad:** a "concept model used for information security" (NASA, 2019) (*see in order footnote 48, Chapter 1 and footnote 16*).

### 4.2.2.1 Use of Small Satellites

At this moment, NATO is basing space-based requirements on satellites being installed in Geosynchronous Orbit (GEO), located at 35,786 kilometres above Earth's equator (Paulauskas, 2020, p. 4). This distance allows satellites to match the rotation of Earth, making them useful for steady SATCOM, monitoring weather, and surveillance. However, NATO MS's interest in integrating the use of *Small Satellites (SmallSats)* increases. SmallSats are typically located in Low-Earth Orbit (LEO) at an altitude of around 1000 km to 160 km above Earth's equator (ESA, 2020). Thereby, UNOOSA states that

> "small satellites and their applications have opened the door for many countries […] with limited funds for space activities to join in the exploration and the peaceful uses of outer space and to become developers of space technology" (UNOOSA, 2015, p. 3).

UNOOSA groups SmallSats into "*mini* satellites < 100 kg, *nano* satellites < 10 kg, *pico* satellites < 1 kg [and] *femto* satellites < 0,1 kg" (UNOOSA, 2015, p. 3). Nano satellites and smaller are as well named *CubeSats,* which are "a class of *research* spacecraft […] built to standard dimensions (Units or "U") of 10 cm x 10 cm x 10 cm" (NASA, 2015) and generally weight around one kilogram. Typical characteristics include "reasonably short development times; relatively small development teams; modest development and testing infrastructure requirements; and affordable development and operation costs for the developers, in other terms 'faster, cheaper and smaller'" (UNOOSA, 2015, p. 3). SmallSats are easily and rapidly developed and deployed on-orbit, and deliver on the rising need for higher quality ISR through enhanced bandwidth and terrestrial coverage (Hallex & Cottom, 2020, pp. 24-25). Additionally, due to their smaller distance to earth, military SATCOM is delivered at a lower latency rate than that delivered by GEO satellites, whilst GEO satellites take longer to be replaced in case of a critical attack or denial (Thomas, 2020, p. 1). As an example, the US DOD's Advanced Research Projects Agency (DARPA) is currently developing the *Blackjack* project, an experimental programme to create a large SmallSat constellation in LEO aiming to provide DOD "with highly connected, resilient, and persistent coverage" (Thomas, 2020, p. 1) by 2021.

Furthermore, Falco points to PNT as a common use case for SmallSats, as commercial companies bring measurements such as enhanced GPS, specific time control systems, or scientific research. Thus, the UK Government is currently preparing to collaborate with the private company *Horizon Technologies Inc*. on enhanced *Maritime Domain Awareness [65] (MDA)* through the usage of

---

[65] **Maritime Domain Awareness:** is the "effective understanding of anything associated with the global maritime domain that could impact the security, safety, economy, or environment" (DHS, 2005). Aim is to collect "the maximum information and intelligence about any ship or vessel in the country's waters. With the collected data, a complete inference can be drawn about all those marine areas that could cause potential damage with respect to safety, eco-system and the economic system. This process is known as *actionable intelligence*" (Maritime Insight, 2018).

Horizon's CubeSat *constellation*[66] AMBER$^{TM}$, being launched in 2021 (Urban, 2020). This public-private partnership (PPP) will provide RF intelligence data from anywhere in the world and in near-real time, and hence would allow for more informed decision-making throughout NATO operations (L3Harris, 2016, p. 3). This includes automated tracking and monitoring of ship-installed transceivers and satellite phones, which could be used by NATO to locate smuggler and piracy communication and detect malicious actors, control borders, and prevent potential terrorist attacks (Urban, 2020).

Hence, investing in a greater number of SmallSats might increase military resilience, as in case of an attack against such a constellation, the extreme disaggregation degree allows for only regional attacks. Additionally, the loss of few space assets would not risk the constellation's availability to function, whilst the replacement of SmallSats is relatively inexpensive (Vasen, 2020b). However, the use of SmallSats comes with significant risks:

-   *Commercial SmallSat components are customarily procured from COTS technology to lower costs*. Hence, due to the wide availability of COTS products, hackers might more easily conduct component analysis and better prepare for attacks (Akoto, 2020, p. 2). Such components draw on open-source technology, allowing hackers to insert vulnerabilities such as backdoors to space asset software. Additionally, these practices cause the reliance on miscellaneous standards, infrastructure, and governmental frameworks and, which render SmallSats less controllable and more vulnerable to cyberthreats than more centralized and aligned SATCOM *interfaces*[67] (*see 5.2.4*).

-   *Commercial SmallSat producers may tend to circumvent cybersecurity controls for financial reasons,* in contrast to technically and strategically more experienced producers of GEO satellites (Livingstone, 2020). This happens especially if the CubeSat mission is cheaper in itself than the implementation of such controls (*see 5.2.6*).

-   *SmallSats are more easily locatable and thus eliminable due to their reduced geographic distance* (Hallex & Cottom, 2020, pp. 24-25). Constellations are more likely to be targeted by cyberattacks, as attackers might affect a whole, interconnected network through only one cyberattack.

-   *The integration and coordination of data retrieved from private Smallsat constellations into existing military GS might be problematic* (Strout, 2020, p. 1). Reasons are, for example, misaligned security or classification levels (*see 5.2.3*). This was observed during the US DOD's *Blackjack*, comprising the processing of data on the ground.

-   *Less redundancy securities and encryption possibilities are available from SmallSats*. This is due to the reduced size of the system, restricting the amount of antennas and frequency bands being built in, as well as stored energy resources (Vasen, 2020b; *see 5.2.3*).

### 4.2.3 Data Link Segment

The *data link segment* describes the connection between satellites and GS, hence the information flow between "firmware, operating systems, protocols, applications, and other software and data components" (NSO, 2020, p. 3). Bander et al. (2020) define it as "the data going back and forth to the satellites", sending commands and preserving communication. The GS personnel "uses the *uplink* to the spacecraft to deliver commands. The spacecraft *downlink* is how data is sent from the spacecraft to a GS that has the necessary antennas, transmitters, and receivers to receive the data" (US DIA, 2019, p. 7). In case of malicious interference, severe outage will be caused. Risks connected to the data link segment include:

- *Data might be destroyed or interfered during its transfer,* suppressing or interrupting the service and rendering reception and/or decryption impossible for its end user (Vasen, 2020b).

- *Encryption key management of Telemetry, Tracking and Command (TTC) systems are oftentimes not handled in a secure manner* (ESA, 1996, p. 3). TTCs are defined as the vital functions of a space asset "allowing data to be communicated between the ground and the spacecraft for spacecraft control and command" (ESA, 1996, p. 3). This exposes space assets to misconfiguration, hacking and encryption or decryption of communication data (Caudill, 2020).

- *Eased direct attacks due to the above stated reasons,* which includes the hacking and re-writing of computer files leading to damage, modification or destruction of mission-critical data (Cimpanu, 2020, p. 1).

---

[66] **Satellite Constellation:** "several satellites working in coordination with each other" (MIT, 2017, p. 1).
[67] **Interface:** Defined as the set of devices „provided by a computer or a program to allow the user to communicate and use the computer or program" (TechTarget, 2020).

### 4.2.3.1   Jamming

*Jamming* constitutes the most common threat to space assets and *Global Navigation Satellite Systems*[68] (GNSS) and can be executed both by means of cyber and *electronic warfare*[69] *(EW)* (Baram & Wechsler, 2020, p. 3). Thereby, Moon states that GNSS and attached civil applications have been oftentimes not designed with cybersecurity in mind (Moon, 2017, p. 8). Signals received from satellites to determine the GPS location of users may be distorted or overridden through commercially-available *GPS jammers*, which is "a typically small, self-contained, transmitter device used to conceal one's location by sending radio signals with the same frequency as a GPS device. When this occurs, the GPS device is unable to determine its position due to interference" (Geotab, 2020). GPS jammers, that are often used to disable track and trace of military trucks and have a reach range of about 100 metres, may cost as little as 20 Euros and are unrestrictedly accessible to less-resourceful malicious actors (Falco, 2020; JammerShop, 2020). As stated by US DIA,

> "*Uplink jamming* is directed toward the satellite and impairs services for all users in the satellite reception area. *Downlink jamming* has a localized effect because it is directed at ground users, such as a ground forces unit using satellite navigation to determine their location" (US DIA, 2019, p. 9).

Additionally, NATO currently bases its navigation solely on US GPS, thus increasing vulnerability if being jammed. As an example, Russian malicious actors reportedly jammed GPS signals during NATO Exercise Trident Juncture 2018 (TRJE18) in Finland and northern Norway (O'Dwyer, 2018; Unal, 2019, p. 6; Baram & Wechsler, 2020, pp. 2-3), whilst NATO officials expressed that Russia may have been actively testing this capability through large-scale exercises, for example during its military exercise ZAPAD 2017, carried out in cooperation with Belarus (Atlantic Council, 2018). Thus, NATO should start creating more redundancies, such as through increased integration of European Galileo *spectrums*[70].

---

[68] **GNSS:** see Footnote 65.
[69] **Electronic Warfare**: whilst cyberattacks are executed via *cyberspace* (*see footnote 9*), EW attacks are take place within the *Electromagnetic Spectrum (EMS)* which is defined as the "range of frequencies of electromagnetic radiation from zero to infinity" (Bommakanti, 2019, p. 10), thus *the totality of wavelengths and frequencies* (*footnotes 36*). Use of Cyber is often also named as "protocol level/layer attack" as in opposition to physical layer attack. *See 6.2.2: ISO 27000.*

### 4.2.3.2 Spoofing

The second most used technique to attack space assets is *spoofing,* which is the "faking [of] signals by broadcasting incorrect GPS signals, structured to resemble genuine ones" (Baram & Wechsler, 2020, p. 3). Similar to jamming, spoofing may be executed both by cyber and EW means (Baram & Wechsler, 2020, p. 3). This involves *tampering*[71] or disrupting frequency signalling, and may as well target unencrypted satellite traffic. Spoofing is an adversarial way to manipulate information on position, location, and condition of satellite systems, being relatively difficult to detect (Moon, 2017, p. 8). Being more difficult to execute than jamming, spoofing attacks are more often performed by nation state actors, and have been "readily used by Russian or Chinese military" (Falco, 2020). Throughout spoofing, CI such as national power grids can be disrupted by inducing false timing data and signals, denying military C2 in times of crisis. This constitutes a more dangerous threat than jamming, as victims usually do not realize the attack. As an example, the 2017 Report on *US Maritime Administration* stated that at least 20 vessels were spoofed and navigated up to 30 kilometres away from their original destination, raising assumptions that Russia possibly experimented on new spoofing techniques (Hambling, 2017; UNCTAD, 2017, p. 101).

### 4.2.3.3 Legacy systems

Since space missions may last decades, space assets are unique in their lack of physical access, decade-long lifespan of on-orbit space segments and hence regularly obsolete legacy systems (Falco, 2018; Koch & Golling, 2016). A legacy system still "meets the needs it was originally designed for, but doesn't allow for growth [and] won't allow it to interact with newer systems" (Talend, 2020). Thereby, once a space segment is being deployed, operators and builders are unlikely to regain direct access for around twelve to eighteen months (Wells & Sielaff, 2020). Basement retro engineering of cybersecurity controls is highly complicated and expensive. Attackers may use this vulnerability to upload malware into such often unencrypted satellites, using more advanced terrestrial technology (Falco, 2020). According risks include

1. *The initial use of proprietary and old IT software and hardware,* thus outdated, *unpatched*[72] IT systems;

2. *Security and reliability problems unfolding through hindered supply of spare parts becoming unavailable over time*;

3. *The inability or failure to conduct software updates regularly,* to remove detected vulnerabilities (Unal, 2019, p. 8).

---

[70] *Spectrum: See footnote 66.*
[71] *Tampering:* to "interfere so as to weaken or change for the worse […] to render something harmful or dangerous by altering its structure or composition" (Merriam-Webster, 2020f).
[72] *Patch:* A patch is "a software update comprised inserted (or patched) into the code of an executable program. Typically, a patch is installed into an existing software program [to] Address new security vulnerabilities, Address software stability issues, Upgrade the software"

However, as put by Falco (2018b), since "space assets are built to last [...] space asset cybersecurity is mission critical; system downtime is not an option" (Falco, 2018b). Therefore, when producing space assets, companies "tend to put as much flexibility as possible into the space asset to cover those legacy issues" (Wells & Sielaff, 2020).

### 4.2.4  Lacking Exchange of Academic Expertise

Among scholars and practitioners, there is high concern over lacking expertise on the nexus between space and cybersecurity, as "how cybersecurity is should be integrated into the design of aircraft, space assets, and into space operations" (S. Lee, 2020). This mindset denies both academics and practitioners the ability to establish knowledge and awareness. Hence, "it is not unusual for the industry to be kind of unfamiliar with cybersecurity and to have a hard time adapting to it when manufacturing" (Lee, 2020). Underlining Lee's statement, Livingstone points to the general lack of education on reasons behind attacks on commercial space assets, influencing commercial judgements "that have to be made between how much governments spend on a commercial offering compared to how much gets invested in cybersecurity of space assets" (Livingstone, 2020).

Additionally, Zarkan points to a significant lack of knowledge and awareness due to the reluctance of NATO MS governments to report cyberattacks against space systems, as this might undermine their credibility, and thus deterrence of attacks. Silverstein states that all NATO "security activities [are] severely undermined by the alliance's inability or unwillingness to openly communicate about space security" (Silverstein, 2020). This is aggravated by the historically widely adopted concept of *Security by Obscurity,* or as put by Lee (2020), "the space asset system I am trying to secure is so weird, unusual or rare, how would anybody ever figure out how to break it or get into it?*"*. Persistent reliance on the space asset's obscurity creates a false sense of security and therefore ignorance, affecting both governments and big traditional commercial players such as Boeing or Airbus. However, this obscurity disappears with the entry of commercial companies to the market through an ongoing revolutionary *industrialisation* of space, increasing the amount of data being stored in COTS products. The lack of exchange across both public and private sectors has been aggravated significantly with the introduction of the EU General Data Protection Regulation (GDPR) on 25th May 2018. As GDPR imposes a fine of at least 20 Million Euro for private entities, or 4 percent of turnover if higher than 20 Million Euro, data losses are oftentimes significantly belated reported throughout the space asset industry (Data Privacy Manager, 2020; GDPR, 2018).

For the above named reasons, barely any official incident statements were published throughout past years except for few cases: In 2014, the US *National Oceanic and Atmospheric Administration* (NOAA) confirmed in a report the hacking of a satellite, blaming Chinese hackers to have done so (Flaherty et al., 2014, p. 1). Similarly, in 2015, a hacker group apparently linked to the Russian government was reported to be using malware to exploit unencrypted data links to communication satellites (Jones, 2015).

### 4.2.5  Lacking Regulation

To this point, there is no international, legally binding instrument to regulate the conduct of state and non-state actors in space, both in general and specifically on cyber means. This increases miscommunication, misuse of the domain, and misperception by nations (Johnson, 2014, p. 1). The same is true for the majority of NATO MS, facing both a lack of national policies in many states and few existing bilateral agreements related to space-capability data-sharing between MS (Hill, 2020).

The following lists most relevant and current international regulation and legal frameworks on the nexus between cybersecurity and outer space.

### NATO ALLIED JOINT DOCTRINE FOR CYBERSPACE OPERATIONS, AJP-3.20

With regards to cyberspace, NATO released in 2020 its *Allied Joint Doctrine for Cyberspace Operations, AJP-3.20* to "plan, execute and assess cyberspace operations (CO)" (NSO, 2020, p. 1). In its opening sentences, AJP-3.20 states that

> "digital networks and systems, therefore, need to be safeguarded against information denial by disruption, degradation or destruction, and manipulation and exfiltration […] freedom of action in cyberspace may be as important as control over land, air *and space*, or sea" (NSO, 2020, p. 1).

Additionally, it acknowledges that all "networks and devices connected by wired connections [and] wireless connections" (NSO, 2020, p. 1) might be potential targets. This adds to command challenges and difficulties in distinguishing between strategic, tactical, and operational levels as well as execution and synchronisation of information, thus compromising NATO's successful functioning.

### NORTH ATLANTIC TREATY

Deterrence of cyberattacks on space assets delivering NATO operation-relevant information is severely hampered by the *North Atlantic Treaty* itself. As the new NATO Space Policy remains *classified*, "the core security benefits of the new space policy are diminished [and] open discourse about allied resolve to protect satellites from hostilities" (Silverstein, 2020, p. 1) is prevented. Also, "by not releasing the terms of the policy, the alliance does not clearly outline if and how [North Atlantic Treaty] Article 5 protections apply to space assets" (Silverstein, 2020, p. 1). Thereby, Article 5, the *Principle of Individual or Collective Defence,* states that

> "an armed attack against one or more [parties] in Europe or North America shall be considered an attack against them all [which then] will assist the Party [by taking] such action as it deems necessary […] to restore and maintain the security of the North Atlantic area" (NATO, 1949).

Whilst it is widely discussed by scholars if an attack via cyber means may be considered an armed attack, as it may be not executed through traditional, physical *armed forces*[73], Article 5 comes additionally solely into action if the requirements of Article 6, the *Nationality Principle* are met. An attack is only recognized if

> "deemed to include an armed attack on the territory of any of the Parties […] on the forces, vessels, or aircraft of any of the Parties, *when in or over these territories or any other area"* (NATO, 1949), applies. However, a "classified internal policy cannot communicate credible resolve to an adversary. NATO has yet to concretely affirm the alliance's commitment to collective defence in outer space" (Silverstein, 2020). Moreover, for COs leading to only temporary disruption or service denial of NATO-used space assets would only be considered as an armed attack if used in the context of a wider, concurrent conventional attack (NSO, 2020, p. 21). Finally, geographic limits as imposed by Article 6 are "incongruent with outer space" (Silverstein, 2020) or cyberspace, as no national borders apply in these spheres and as cyberattacks are difficult to attribute to a specific country (Hill, 2020). Thus, to this point, there is no comprehensive, unclassified deterrence strategy for NATO MS space asset security, whilst its "declaratory policy does not delineate space-based thresholds or actions that would trigger NATO to act in collective defence" (Silverstein, 2020). This is especially aggravated by the fact that NATO itself states that collective defence commitments must be "clearly and unambiguously communicated to avoid misunderstanding and miscalculation by any potential adversary" (NATO, 2016).

---

[73] *Armed Forces:* A term used to denote "members of the Army, Marine Corps, Navy, Air Force, and Coast Guard" (OECD, 2003).

### TALLINN MANUAL 2.0 ON THE INTERNATIONAL LAW APPLICABLE TO CYBER OPERATIONS

The *Tallinn Manual 2.0,* prepared in 2017 by the International Groups of Experts at the invitation of NATO CCDCOE is the second edition of the CCDCOE's analysis on the application of international law to cyberspace. The analysis emphasises on the "idea that cyber operations do not occur in a legal vacuum, and pre-existing obligations under international law apply equally to the cyber domain" (National Security Archive, 2019). Chapter 10 emphasises on Space Law, which is the application of MS and international treaties governing on the use of outer space to cyber operations (CCDCOE, 2017, pp. 270-283). The Tallinn Manual 2.0 considers space, particularly space assets, as both a target and platform for cyberattacks. It remarks MS's responsibility to supervise non-state actor activities and state liability for cyber operation outcomes involving space assets. Hence, it states that

> "In the context of a non-State actor's actions, a State's preponderant or decisive participation in the 'financing, organising, training, *supplying*, and equipping […] the planning of the whole of its operation' has been found insufficient to reach the 'effective control' threshold" (CCDCOE, 2017, p. 97).

Referring to International Air Law, the Tallinn Manual 2.0 hints to Annex 17 of the 1944 *Chicago Convention on Civil Aviation*, stating that states should

> "ensure that measures are developed in order to protect critical information and communications technology systems [including] security by design, *supply chain security, network separation*, and remote access control, as appropriate" (CCDCOE, 2017, p. 269).

However, due to its nature, the Chicago Convention does not include any further guidance on the nexus of space assets, cybersecurity and space law but rather focuses on civil aviation regulation.

CCDCOE just announced its "commitment to host a project to revise and expand the influential Tallinn Manual 2.0 on International Law Applicable to Cyber Operations [which] will be updated in the light of emerging State practice" (CCDCOE, 2020). This Tallinn Manual 3.0 will be a five-year project, involving an update of all chapters and adding new topics addressing evolving cyber operations, including State practice, international law, academic publications as well as multi-stakeholder initiatives between governments, civil society players and industry. Hence, the Tallinn Manual 3.0 presents a fitting opportunity to address gaps.

### EU CODE OF CONDUCT

In 2007, 2008 and 2014, the EU *External Action Service* (EEAS) authored drafts for a voluntary international *Code of Conduct (COC) for Outer Space Activities* (Johnson, 2014, p. 1). The most recent draft aims at enhancing the security, safety, and sustainability of outer space activities through so-called *Transparency and Confidence-Building Measures (TCBMs)*. This underlines the principle of national responsibility to "take all appropriate measures and cooperate in good faith to avoid harmful interference" (DRAFT International Code of Conduct for Outer Space Activities, 2014, 26-28). Yet, none of the major space players outside the EU, such as the US, Russia, or China, signed this draft. This was justified by voiced concerns about its applicability, "vague terminology and lack of definitions, [and therefore] the degree to which it is legally binding" (Johnson, 2014, p. 2), and facing a lack of consensus on mishaps and best practices.

### EU CYBERSECURITY STRATEGY FOR THE DIGITAL DECADE

On 16 December 2020, the EU High Representative of the Union for Foreign Affairs and Security Policy and the European Commission presented the new EU Cybersecurity Strategy JOIN(2020) 18 final (European Council, 2020). This document points specifically to space as being

> "heavily reliant on network and information systems that are increasingly interconnected [whilst stating that] the threat landscape is compounded by geopolitical tensions over the global and open Internet and over control of technologies *across the whole supply chain*" (European Council, 2020, p. 1).

Furthermore, it points to the need for an ultra-secure communication infrastructure, with EU Governmental Satellite Communications providing cost-efficient and secure space-based capabilities to ensure safety- and security- critical EU and MS missions and operations including national security actors and EU institutions bodies and agencies. Through the deployment of a secure quantum communication infrastructure (QCI), the EU aims to create a "new way to transmit confidential information using an ultra-secure form of encryption to shield against cyberattacks, built with European technology [including] linked space satellites covering the whole EU, including its overseas territories" (European Council, 2020, p. 7). Additionally, it envisions strengthening the Galileo Security Monitoring Centre and cybersecurity of critical space infrastructures though creating a Commission Action Plan on synergies between the defence, civil, and space industries (European Council, 2020, p. 18).

## 4.3  Summary

Cybersecurity gaps in the design and operation of space assets are a significant threat to the delivery of mission-critical data. Actors behind attacks may be classified as either *nation-state actors*, *professional or amateur hackers*, *organised criminals*, or *insiders*. However, attribution and localization are significantly hindered by the ambiguous nature of cyberspace. Furthermore, if taking out a single space asset segment, thus either GS, space segment or data link, the whole of the space asset system is likely to become inoperable. Accordingly, main cyber-related risks include:

Table iv: Cyber Vulnerabilities per Space Asset Segment

| Space Asset Segment | Cyber Vulnerability |
|---|---|
| *Ground Segment*<br><br>*Space Segment* | Legacy and/or unsegmented soft- and hardware being oftentimes installed in ground and space segment's systems, allowing for eased interception of data and malware attacks such as through phishing, spearphishing, ransomware, identity theft, social engineering, or DoS attacks; |
| *Data Link Segment* | Increasing use of COTS technology for SmallSats and insecure handling of TTC and ISR encryption management, tampering NATO mission-critical data during its transfer and allowing for jamming and spoofing of data; |
| *Organizational Handling* | Lack of expertise on the nexus between space and cybersecurity;<br><br>Reluctance of NATO MS governments to report cyberattacks against space systems, hindering the development of knowledge and awareness;<br><br>Absence of international instruments to regulate the conduct of state and non-state actors in space; geographic limits by NATO Treaty Article 6 being incongruent with space and cyberspace, preventing the triggering of Article 5 (Principle of Individual or Collective Defence). |

Hence, *hypothesis one (H1) is confirmed*. However, it is insufficient in its assumption as it has become apparent that outsourcing of space capabilities to private companies, the use of outdated software as well as the almost complete lack of international laws and regulations are not the only drivers for cyber vulnerabilities along the cyber SASC. Thus, cyber vulnerabilities become ever more threatening for NATO as the influence and power of nation states in space are substituted by those of commercial, competition-oriented non-state actors. For NATO, this is crucial as a rapidly increasing number of manufacturers, globally widespread, becomes involved in building and proliferating the various components of those space assets that NATO retrieves its mission-critical data from. The industrialization and globalization of space assets multiplies to an uncontrollable number of possible cyberthreat actors and vulnerabilities. This might crucially impact NATO as well as its staff on the ground, including the evocation of casualties during training and missions. The following *Chapter 5* aims to identify cyber SASC habits throughout NATO, its MS, and commercial partners, as well as threats, vulnerabilities and organizational gaps being caused by these structures.

# 5 NATO SPACE ASSET CYBER SUPPLY CHAIN

## 5.1 Background

The multiplied involvement of suppliers and manufacturers in delivering mission-critical data and services for the military and NATO opens up numerous cyber intrusion threats (Falco, 2018b; Koch & Golling, 2016). As underlined by Akoto, this issue is aggravated by obscure SCs, as the extremely complex technical nature of satellites involves multiple manufacturers in building its components and launching those satellites. Even "once satellites are in space, day-to-day management is oftentimes outsourced to other companies" (Akoto, 2020, p. 2). This may lead to utilizing "third-party intellectual property and/or open-source software or firmware with unknown vulnerabilities or *implants[74]*" (Bailey, 2019, p. 7). The following chapter aims at identifying and analysing the so-caused vulnerabilities as well as gaps in NATO's handling of its cyber SASC.

### 5.1.1 General

Boyson et al. coin the expression of *cyber SC assurance*, with *cyber SC* being defined as

> "the entire set of key actors involved with/using cyber infrastructure: system end-users, policy makers, acquisition specialists, system integrators, network providers, and software/hardware suppliers [and] these users/providers' organizational and process-level interactions to plan, build, manage, maintain, and defend cyber infrastructure" (Boyson et al., 2009, p. 5).

This includes the "integration of computation and physical processes […] and the linking together of devices and sensors for information sharing" (Pandey et al., 2020, p. 3). Thereby, *cyber SC management* describes "everything from product development, sourcing, production and logistics as well as the information systems needed to coordinate these activities to […] achieve a sustainable competitive advantage" (Pandey et al., 2020, p. 3). However, to deliver on such aims, organizations are increasingly digitising operations to improve efficiency, sustainable growth, and cost of processes. This creates new *cyber SC risks,* caused by adversarial motivation to compromise the SC's integrity, or introduce malicious functions to it. Cyber SC risks may occur in three forms: *Supply* Risks are "failures from supplier (s) or the supply market [as] consumers are not able to know how their product's particular components are built" (Pandey et al., 2020, p. 14), being aggravated by inappropriate manufacturing standards. *Operational* Risks are defined as "serious breach[es] of the access controls […] allowing the attacker to cause the plant to malfunction and resulting in physical damage and operational disruption" (Pandey et al., 2020, p. 14). Such breaches may be perpetrated by perpetration through external entities or company insiders such as employees, breaching regulatory frameworks, manipulating documents, and providing authentication passwords via internet or electronic devices. Finally, *Demand* Risks are defined as "demand disruptions […] because of a lack of trust across SC companies" (Pandey et al., 2020, p. 14).

---

[74] To be understood as the covered implementation of malware before delivery/ throughout the SC.

A cyber SC is compromised if an adversary successfully interferes with the CIA of an attached system or stored or transmitted information (CSC, 2020, p. 138). The US Cybersecurity and Infrastructure Security Agency (CISA) acknowledges that

> "if vulnerabilities in the ICT SC - composed of hardware, software, and managed services from third-party vendors, suppliers, service providers, and contractors - are exploited, the consequences can affect all users of that technology or service" (CISA, 2020, p. 1).

However, as stated by the US National Institute of Standards and Technology (NIST), increasing globalization accelerates SC complexity, creates prolonged SC ecosystems and thus diminishes control and visibility (NIST, 2020d). In turn, cyberthreat actors increasingly target customer and supplier networks through so-called *island hopping*, which is "an attack that focuses on impacting not only the victim but its customers and partners, especially if these partners have network interconnections" (NIST, 2020b, p. 1). A *cyber SC attack* is defined as "any deliberate action or assault on the SC system with the intent to compromise its processes, procedures, and delivery of electronic products, information flows and services" (Pandey et al., 2020, p. 9). Such attacks made up over 50 percent of incidents in 2019, having increased by 78 percent over the same year and successfully caused significant damage (Symantec, 2019, p. 9). SC attacks "are most effective when they find targets with the broadest reach possible" (Townsend, 2019). *Downstream cyberattacks* are directed against customers in the chain, whilst *upstream cyberattacks* affect suppliers, third-party vendors and external organizations remotely accessing the SC system (Pandey et al., 2020, p. 9). This may include information distortion attacks on SC members, or the supply of false information such as business licences, product test certificates, or trade documents, illegally attracting customers by stealing a well-known company's identity. *Malicious tampering* of assets may be caused by pre-installing and hiding malware at one point of a product's SC. Thereby,

> "vulnerabilities may be introduced during any phase of the product life cycle: design, development and production, distribution, acquisition and deployment, maintenance and disposal [through] the incorporation of malicious software, hardware, and counterfeit components; flawed product designs; and poor manufacturing processes and maintenance procedures" (CISA, 2020, p. 1).

Throughout 2020, NIST criticised that several SC security controls are rarely implemented across the private sector. This includes inconsistent monitoring of supplier compliance, incomprehensive enforcement of anti-malware protection mechanisms and a lack of logging of "critical suppliers accessing networks" (NIST, 2020a, p. 16).

## 5.1.2  NATO Cyber Supply Chain Handling

It is  key to trace the NATO handling of its cyber SC throughout the organizational system and mechanisms, to locate potential cyber gaps and vulnerabilities when utilizing mission-critical information retrieved from space assets, and accordingly assign responsibilities. Within NATO, the term *supply* covers "material and items used in the equipment, support, and maintenance of military forces [including their] provisioning, distribution and replenishment" (NATO, 2017, p. 2). As *enabling functions*, NATO lists

1. *Logistic information management*, which couples IT to logistic processes to meet information requirements by NATO commanders and allied MS;
2. *Reception, staging and onward movement,* or the operational transition of personnel, units, materiel, and equipment from arrival to final destination;

3. *Contracting* with external suppliers*, which has "become increasingly important to the conduct of operations, especially when operating beyond NATO territory [to] gain quick access to in-country resources by procuring the supplies and services that the commander requires" (NATO, 2017, p. 2); and

4. *Host nation support*, to provide NATO commanders with arrangements and national partnerships (NATO, 2017, p. 2).

Within NATO, *production logistics*, also called *acquisition logistics*, are overseen by the NATO *Conference of National Armament Directors (CNAD)*, largely include the industrial domain and concern

"the planning, design, development and procurement of equipment and therefore […] standardization and interoperability, contracting, quality assurance, acquiring spares, reliability and maintainability analysis, safety standards for equipment, specifications and production processes, trials and testing, codification, equipment documentation, and configuration control and modifications" (NATO, 2017, p. 2).

The *NATO NDPP* emphasises on cyber defence as a both immediate and long-term challenge. Thereby, NATO states that "National and NATO logistic plans must ensure that logistic resources of sufficient quantity and quality are available […] to support forces as needed" (NATO, 2020, p. 3). As stated by Vasen, NATO procures commercial services related to cyber sustainment of commercially procured SATCOM services via the NCIA (Vasen, 2020b). *Cyber sustainment* is described by AJP-3.20 as supporting logistics being "dependent on cyberspace with their reliance on computerised networks, for example, to share information; order material and medication; process databases; or have a logistical overview" (NSO, 2020, p. 10). It aims to ensure functionality and security of NATO systems and network, unobstructed access to them, maintenance of ability to perform, and preservation of operational effectiveness.

In 2018, the *European Defense Agency* (EDA) stated that it aligned its annual *Capability Development Plan* (CDP) with the NDPP, to avoid unnecessary duplication (EDA, 2018, p. 4). However, Widmann (2020) sees a major difficulty in aligning NATO's and the EU's interests in the cyber SC field: whilst the EU rather focuses on economic and political goals, NATO encompasses defence and national security issues. Thus, he states that

"what happens is, the NDPP spurs further competition between private companies like Airbus or Boeing to get awarded defence contracts. There arise political considerations which can affect cybersecurity decisions" (Widmann, 2020).

Furthermore, alignment of MS cybersecurity of their space assets' SC is hindered, as it is handled by each MS independently (*see *).

## 5.2  Threats and Vulnerabilities

As stated by Falco (2020), SC intrusions are currently considered to be the most important threat to space asset security right now. Meanwhile, concerns on cybersecurity of space assets

"have historically focused mostly on electronic warfare threats such as jamming [or] spoofing […] Better understanding of cyberthreats has led to a realization that […] motivated adversaries may develop highly targeted malware, assumptions about isolated networks may be invalid, and adversaries may breach development environments and SCs" (Bailey, 2019, p. 9).

Therefore, Caudill (2020) advises to generally assume "that every single component of a space asset was already maliciously modified throughout its SC". This is caused by members in the space community, using for example non-authorised vendors or circumventing supplier *due diligence*[75] (Lister, 2020b). In 2020, the phrase '*SC security*' became closely interconnected with the use of non-proprietary, open-source software being poorly understood by policy-makers due to heightened opacity of implemented software, as well as Chinese intrusion for intellectual property theft (Access Partnership, 2020, pp. 12-14). Thereby, such SCs are

> "thought to be unsusceptible to cyberattacks due to some of the same reasons as spacecraft: closed supply chains, unique embedded hardware/software systems, and physical protections [...] Compromising the hardware and software supply chains [was] successfully executed in the IT-embedded world, and space systems could fall victim to similar attacks if proper protections are not taken" (Bailey, 2019, p. 9).

### 5.2.1  Dual-Use Space Assets: NATO Dependency on the Private Sector

#### 5.2.1.1  Background

As stated by NATO NSO, especially in cyberspace and outer space, space assets may be of dual-use, thus "can be used for both civilian and military purposes" (NSO, 2016b; Unal, 2019, p. 7). Most space assets serve multiple commercial functions, whilst "civil satellites, operated by private companies, may be used to fulfil specific missions in locations where NATO allies do not have their own space equipment" (Unal, 2019, p. 8). MS' procurement policies typically require to commercially "procure equipment and software to be integrated into their national defence architecture, which becomes part of the overall NATO capability. The commercial supply chain is embedded in nearly every aspect of military equipment" (Unal, 2019, p. 8).[76] Dual-usage of space assets is not a new phenomenon and coins the term *New Space,* which is

> "the growing use of civil technologies for space missions and the military working with private space companies, thus private companies operating in places that just until a few years ago were occupied only by government agencies" (Baram, 2020).

Hence, "the number of private companies that have space assets, and military and national security dependence on those systems is much greater than people realise" (Widmann, 2020).

#### 5.2.1.2  Dual-Use SATCOM Assets

In the field of military communications, the highly secret SATCOM is traditionally provided by government satellites operating in GEO. However, such satellites are not competitive with new, commercial services by SmallSat constellations in LEO (Hallex & Cottom, 2020, p. 21). Newer LEO constellations provide lower latency and higher bandwidth, making enhanced SATCOM accessible to mobile users and high-latitude areas previously being outside GEO SATCOM coverage, such as Alaska, Russia, northern Canada, or Scandinavia. As a result, there is growing military demand for such services. Examples are services provided by the US company *SpaceX Starlink*, having contracted with the US government to launch almost 12000 satellites into space, or the UK company *OneWeb*, having just been acquired by the UK government to expand with 2000 satellites by 2020 (Hallex & Cottom, 2020, p. 21; Henry, 2020). Within NATO NCIA, IT and SATCOM equipment is generally bought as COTS from private firms and subsequently being secured - *hardened* - as "there is no meaning to produce a modem for military purposes, unless you need encryption and decryption" (Conti, 2020). As emphasised by Conti, "SATCOM is really a commercial, not a military system. What makes it military, is the subsequent encryption and decryption of data" (Conti, 2020).

---

[75] ***Due Diligence:*** describes "the detailed examination of a company and its financial records, done before becoming involved in a business arrangement with it" (Cambridge Dictionary, 2020b).
[76] Whilst traditionally, there used to be an innovation spin-off from military to industry, thus civilian goods being the collateral result of military research, it became more of a spin-in on the reverse path (NASA, 2020).

### 5.2.1.3  Dual-Use ISR Assets

On the ISR side, NATO MS governments are increasingly utilising commercial assets to provide data to the military, due to *lower cost and higher access,* as commercial providers are confronted with less limitations to match intelligence needs and requirements (Bander et al., 2020)*.* Using commercial space assets for ISR a) *frees up more secure governmental intelligence satellites* for the most important military targets; and b) *eases the sharing of commercially gathered imagery with allies* as this is generally registered at a lower level of classification*.* US DOD estimated that around 80 percent of DOD SATCOM used commercial lines during the last decade (Bander et al., 2020)*.* This is especially true so for weather services and in the *earth observation* field, using commercial SmallSat constellations to retrieve high-revisit, optical imagery, all-weather and night-time services (Hallex & Cottom, 2020, p. 22). Therefore, such assets may significantly facilitate and logistics, mapping, and emergency rescue and search. The US *National Geospatial-Intelligence Agency* (NGA) is the largest customer for these services, profiting from lower prices for industrial monitoring, marine transportation analytics, business intelligence, resource management, and other decision-making, data-driven practices.

### 5.2.1.4  Risks

The indistinguishable and intertwined use between military and civilian space asset applications poses a key challenge to monitoring and regulating space (Moon, 2017, p. 4). Currently, around 40 percent of active orbit space asset are military satellites or contain some sort of military application. The integration of dual-use objects is not necessarily a vulnerability,

> "as long as commercial equipment is designed to military standards and is secure. However, if military standards are not met, items procured from commercial industry with design flaws may expose NATO's systems to additional vulnerabilities" (Unal, 2019, p. 8).

Thereby, Wells and Sielaff add that

> "there is a whole series of protection measures that are completely different from commercial systems but must be included in military space systems, such as anti-jamming and -spoofing capabilities, or designing the space assets to specifically protect the data, which includes commands being transmitted from the spacecraft" (Wells & Sielaff, 2020).

Commercial space assets are being considered a high-value and easily breakable target, as military space assets are typically designed more carefully (Hallex & Cottom, 2020, p. 22). In case of exploiting such dual-use space asset vulnerabilities, hostile state and non-state actors could track and subsequently attack targets with heightened effectiveness.

This is aggravated by a lack of awareness and understanding, with MS governments rather emphasizing on ensuring data availability and compatibility in the field, and thus having a *warfighter perspective*. This is where the military in the field would be less concerned about a data's source but rather making sure "that it is a) coming, b) not going to be interrupted, and c) not going to be corrupted or interfered with" (Bander et al., 2020). Underscoring this, Conti states that with the increasing involvement of commercial technology, requirements for NATO technicians changed. Whilst in the past, technicians were required to possess specific knowledge and understanding of modems used such as for ICT or air conditioning, increasingly, the expertise need has become solely about *operating* such assets*:* "we don't touch the equipment anymore. If something is wrong, we take it, we pack it, and send it back to the company instead of fixing it by ourselves" (Conti, 2020). This causes long-term reliance on the asset's vendor for security issues. Falco criticises the "limited visibility that militaries have into the parts that the space asset is composed of" (Falco, 2020). However, there is in contrast too little concern to increase vetting of suppliers or restrict outsourcing of critical functions.

Potential attacks on space assets throughout their SC might include the substitution of hardware or software components by counterfeit parts, implementing security backdoors by software vendors or execute invisible changes during the manufacturing (Falco, 2018a; Unal, 2019). Past examples of so-caused cyber SASC incidents exist: in 1998, the US-German ROSAT X-Ray satellite was hacked by infiltrating computer systems at the US Maryland Goddard Space Flight Centre, instructing the satellite to direct its solar panels directly towards the sun, hence burning out its batteries and leaving the satellite defunct (Akoto, 2020, p. 4). In 1999, hackers took control of the UK 's SkyNet satellites, holding it for ransom. Over recent years, cyberattacks on space assets became more recurrent: in 2008, hackers, suspected to be Chinese, reportedly took full control over two NASA satellites for two to nine minutes (Akoto, 2020, p. 5). In 2018, another group of reportedly Chinese and Iranian state-backed hackers attacked defence contractors and space asset operators. In June 2020, ZDNet reported the DopplePaymer ransomware gang to have a breached *DMI*, one of NASA's IT contractors and a major US IT and cybersecurity provider, through having infected its networks via a SC contractor (Cimpanu, 2020, p. 1). Whilst it seems to be still unclear how deep networks were breached, it appears that the DopplePaymer gang achieved to breach DMI's NASA-related infrastructure and obtain NASA-related files. The group boasted about its breach by publicly posting twenty NASA archive files, revealing details on project plans, Human Resource (HR) documents including employee details matching public LinkedIn profiles. Furthermore, it posted a comprehensive list of workstations being part of DMI's internal network, which the ransomware gang encrypted and hold for ransom.

### 5.2.2  Competitiveness Leading to Insufficient Cybersecurity Controls

NATO MS may profit from increased space capacity when outsourcing to the commercial sector, however they do so in exchange for *loss of control* (Bander et al., 2020). This is rooted in

a) *Cost minimisation*: security issues may be caused by pursuing a traditional outsourcing approach, thus procuring the lowest-cost option at a still-acceptable quality (Bratton, 2020b, pp. 2-6). This approach is likely to cause a trade-off in SASC reliability and built-in cybersecurity measures.

b) *Competition race between nation states:* In 2020, the US Cybersecurity Solarium Commission acknowledged that MS fell so far behind Chinese Tech Industry leaders, that regaining competitiveness and competence may become impossible (CSC, 2020, p. 119). To solve this challenge, governments strive to collaborate with large prime contractors, small to medium sized companies and subcontractors (CSC, 2020, pp. 8-10). Also, this is driven by the desire to *deter* through strongly competing with for example Russian and Chinese actors (Moon, 2017, p. 3): the US and other Western actors, as well as China's Aerospace Science and Industry Corporation, Russia and India are actively developing own small satellite projects in partnership with commercial suppliers to access more innovative and effective space asset-enabled services (Hallex & Cottom, 2020, pp. 4/ 23).

Hence, NATO becomes more dependent on space assets to retrieve ever better ISR, process and analysis information and stay competitive (Baram, 2020). Thereby, Widmann states that outsourcing "is the better way to get innovations for militaries. However, the time to actually go through the acquisition process for new emerging technologies puts NATO at a disadvantage to adversarial nations" (Widmann, 2020). He points to the increasing *military trade-off between thorough risk analysis and resource investment,* since vetting items is time-consuming and expensive once procured from a commercial vendor (Widmann, 2020). Hence, Livingstone criticises that in this rush to tune military and intelligence space assets, dependency on the civilian sector steadily increases. Even though it might be acceptable to delegate some of the space-asset requirements to civilian capacities, governments must be aware that these are "not nearly as secure as your military and intelligence space capacities" (D. Livingstone, 2020). As an example, Falco states that even in 2017, NASA's laboratory used to work with a catalogue of asset parts for science exploration projects with factories mostly located in China and containing Chinese-manufactured codes.

On the  commercial side, companies do "not only supply products to governments, but they also compete commercially" (US DIA, 2019, p. 7), leading to a trade-off of cybersecurity controls. Caudill also points to CIA of data being disregarded by many military space asset developers and suppliers:

> "small space asset operators do not really care about CIA. In contrast, state actors do and should care because of competition: you don't want another nation state to be able to hack your weather or GPS satellite constellation" (Caudill, 2020).

In consequence, targeting space assets is more easily facilitated than ever before (Lister, 2020b), "with the wider availability of lower cost, small satellites and with the prospect of large constellations consisting of thousands of satellites [,] technological and cost barriers have fallen" (US DIA, 2019, p. 7). Such market forces pressure many companies into cutting costs and speeding up production and development - at the cost of suppressing cybersecurity concerns (Akoto, 2020, pp. 1-8). Hence,

> "the private sector's motivation at the end of the day is to make *profit,* which is misaligned with the cause of the federal government or cooperating states to secure space initiatives for all of their citizens when outsourcing ideas, technology and products" (Lister, 2020b).

Additionally, since technical capabilities steadily lower in price, the market opens to a broader range of customers, and hence potential cyber threat actors, whilst security controls of the cyber SASC decrease (Livingstone, 2020). This becomes a risk to NATO, as "we now have commercially available software such as hacking tools on the market, being updated every day, that make it easier to target space based assets and the GS that control them" (Lister, 2020b). Finally, Lee (2020) criticises that "even renowned companies are famous for doing things very economically, trading off the consideration of cybersecurity safety when adapting terrestrial components to space applications [and thus] cutting a significant corner in cybersecurity".

### 5.2.3 Classification Misalignment Leading to Lacking Cooperation

MS classification restrictions frequently hinder support to NATO operations. As outlined in AJP-3.3, since many space systems are determined by complex relationships,

> "preparatory planning and agreements should be in place […] made *as transparent as possible* to the supported commander [to] provide clarity of the relationships to NATO strategic and operational commands and clearly define the process for requesting and disseminating the products and/or services" (NSO, 2016, p. 5-9/-10).

However, there is "no one overarching framework that covers both cybersecurity and SC, as each MS will have its own classification approach" (Wells & Sielaff, 2020). As agreed upon by Wells, Sielaff and Vasen, the core problem is with the challenge of exchanging data, for example by using diverging formatting of data encryption (Wells & Sielaff, 2020). Silverstein points out that

> "unfortunately, sharing and consultation do not address the alliance's most conspicuous vulnerabilities in space […] Notably, some NATO allies have reservations about ceding control of space systems to foreign commanders during a crisis. This might drive NATO members away from beneficial redundancies and toward inefficient replication" (Silverstein, 2020, p. 5).

Such redundancies are needed to maintain multiple ways of communication, to ensure resiliency if systems are attacked or need to be suppressed occasionally for covert military operations (T. Vasen, 2020). This includes making use of multiple navigation system providers, such as the EU's GALILEO and other GNSS in addition to US GPS. In regards to NATO's new Space Policy, it is unclear if "the current NATO space ecosystem is adequately layered with redundancies to realize the benefits […] a dearth of sophisticated redundancies would make those capabilities - unsupported by substitute systems as they are - valuable targets" (Silverstein, 2020, p. 5).

Both Heren and Vasen specify that such classification restrictions cause compatibility and collaboration issues, hindering operating at the required speed of a given operation. Thereby, Wells points to standard frequency bands being used by several European countries as well as Russia and China, and thus allowing for a common set of interoperability and sharing tools. In turn, when creating a single network, the risk for a cyberattack's impact increases due to heightened interconnectedness (Bates, 2020, p. 4). Hence, NATO MS military forces tend to remain with operating different data protection and segmentation regimes of classified and unclassified data on computer systems (Heren, 2020). This is reinforced by segmentation not being a required standard throughout the private cyber SASC. In sum, this leads to a lack of control by NATO over the integrity and source of data and how information is processed.

#### 5.2.3.1 Intelligence

For NATO and its MS, it is especially difficult to share actionable threat intelligence with acquisition companies as the central challenge "is to protect sources and methods" (Eversden, 2020, p. 1). Hence, intelligence sharing requires *trust,* whose establishment and maintenance

> "is a long-term commitment being hard to do, not just from government to commercial, but as well between governments. We don't like to reveal sources, methods and tactics in the intelligence community for a good reason, but yet there has got to be a way in which we are able to share more vital, actionable intelligence, especially for cyber" (Yingst, 2020).

Furthermore, Yingst states that

"many times, commercial vendors are reluctant to reveal information because they are afraid of either attribution, giving up proprietary information or potentially being penalised if they are perceived to be in violation of a regulation. We have to develop that trust between governments and commercial vendors, as there has to be a sense of safety and security that what companies are sharing is for the mutual benefit of both parties, and that they are going to be protected" (Yingst, 2020).

*Insider threats* are the main concern to not share intelligence: both Wells and Sielaff criticise that only a handful of people in crucial space asset companies are cleared. As stated by Sielaff, "companies can and should protect themselves technically and physically, but in the end, if someone inside the SC leaks sensitive information, intelligence gets exposed" (Wells & Sielaff, 2020). Hence, Yingst points to the importance of increasing the issuing of security clearances to "enable a great deal more open discussion to take place" (Yingst, 2020), as NATO operates based on a *need-to-know policy*. This includes information sharing "only if the employee has a need-to-know to perform his or her job" (NATO ACT, 2019) and is detailed in *C-M(2002)49: Security Within the North Atlantic Treaty Organisation*:

"classified information shall be disseminated solely on the basis of the principle of need-to-know to individuals who have been briefed on the relevant security procedures; in addition, only security cleared individuals shall have access to information classified CONFIDENTIAL and above" (NATO, 2002, p. 9)*.

However, "it costs a lot of money for these companies to have security clearances" (Yingst, 2020). As governments oftentimes do not possess sufficient manpower, and clearing processes are typically stretched over an extended period of time, which "poses an opportunity for adversaries to start recruiting" (Wells & Sielaff, 2020). Across the board, interviewees state that the level of security needs to be increased through implementing more management and access controls, higher levels of clearance for personnel handling more sensitive data, and a greater number of background checks within the government to establish confidence in SC partners (Yingst, 2020).

### 5.2.4  *Lacking Legal Regulation*

As stated by Caudill, *New Space* is growing rapidly, whilst few security measures and even less enforced regulations are in place (Caudill, 2020). Thereby, dual-use objects are being with more difficulty legally identified as legitimate military objects and thus face fewer security restrictions (NSO, 2016b). Worsening, private space asset component providers are reluctant to regulate their SCs, as players are seeking flexible options to ensure workflows and keep costs low (Baugh, 2020, pp. 1-7). As stated by Lister,

"we could assume that there are stringent governmental cybersecurity controls in place for those companies building sophisticated space-based assets. However, what we have seen with regulations so far in 2020 is that there has not been a specific guidance on what exactly this should look like, for example, whether or not the providers should be encrypting all the satellite links, or any other specific guidance on securing sensitive or operational data" (Lister, 2020b).

This causes several political and regulatory issues. Silverstein notes that

> "NATO members may not have comparable capacities to observe or monitor space activities, leaving allies unable to assess independently collected data […] Without common methods or an established procedure for space observation and shared threat-assessment tools, NATO may be unable to arrive at a conclusion and may cede initiative back to an adversary" (Silverstein, 2020, p. 7).

Due to lacking intergovernmental cyber SC standards, private space asset suppliers remain responsible for their own cybersecurity, which leads to complacency, a lack of clarity on responsibility and liability in case of cyber breaches, and hence missing cybersecurity efforts (Falco, 2018b, pp. 11-12). Therefore, analysts have begun "to advocate for strong government involvement in the development and regulation of cybersecurity standards for satellites and other space assets" (Akoto, 2020, p. 7). However, the pace of doing so remains low.

Furthermore, a major concern amongst scholars and practitioners is COTS regulation along the cyber SASC (Akoto, 2020; Falco, 2018b; Koch & Golling, 2016; Ziolkowski et al., 2013). As stated by UNOOSA, SmallSat missions "are not always conducted in full compliance with international obligations, regulations and relevant voluntary guidelines [,] due to inexperience or unfamiliarity with the national and international regulatory framework" (UNOOSA, 2015, p. 3). Thereby, Falco states that it is impossible to simply apply already existing cybersecurity frameworks, as SmallSats are much more complicated in their technological structure than other space assets.

Within NATO, regulation and strategic NATO guidance of SC cybersecurity for space assets is, beyond minor space-related subjects, nearly non-existent because it follows the individual regulations of the operating nation (Vasen, 2020b). In AJP-3.20, NATO emphasised on the need to further

> "protect or ensure the continued function and resilience of capabilities and assets, *including [...] supply chains* critical to the execution of NATO mission-essential functions in any operating environment or condition" (NSO, 2020, p. 4).

Due to bureaucracy and high classification restrictions between MS, it remains difficult for NATO to align MS positions on space asset security (Vasen, 2020). As stated by Falco,

> "every country is currently pursuing their own version of different standards for security systems, either because of nationalism or because of national security concerns. However, space needs an international perspective for innovation and scientific exploration. We need to have a standard platform fully vetted by the community to provide diligence for security frameworks" (Falco, 2020).

### 5.2.5  *Lacking Awareness*

At the strategic level, Livingston criticises the high degree of *stove-pipe thinking* within the military sector, being the restricted holistic view and vertical flow of information when considering the security of data delivered from space assets (Livingstone, 2020). This lack of reporting and dialogue is aggravated by a high degree of military bureaucracy. As criticized by Lee, the military space asset sector

> "is very regimented and locked down. When you start having any conversations with people who are in charge of building military or intelligence satellites, they feel like they have cybersecurity completely under control" (Lee, 2020).

This is underscored by the current state of *military stove-pipe system* across NATO and MS, which leads to limited agility and prevents "to effectively exchange information [which] has been a long-time weakness in the acquisition of MILSATCOM systems" (US Space Force & US Air Force, 2020, p. 8). Referring to recent operational NATO exercises like the Trident series, Widmann doubts that awareness of SC threats was operationally trained as much as it could have been, including the strategic planning process that it takes to secure a SC. Moreover, further security vulnerabilities are caused by the fundamental difference between the *high speed of cyberattacks* as opposed to the *reaction speed* of MS governments and military entities. This is aggravated by the vast volume of data received on a day-to-day basis from space assets. This makes it difficult for the military to determine if, to what extent, and at what point of the SC, MS may have been compromised (Tucker, 2019). Livingstone points out that "there is a less than perfect understanding of governments and space asset manufacturers of SC threats" (Livingstone, 2020). This enables cyber criminals to corrupt reliability and accuracy of data with a low probability of being discovered (Livingstone & Lewis, 2016).

As NATO does not operate its own satellites but is depending on MS' capabilities, it does not have any oversight but only receives data, products and services from the nations (Heren, 2020). Regarding NCIA, Conti states that it is *difficult for NCIA to keep the overview over the high number of companies NATO deals with,* especially for an individual security officer. Yingst and Lister stress the problem of managing an *uncontrollable number of supplier tiers[77],* especially if those are circumventing due diligence processes. Lister emphasises on this obscurity of the SASC, making it highly attractive to hackers:

> "A lot of manufacturers use very specialised technologies procured through an unknown number of manufacturers, and an unknown number of integrators, opening an unknown number of opportunities for attackers to gain access. Not all companies even understand from where their developers are pulling information from. So that's very concerning from an operational standpoint" (Lister, 2020b).

Yuval also points to market failure due to a lack of

> "uniform language [as] most customers develop alone; poor maintenance; one size fits all; difficulty to manage the process [as] resources are often lacking; limited scope of testing; lack of technical knowhow; and lack of responsiveness of suppliers" (Yuval, 2019, pp. 3-5).

The ongoing COVID pandemic has exposed further SC volatilities within the space industry, such as product shortages or lagging timelines (Lister, 2020, pp. 2-5). As the pandemic proceeds, the US Pentagon is discussing space SC concerns, especially for commercial start-ups, smaller suppliers, and other non-traditional defence industry contractors. Thus, Lister states that

> "from an outsourcing perspective on the cyber SC, the pandemic puts new pressure on the space community to find and locate vulnerabilities, conduct supplier due diligence or identify alternative suppliers that meet quality and quantity timelines. This temps some companies to circumvent supplier due diligence in order to save time and resources" (Lister, 2020b).

---

[77] ***Tiers:*** In a SC, a tier one company is "the most important member […] tier two companies supply companies in tier one; tier three supplies tier two, and so on. Tiered supply chains are common in industries such as aerospace […] where the final

### 5.2.6 *Understaffing and Lacking National Financial Resources*

Handling security of data and services retrieved from or transmitted through NATO MS space assets remains complicated due to *significant understaffing*. Thereby, processes can be significantly delayed as SpSC is manned with only one single permanent SME, hence being incapable to operate in a 24/7 availability (Vasen, 2020, pp. 23-25). Moreover, on the tactical level, no permanent SpSC has been foreseen, even though TRJE18 showed a clear demand for at least one tactical Space SME both in times of conflict and peace. Such an expert is needed to educate, advise, and train HQ representatives on space support limitations. Whilst there exists a need to stock up the current total 20 dedicated NATO staff to at least 150 positions, at present a lot of positions are *double or triple hatted* between air, cyber and space divisions. These organizational structures create disjointed units that are unequipped with regard to the right level of expertise, experience or training (Heren, 2020). This issue is aggravated by most relevant NATO MS currently prioritizing their *own* space capabilities and internal reorganisation and hence hesitating to contribute rare, specialized personnel to NATO (Vasen, 2020). However, with the creation of a new NATO Space Centre at Ramstein, it is safe to assume that an yet undecided number of additional Space professionals will be added to the NATO Command Structure (Heren, 2020). Furthermore, such a centre will, once established and fully staffed, close the 24/7 gap in the NATO SpSC functionality.

Additionally, NATO MS are not equal in terms of financial capability to secure SASC cybersecurity. Zarkan (2020) points to the challenge for states and companies to invest enough money into the development of secure space assets, and to finance technology and research. The high costs of guaranteeing each component's security may be prohibitive for companies to make cybersecurity a high priority, which therefore might be ready to take the risk of keeping poor protection and facing the consequences. Especially since the start of the COVID pandemic,

> "smaller suppliers and commercial space companies may not have infinite resources focused on cybersecurity. There is the concern that these companies will start circumventing any due diligence processes for their SC and their vendors, driving the need for prioritising risk management on hardware and software services and the people that support both" (Lister, 2020b).

As stated by Timm, the most difficult challenge is to get lower SC tiers to comply with cybersecurity requirements, since, even if companies may get reimbursed for cybersecurity costs over the course of a governmental contract, this may take several years (Bander et al., 2020). Thus, smaller companies lack the incentive to enhance security efforts and prevent vulnerabilities (CSC, 2020, pp. 8-10). This is more so the case for low-cost space operations during which the cost of implementing cybersecurity might exceed cost of COTS components itself. As stated by Caudill, "nobody along the way has the financial intention to secure the cyber side of space missions" (Caudill, 2020).

## 5.3 *Summary*

Numerous cyber intrusion opportunities threaten the delivery and integrity of NATO mission-critical data. These are caused by the involvement of manifold suppliers and manufacturers along the cyber SASC. Outsourcing to such external public and private companies is needed to stay competitive, deterring and up to date with innovative technology enhancing the outcome of NATO's operations to

---

product consists of many complex components [...] that must comply with stringent quality, manufacturing and business standards" (Linton, 2019).

the best possible extent. Cybersecurity threats are aggravated by the increasing obscurity of SCs, easing the implementation of third-party vulnerabilities into relevant soft- and hardware. Driving causes behind such threats are:

Table v: Cyber SASC Threats and Causes

| Cause | | Cyber SASC Threat |
|-------|---|-------------------|
| *NATO's dependency on unregulated unsecured dual-use space assets* | - | leaving open the quest for liability and responsibility in case of a cyberattack along their SC; |
| *Market pressure which increases NATO's need to stay competitive and thus deterrent* | - | diminishing the focus on cybersecurity by both public and private actors along the SC; |
| *Classification restrictions between member states and private companies* | - | worsening cyber insecurities and limiting the sharing of essential threat intelligence to prevent cyberattacks; |
| *Lacking awareness of the significance of securing the cyber SASC* | - | leading to a lack of risk management training, organizational overview and hence performance during crises such as the ongoing COVID pandemic; |
| *Significant understaffing and lack of clarity in coordinating space-related operational tasks between NATO entities* | - | being aggravated by some MS' lack of resources and/or willingness to focus on cybersecurity enhancement and prevent vulnerabilities. |

The following and final chapter aims at identifying policy recommendations to reduce such vulnerabilities and gaps in NATO's handling of its cyber SASC, as identified through the preceding literature and field research.

# 6 Policy Recommendations

As outlined by NATO Secretary General (SG) Stoltenberg, NATO's new Space Policy will allow NATO to leverage its collective MS' portfolio "of space-based technologies to supplement lost capabilities and negate adversarial interference with space systems" (NATO, 2019). Stoltenberg also states that the Space Policy enables NATO to establish a MS consultation and conversation forum, formally within NATO and informally between its MS (NATO, 2019). Through these efforts, NATO lives up to its goal to provide cooperative security and collective defence to its MS regarding space and cyberspace-related challenges (Baram & Wechsler, 2020, pp. 5-6). Livingstone underlines the solidity of the NATO alliance over the past 70 years, "which makes it an established, preformatted, trusted alliance. NATO provides the required platform to invite all those concerned in the governmental and commercial space asset sector, instead of creating some new cooperation platform" (Livingstone, 2020).

Based on these conclusions, NATO should consider two major collaboration initiatives: a) *Raising Awareness* throughout the whole of the NATO system, and b) *Pushing forward the creation of Regulation* through a standardized NATO security framework on SASC cybersecurity. Doing so would allow NATO and its MS to recognise cyberthreats to mission-critical data earlier on along its SASC. Furthermore, a standardized NATO security framework would allow the alignment of working language, external and internal assessments, as well as incident reporting and knowledge on risks, vulnerabilities and consequences (Widmann, 2020). The following chapter aims at detailing those approaches further, transforming them into concrete policy recommendations for NATO, its MS and private sector entities.

## 6.1 Awareness

### 6.1.1 Initiate Intersectoral Collaboration

As emphasised on by Falco, international collaboration is crucial for designing safe space and streamlining differing security and ethical standards between countries. Thereby, Falco states that

> "it seems like none of the international bodies that we have today are effective in coordinating such efforts, but there needs to be international communication because if not geographically defined systems in space are attacked, there is potential for hostile impact on everybody" (Falco, 2020; *see 4.2.5*).

Urban (2020) also designates commercial and governmental cooperation across borders as compulsory and inevitable to remain competitive. Using NATO as a collaboration platform on cyber SASC security would allow for an inclusive and comprehensive approach due to its transatlantic MS: European MS rely heavily on US industry-produced space technologies and components. Trans-Atlantic agreements and interoperability may be eased through NATO, mediating, and coordinating security efforts, common compliance, and standards. Unal points to the NATO as the

> "fastest and least expensive ways to increase […] cyber resilience, improve incident handling and mitigate vulnerability to attacks [which] should foster timely information-sharing on cyberthreats, allowing stakeholders to enhance situational awareness and better protect their networks [and] facilitate rapid and early bilateral exchange of non-classified technical information related to cyberthreats and vulnerabilities" (Unal, 2019, p. 26),

Thus, NATO should advance the creation of an *effective multi-stakeholder approach* that includes as many as possible public and private sector SASC partners (Akoto, 2020, p. 7; Unal, 2019, p. 26). As stated by Lister, the community of MS and private space asset companies

"is not going to be able to prevent targeting of their systems, but it can proactively *work together to prevent, detect and prepare for how to respond to those incidents,* by developing resources such as white papers and working groups focused on how emerging technologies can help protect the space community from cyber incidents which provide their cyber SC perspective, as currently there is a lack of care and coherent rules to help ensure that supply risk management is consistent" (Lister, 2020b).

This should include at least one *annual Cyber SASC Risk Management[78] symposium,* involving SC executive representatives from both the commercial and governmental side to exchange on practical recommendations, status updates, current challenges, and estimated organizational impact of attacks (NIST, 2020, pp. 8-13). Furthermore, this includes sharing organizational best practices to identify, respond to and prioritize cyber SC risks, ensure cyber resilience and minimize SC incident impact on services (NIST, 2020, pp. 9-13). These symposiums should be led by NATO's *Industry Advisory Group* (NIAG), NATO's *Industry Cyber Partnership* (NCIP) as well as the NDPP MS Delegations and SME's to facilitate the incorporation of results. The NATO *Parliamentary Assembly Defence and Security Committee* (PA DSC) and the Board of Directors of commercial organizations, at least of more crucial tiers, should be involved to increase the cyber SC literacy of executives. Lister states that such recurring events would pave the way to achieve an uniform, high quality SC security, with NATO and its data suppliers continuously coaching each other up- and downstream and sharing lessons learned from SC incidents (Lister, 2020b).

Intersectoral collaboration should emphasise on establishing further PPPs along the SASC. As an incentive, "there has to be a cost-benefit to both sides, they have to see or feel that there is going to be a return on investment to build up trust" (Yingst, 2020). The first efforts are in progress. In June 2020, the Germany-based aerospace, maritime and cyber IT solutions company *AMC Solutions* held a virtual event on *Governance and Security in Outer Space*, gathering approximately 60 participants from 25 nations including NATO's CCDCOE (Space and Cybersecurity Project Group et al., 2020, p. 1). Thereby, participants acknowledged the connection between risks from cyber and outer space, current trends in security, governance and sustainability, and stability in space. Additionally, participants recognized the need for more open collaboration through starting to create internationally agreed documents, such as law, comprehensive frameworks and guidelines linking cyber and outer space, as well as a platform to discuss risk mitigation and jurisdiction in outer space. Subsequently, appropriate international collaboration platforms bringing together public and private space asset suppliers will be introduced as potential cyber SASC collaboration partners for NATO.

---

[78] *CSCRM* is defined as "a systematic process for managing [cyber] SC risks by identifying susceptibilities, vulnerabilities, and threats throughout the SC and developing mitigation strategies to combat those threats whether presented by the supplier, the product and its subcomponents, or the SC itself" (CSC, 2020, p. 138).

### 6.1.1.1  US Aerospace Industries Association (AIA)

Following Galer, another cooperation partner could be the US-based *Aerospace Industries Association (AIA)*, representing manufacturers and suppliers of civil, military, and business space systems, aircraft, and related equipment, services, and IT, beneath others. It aligns its member companies interests, from the biggest primes down to the individual supplier level, and provides more education and understanding on the growing interest on the nexus between space and SC security (Bander et al., 2020). Therein, following Timm, the *AIA Cybersecurity Committee* handles defence acquisition matters, beneath others for the DOD. It mostly analyses the infrastructure of the company, thus access to the networks, physical security and protection of intellectual property and aims at securing the SC, the actual product, and the mission systems themselves (*see 6.2*).

### 6.1.1.2  CCSDS and ISO

The *Consultative Committee for Space Data Systems (CCSDS),* currently consisting of eleven member agencies, 140 industrial associates and twenty-eight observer agencies, was formed in 1982 "by the major space agencies of the world to provide a forum for discussion of common problems in the development and operation of space data systems" (CCSDS, 2020a). This includes the

- *Agenzia Spaziale Italiana (ASI), Italy;*
- *Canadian Space Agency (CSA), Canada;*
- *Centre National d'Etudes Spatiales (CNES), France;*
- *China National Space Administration (CNSA);*
- *Deutsches Zentrum für Luft- und Raumfahrt (DLR), Germany;*
- *European Space Agency (ESA);*
- *Japan Aerospace Exploration Agency (JAXA), Japan;*
- *National Aeronautics and Space Administration (NASA), US;*
- *National Institute for Space Research (INPE), Brazil;*
- *Russian Federal Space Agency (RFSA);*
- *UK Space Agency.*

Its aim is to actively develop "recommendations for data- and information-systems standards to promote interoperability [… and] reduce the cost burden of spaceflight missions by allowing cost sharing between agencies and cost-effective commercialization" (CCSDS, 2020a).

CCSDS cooperates since 1990 with the *International Standards Organization* (ISO), an international non-governmental organization based in Geneva, Switzerland, bringing together 165 national standards bodies and "experts to share knowledge and develop voluntary, consensus-based, market relevant International Standards" (ISO, 2020). In collaboration, ISO and CCSDS formed the *ISO Technical Committee 20 Subcommittee 13* [*ISO TC 20/SC 13*] on designated "Space Data and Information Transfer Systems", to allow "completed CCSDS standards to be processed and approved as ISO standards" (CCSDS, 2020b). NATO should profit from this established platform on inter-governmental policy making and start collaborating on interoperable space asset regulations (*see 6.2*)

### 6.1.1.3  Space Information Sharing and Analysis Center (ISAC)

Yingst emphasises on collaborating with sector-specific *Information Sharing and Analysis Centres (ISACs)* to be able to share sensitive information, bringing together national cybersecurity information coordination centres as well as representatives from various private sectors. As defined by the EU Agency for Network and Information Security (ENISA), ISACs are

> "non-profit organizations that provide a central resource for gathering information on cyberthreats (in many cases to critical infrastructure) as well as allow two-way sharing of information between the private and the public sector" (ENISA, 2020a, p. 7).

In 2019, a new Space ISAC was sponsored and launched by US NASA, the US National Reconnaissance Centre and US Space Force (Space ISAC, 2020). On the international side, *SES* became a founding member, being already connected to NATO via LuxGovSat and thus offering an eased opportunity to start partnering (*see 3.2.3*). Thereby, the Space ISAC considers SC concerns as a research priority, whilst Erin Miller, VP of Operations, states that one of the Space ISAC's primary objectives

> "is to give its members, no matter how small they are, access to current threat intelligence, as not all have the resources to do it by themselves [whereas] we have started efforts to manage security in the supply chains and develop recommendations on policies, standards and security requirements" (Miller, 2020).

Such access to cyberthreat intelligence is oftentimes important to increase companies' awareness of vulnerabilities in the first place. The documentation of cooperation outcomes will be divided into confidential and public, as decided by the Space ISAC's members, which allows to share more confidential information. Additionally, Miller points to the Space ISACs member-exclusive *Priority Intelligence Requirements Document,* identifying current threats and adversaries and creating a "high-trust environment that has to exist in an ISAC for making it successful and create a common understanding" (Miller, 2020). Information is shared via a common online threat intelligence platform, in member meetings, alerts and reports around threats, which are released at a partnership level. This is complemented by a strong vetting process of potential ISAC members, to ensure the highest possible level of trust between partners.

### 6.1.1.4  CERTs, CSIRTs and ENISA

As recommended by Widmann, NATO should start collaborating and unifying specific *cyberthreat intelligence* centres such as *Computer Emergency Response Teams* (CERTs) from different countries. CERTs are defined as internationally distributed coordination centres responsible for "analyzing and reducing cyberthreats, vulnerabilities, disseminating cyberthreat warning information, and coordinating incident response activities" (CISA, 2020b). This should include the permanent *CERT for the EU institutions* (CERT-EU) and the *US National CERT* (NCERT) to work with each other on one platform, making SC protection an intelligence priority (Widmann, 2020). Additionally, EU MS are assisted by ENISA with the establishment and maintenance of currently more than 500 business *Computer Security Incident Response Teams* (CSIRT) to spur proactive detection, prevention, and analysis of general cybersecurity threats and incidents (ENISA, 2020, pp. 4-5; Vasen, 2020). This effort aims "to contribute to developing confidence and trust between the Member States and to promote swift and effective operational cooperation" (ENISA, 2018), and strengthen the EU's overall cyber resilience. Additionally, ENISA gathers information on currently used or planned incident response implementation, available measures and future growth and improvement areas. Hence,

ENISA aims to close gaps in tooling, differing data formats or standards in the interaction of different MS systems.

### 6.1.1.5  Orbital Security Alliance (OSA)

The *Orbital Security Alliance* (OSA) was formed to fill the gap of lacking cybersecurity guidelines for the space industry and create a common understanding of cybersecurity incidents in space (Orbital Security Alliance, 2020, p. 1). Therefore, it brings together representatives and experts from government, academia and industry of all sizes and missions. In 2020, OSA published the revised version of its *Commercial Space System Security Guide*. Therein, OSA discusses SC threats such as adversarial physical and espionage attacks, and states that "effective practices from other industrial sectors – augmented by measures tailored to new developments and specific space operations practices and needs – can form the basis for effective space sector supply chain cybersecurity" (OSA, 2020, p. 24). Furthermore, OSA recommends that "critical parts and software should be sourced from these trusted vendors and checked for signs of counterfeiting or malicious content" (OSA, 2020, p. 28). Finally, it strongly recommends a SC risk management programme

> "to ensure that each of their vendors handles hardware and software appropriately […] with an agreed-upon chain of custody [and] define approaches to ensure that vendors are trusted suppliers" (OSA, 2020, p. 28).

Such a SC risk management should be endorsed on a NATO strategic level, as suggested throughout the following subsection.

## 6.1.2  Increase NATO Strategic Level Engagement

As stated by Moon, one Ally's space assets being attacked will impact each other's MS's security (Moon, 2017, p. 8). Hence, the strategic coordination of space asset cybersecurity is vital for all NATO pillars of defence, whilst increased involvement of CSCRM should become a top priority for NATO's strategic level executives. Eventually, this would allow for a whole-of-alliance approach to deter space-based threats through showcasing the importance and visibility of the cyber SASC (Yuval, 2019, p. 6). Following AJP-3.20, NATO commanders are required to

> "continuously monitor their areas of interest to anticipate potential crises and allow them to assist the strategic level in understanding [which] includes an analysis of cyberspace as part of the overall understanding of the operating environment [and] potential risks to friendly or neutral usage of cyberspace […] against NATO, Member States, or neutral nations/entities (NSO, 2020, pp. 23-24).

This implies the creation of mission-specific guidance, agreements, arrangements, and security considerations. Aligning, NIST emphasises on CSCRM as

> "a critical capability required for organizations to reduce the risk of business interruption if a cyber incident were to occur [requiring] close integration across functional and business lines, engage executive leadership effectively […] foster close supplier relationships, and leverage industry standards throughout the supply chain lifecycle" (NIST, 2020, p. 5).

To integrate the strategic level more effectively, executive-level SC leadership councils should be established throughout NATO (NIST, 2020, p. 5). Education on cyber SASC risk management should be introduced to NATO CCDCOE's recurrent *Executive Cyber Seminars*. Widmann considers those seminars as the most appropriate level to discuss SC issues and cyber incidents, and consolidate strategic understanding of space asset cybersecurity and NATO's critical dependence on it (Wells & Sielaff, 2020). The aim is to "provide the senior leadership with a baseline of information, so that if they are faced with a cyber issue in the SASC, they would have the information that they need to make good decisions" (Widmann, 2020). Furthermore, as stated by Unal, the annual *NATO Information Assurance Symposium (NIAS)* could focus on space in the upcoming years. In doing so, NIAS could support the clarification on responsibilities and protection measures, and suggest higher-grade military cyber protection as well as security specifications for critical civilian space applications used by the military (Unal, 2019, p. 16). Such executive-level gatherings will showcase commitment, formalize responsibilities and therefore effectively ensure CSCRM (NIST, 2020, pp. 8-13). Concrete strategic decisions to timely increase resources, which might be decided upon during such seminars, include:

- *The overall review and extension of space management within the NCIA, and SpSC's role;*

- *Forming an organic SpSCE at each tactical-level HQ*, being permanently staffed by at least one SME or liaison officer (Vasen, 2020, pp. 21-23);

- *Establishing a Direct Liaison Authority (DIRLAUTH)* between the operational level and national space capacity providers to reduce requesting times (Vasen, 2020, pp. 23-25); and

- *Establishing a stand-alone space operations AJP* to properly acknowledge and consider increasingly complex space capabilities.

### 6.1.3  Spur Exchange of Academic Expertise

NATO dependency on untrusted technology should be counteracted by strategic investment in space asset Research and Development (R&D) in collaboration with its MS, and hence the development of more viable alternatives (CSC, 2020, pp. 8-10). This might be done in collaboration with the existing NATO Science and Technology Organisation (STO) Information Systems Technology Panel (NATO STO, 2020). As stated by Silverstein,

> "NATO has a unique advantage as a collective organization to convene strategists from like-minded but diverse military backgrounds […] a diverse cadre of experts would support the development of a comprehensive space strategy" (Silverstein, 2020, p. 7).

Livingstone suggests inviting outside organisations and cybersecurity experts to provide expertise and differing perspectives on the topic and focus more on including cybersecurity into space asset's designs. This might include for example to further spur cooperation between NATO and Europol, in particular between NCIA and the European Cybercrime Centre (EC3) via Europol's *Platform for Experts* (EPE). EPE is a "secure, collaborative web platform for specialists [which] facilitates the sharing of: best practices; documentation; innovation; knowledge; non-personal data on crime" (EUROPOL, 2020). Exchange and collaboration are enabled via virtual communities, sharing a wide range of expertise, know-how, and information, to foster an environment of trust and online collaboration. Thus, "governments should not compete with industry but instead take advantage of the expertise in the industry and help to spur innovation going forward" (Bander et al., 2020). Hence, NATO will be allowed to *encourage innovation and competition* and *drive excellence* on the market, whilst ensuring that *the appropriate levels of security are met*. This will prevent companies from striving for private capital outside MS and may help to overcome industry competitors' reluctance to share knowledge (GSA, 2019, p. 2). Also, common R&D will help to draft a common terminology within NATO and beyond, as technical cyber-speak should be adjusted and translated for the political space community (Waterman, 2019).

NATO currently reviews the establishment of a *NATO Space Technology Centre* (Waterman, 2020, pp. 1-6). Such a centre would

> "bring together the various space-related activities of the NCIA, from SATCOM and research to operational support […] summarized in a so-called "virtual hub" for space know-how, projects and tools" (Kanig & Forkert, 2020, p. 1).

This is important for the needed acceleration of NATO space awareness and the fielding of new military capabilities, facing for example the rise of *Artificial Intelligence*[79] (AI), further shifting the cyber landscape for space systems as space assets increasingly employ AI capabilities. Not only will NATO need to worry about cyber doctrine and how it interacts with space, but AI doctrine and how it intersects with cyber and space. Furthermore, the centre should be led by or involve NCIA in cooperation with other COEs like CCDCOE, to "give advice on relevant cyberthreats and vulnerabilities, such as those related to the integrity or security of supply chains" (Unal, 2019, p. 26). This involves research and recommendations for agreement on what exactly constitutes space as a domain, what type of similarities and divergence space shares with cyber, and finally what kind of SC security approach should be modelled for NATO (Unal, 2019, p. 28). In that way, technological advancements can be more securely incorporated in future NATO procurement of space asset services (Zarkan, 2020).

### 6.1.4 Conduct Mission-Specific Cyber SASC Risk Assessments

NATO needs to make sure to know the procurement and SC environment around each specific mission, sourcing the materials and services to the level of integrity required (Livingstone, 2020). External partners should be analysed based on an exact understanding and agreement on each mission's unique risk tolerance and critical assets, to better ensure responsibilities and liabilities (Lister, 2020). Risk assessment policy recommendations between MS and NATO are highly needed "so that everybody agrees with the same concept of diligence, safety, and risk management" (Lee, 2020). Silverstein points out that

> "without agreed-upon threat-assessment processes, allies may arrive at different conclusions about threats to space systems, based in part on their differing abilities to collect and analyse data. This directly impacts the alliance's ability to come to a consensus decision" (Silverstein, 2020).

NIST encourages the move "from detective to preventive capabilities in managing third-party cybersecurity risk" (NIST, 2020, p. 10) through establishing a proper SC cyber risk assessment framework. Doing so would ensure cyber SC risk accountability across the organization through manifesting policies, governance, procedures, tools and processes, as well as integrity, visibility and control over data sent to and received by any supplier (NIST, 2020, pp. 9-13). Furthermore, risks assessments are essential to "identify [,] determine [and] estimate the likelihood of the potential losses, [and] define the relative likelihood and consequence of various risks" (Pandey et al., 2020, p. 16). A proper NATO SASC cyber risk assessment framework should therefore include:

**1.** *Assessment of NATO-procured ICT components, such as hardware, software, and services:*

---

[79] *Artificial Intelligence:* Defined as „the study of how to produce machines that have some of the qualities that the human mind has, such as the ability to understand language, recognize pictures, solve problems, and learn" (Cambridge Dictionary, 2020).

Common risk assessment includes collaboratively and proactively using software to monitor SC interruptions, implement recall policies, or allocate resources early in the space asset development lifecycle to determine SC risks and their feasibility for NATO. Also, this includes involving independent cybersecurity experts for up-to-date analyses on procurement samples; considering "known vulnerabilities that could be exploited to carry out an attack or otherwise cause adverse events" (CCSDS, 2019, p. 12); and subsequently scoring potential vulnerability impacts on NATO (NIST, 2020a, p. 13). Unal states that "any digital system that relies upon near real-time information is vulnerable to cyberattacks […] In order to understand the value of each space-dependent capability, it is important to analyse the consequences of cyberattacks on each" (Unal, 2019, p. 16). Bridging the gaps between the diverging areas of SC, cybersecurity and space assets present a "fascinating opportunity for different states to work together, because they are providing completely different perspectives, data and evidence and thus encourage out-of-the-box thinking when anticipating different types of attacks" (Lister, 2020b). Additionally, NCIA could further take advantage of collaborative open-source tools such as the *Malware Information Sharing Platform (MISP)* to promote rapid "cooperation and information-sharing among allies [as being] one of the most effective defences in cyberspace" (Unal, 2019, p. 26).

## 2. *Analysis of upstream suppliers, suppliers' sources, and the larger SC ecosystem:*

It is key for NATO to identify and well manage *critical space asset suppliers,* which "if disrupted, would create a negative business impact on the organization" (NIST, 2020b, pp. 8-14). This includes implementing preliminary review processes and rank suppliers based on their criticality to NATO, following a supplier criticality score prioritizing SC risks; and automate supplier risk monitoring and its mitigation. Firstly, mission-critical systems, assets, data, and processes need to be identified and prioritized, secondly, those suppliers having access or providing infrastructure for such capacities. Additionally, supplier practice assessment protocols should be established, and supplier self-assessment questionnaires be implemented to determine whether key, agreed-upon controls and requirements are being met and or improvements are required. Such steps will enable NATO to "understand how NATO can partner with developers, suppliers and vendors, maintain cybersecurity standards and ensure that validity, integrity and availability of the SASC are secured" (Lister, 2020).

## 3. *Dedication of specified NATO personnel to effectively assessing and communicating supplier risks to NATO's executive leadership:*

A centralized space asset SASC cyber risk assessment team and cross-functional SC Risk Councils should be established to proactively review and mitigate relevant risks (NIST, 2020, pp. 5-8). These teams should be responsible for providing and enforcing oversight and guidance on SASC cyber risk assessments and requirements to business units, approving SC changes such as new suppliers or contract renewals, and providing joint post-incident analysis and recurring seminars on the current SASC cyberthreat landscape. Thus, this team would serve as the central resource for SC threats (CSC, 2020, pp. 8-10). Such a new task force would significantly improve SASC risk information sharing and appropriate resource funding to aggregate all-source information by public and private partners. It would allow to rapidly address SC incidents and simplify supplier management below the executive leadership level. Finally, it would set an increased focus on heightened SCSRM support for IT service, software, and hardware SCs, pointing out particular vulnerabilities of each MS's SC.

### 6.1.5  Train Space Asset CSCRM throughout NATO Exercises

NATO should include space asset CSCRM in every NATO exercise, to support NATO's forward-leaning, collective defence approach (Vasen, 2020, pp. 21-23). Such a thorough joint training would enhance

> "allied capacities to prevent, defend against, and recover from attacks on space infrastructure. In turn, these activities will demonstrate allied resolve to protect and leverage space systems in broader transatlantic security missions" (Silverstein, 2020, p. 9).

Measures should include "training for relevant stakeholders' organization-wide, such as in the departments of SC, […] legal, as well as key suppliers" (NIST, 2020a, p. 16). Lister pledges to include therefore relevant private sector suppliers and warns that

> "if the incident response to a cyber SC attack is not rehearsed by members of the space community beforehand, it is going to be too late to apply lessons learned or respond most efficiently and effectively when facing the first major cyber incident. Also, this allows NATO to understand all of the stakeholders of such an incident prior to responding to an incident, which can help save valuable response time during an actual incident" (Lister, 2020b).

Training the individual end user of space assets is key, as a lot of military staff are frequently redeployed (Wells & Sielaff, 2020). Such training will help to identify responsible NATO functions, audit suppliers and incorporate feedback and thus enhance the mutual and clear understanding of space assets' potential contribution to operations, as required in AJP-3.2 (CISA, 2020, p. 1).

Furthermore, Heren and Caudill advocate for engaging *white hat hackers*[80] to test for possible vulnerabilities throughout NATO training (Caudill, 2020).[81] This may involve issuing rewards for hackers who hack a satellite and disclose their method. Some other training steps may involve phishing simulation exercises, mapping upcoming cyber SC threats and priorities, or tabletop exercises to determine appropriate stakeholders and their responsibilities throughout incident response. Subsequently, training and test reports should be appropriately shared not only to the military and intelligence community, but as well the commercial sector (Livingstone, 2020). Such collaboration will help to "closely collaborate with […] key suppliers; include key suppliers in […] resilience and improvement activities [and] assess and monitor throughout the supplier relationship [and] plan for the full lifecycle" (NIST, 2020b, p. 6).

## 6.2  Regulation

The significant need for security regulations on all space asset segments is aggravated by severe difficulties in resolving vulnerabilities once on-orbit assets are in space. Highest-level advocacy within NATO is needed to reach similar sophistication of written guidance for space as it exists for

---

[80] **White hat hackers**: also known as *ethical hackers*; "employees or contractors working for companies as security specialists that attempt to find security holes via hacking" (Norton, 2020).
[81] This might be done in form of **penetration tests**, thus "to find, exploit and thus determine the risk of architecture vulnerabilities" (Privasec, 2019) or a Red Team assessment, being more targeted "to test the organisation's detection and response capabilities" (Privasec, 2019).

Air, Land, Sea and Cyber (Vasen, 2020, pp. 22-23). Hence, NATO should adopt a comprehensive regulatory and standardization framework for its SASC suppliers. This should include mandatory reports of all space asset cyber breaches, and guidance on prioritization and determination of critical components and responsibilities (Akoto, 2020, p. 7). Aim should be to substitute the principle of *Security by Obscurity* by a principle of *Security by Design and Default* both for technology and SCs, turning NATO into a cybersecurity-aware, constantly learning organization, analysing but not punishing incidences and celebrating good and best practices.

### 6.2.1 *Streamline Data Standardization and Classification Levels*

NATO MS should discuss "how national capabilities can work together with others, how they can provide national ISR for the common good and share that information with other MS" (Hill, 2020). Silverstein states that

> "many NATO members are undeniably proficient in space technology and can contribute to sharing agreements [on] unique data and capabilities, and also build beneficial redundant layers within the NATO space ecosystem. These redundancies help bridge the gap between growing military reliance on satellites and the inherent fragility of space objects" (Silverstein, 2020, p. 7).

Wells points to *standardized encryption of all data* throughout the NATO SASC having to become obligatory, as "one has always to operate under the assumption that somebody tapped in" (Wells & Sielaff, 2020). However, as suggested by Falco (2020), not all systems and organizations may be able to afford the processing around this, especially when considering legacy systems. Thus, it is important to consider the prioritization of risk mitigation techniques, encryption being included. Furthermore, Vasen suggests introducing *formatting standardization* throughout NATO's cyber SASC, such as through following specific frequencies for SATCOM or software, or streamlining data formatting such as for timing. This will simplify the integration of given data within NATO's processing systems. Such formatting standardization should be continuously undertaken by each MS before forwarding data to NATO, to decomplicate its later use.

Additionally, classification levels for key members in the commercial sector should be streamlined, to reduce "barriers to sharing space systems and data" (Silverstein, 2020, p. 5). Therefore, NATO should invest in unified secure platforms for exchanging information with suppliers, allowing them to involve the military end-user in the development and production supply chain to give immediate feedback. As an example, the US *Office of the Director of National Intelligence* (ODNI) currently aims to declassify information such as through one-time read-ins for permitted SC members, influencing supplier conduct through contractual security requirements. Such streamlining may resolve military stove-pipe thinking through allowing for

> "agile acquisition processes capable of absorbing innovative capabilities, commercial and military [,] more flexible, agile and resilient […] acquisition processes and faster command and control constructs to maintain the advantage in any conflict" (US Space Force & US Air Force, 2020, p. 8).

Finally, as stated by Yingst, streamlined "vetting of staff to avoid insider threats is critical to provide security and trust, whether it's the service provider or those who are providing assets, hardware or software to the space infrastructure or the operators of the space infrastructure" (Yingst, 2020). Sielaff states that employees need to undergo stricter clearances and have a good operational understanding of interacting with commercial representatives and users. Thus, CCSDS suggests developing an *enterprise security plan*, which is an

> "assurance that any shared resources (e.g., organizational processes, networks, and physical facilities) are adequate for the highest criticality and sensitivity handled [… which] should take into account regulatory authority, trust relationships, supply chain, and line management authority" (CCSDS, 2019, p. 16).

### 6.2.2 Enhance Procurement Contract Requirements

A *standardized security framework for SASC cybersecurity* should be created for NATO procurement processes to enhance contract requirements. As stated by Falco, "we need to move forward and increase accountability of manufacturers and space asset component developers for software and any other part of the SC" (Falco, 2020). This includes streamlining working terminology, external and internal assessments, incident reporting and communications across all partners and entities, as "there seems to be a lack of organised structure as far as regulation within the community" (Lister, 2020). Thereby, Timm states that "when you look at the actual SC cybersecurity itself, everything comes to play through a contract. You have to install controls in the contracts down the SC, to increase responsibility and knowledge about adverse effects" (Bander et al., 2020).

Hence, NATO should, at a strategic level, further focus on enhancing its supplier contracts to allow for more security, integrity, and quality of data delivered (CISA, 2020, p. 1). This should be done through integrating or partially adopting industry standards and procedures, to provide for an understandable and manageable structure of the SASC. Within NATO, Standardization Agreements (STANAG) or NATO Standardization Recommendations (STANREC) might be used as cover agreements. This would reduce disaggregated, smaller space asset contract series to a centralized industry procurement framework (Waterman, 2020, pp. 1-6). Thereby, as suggested by US Space Force and US Air Force, through creating "a single-entry point for all SATCOM requirements, enterprise SATCOM needs can be acquired in a deliberate and efficient manner, *avoiding the stovepipes of the past* […] using multi-year, *pooled-resource* contracts when possible" (US Space Force & US Air Force, 2020, p. 8). Finally, such an approach will simplify senior leadership's oversight through a common playbook, single incident-handling processes, and will prevent duplications to the greatest possible extent. Appropriate regulation enhancement measures include:

1. *Adopting a standards-oriented, streamlined supplier risk approach to SASC cyber risk assessment processes;*

2. *Incorporating strengthened insurances, such as for data confidentiality, into negotiations with suppliers;*

3. *Better monitoring progress, incidents and operations via customer and supplier assessments to determine SASC cyber risk assessment maturity.*

However, NATO should as well ensure "that commercial contracts meet military protection standards, in order to mitigate the risk posed by the military's use of commercial space assets" (Unal, 2019, p. 26). Commercial standards complying to those requirements, which could be partially or completely adopted by NATO, exist and are well-trusted (Wells & Sielaff, 2020). Such standards include:

### DOD CYBERSECURITY MATURITY MODEL CERTIFICATION

Across the board of interviewees, the US DOD *Cybersecurity Maturity Model Certification* (CMMC) is seen as currently the closest regulation to secure NATO mission-critical data retrieved from space assets (Caudill, 2020; Widmann, 2020). CMMC was released in February 2020 and aims to prevent foreign information theft and hacking throughout contractor SCs. It is the US government's most ambitious effort so far to shore up cyber vulnerabilities (Tritten, 2020, pp. 1-5). CMMC is imposed as an universal auditing measure, threshold condition for governmental bidding contracts as well as a strict standard on COTS security. Miller states that the CMMC makes

> "the transition to a model in which every company that does business with the federal government is being held accountable to implementing security. This securing of the entire defence SC is an enormous undertaking and will be a multi-decade effort" (Miller, 2020).

Based partly on *AIA National Aerospace Standard NAS9933*, CMMC is a sophisticated model with different requirement and classification levels: it proposes five different levels of cybersecurity controls, structures and methodologies (Bander et al., 2020). Whilst Level 1 imposes the lowest level of cybersecurity requirements on a contractor, Level 5 requires the most stringent controls for the most sensitive operations. CMMC Levels 4 and 5 relate directly to NIST 800-172 (*see below*). Thereby, Timm states that

> "every company in the SC of a prime, who will ultimately have CMMC on their contract, will be certified at least at CMMC Level 1. This basically includes 15 controls directly taken from the US Federal Acquisition Regulation (FAR), called the cybersecurity framework. Additionally, those 15 controls relate to 17 out of the 110 controls in NIST 171, being compliant to very basic cyber hygiene" (Bander et al., 2020).

The CMMC is expected to be specifically important for NATO due to the high number of US-NATO common weapons systems and components, and therefore intermingled military procurement and SC issues (Lee, 2020). The same is true for the use of commercial software running on NATO systems such as basic Microsoft Office365 software. Thus, for space asset-retrieved data, Lee and Timm suggest requiring a minimum of CMMC Level 3 of suppliers, managed processes, and good cyber hygiene, which would allow "some level of assurance and create trust that suppliers actually supply a product which does what it is supposed to do and does not have any malware implemented" (Lee, 2020). Additionally, Lee suggests imposing even a higher CMMC level for contractors producing GS components (Lee, 2020, *see 4.2.1*). Widmann warns that it will be easier for the US to install such a stringent regulation as it is a single federal government, however in case of NATO, the high number of different MS will significantly prolong the process.

### ISO 27001

Interviewees across the board cited the *ISO 27001 Standard* as being one the most important and basic standard to regulate the nexus between space, cyber and SC. ISO 27001 was developed in collaboration with the *International Electrotechnical Commission* (IEC) in 2008, and continuously updated until 2016. It bases on the unification of British and German IT security standards, and therefore draws on expertise of many different industries and application domains from different cultural backgrounds. The standard aims to

> "keep information assets secure [through] a framework for policies and procedures that include all legal, physical, and technical controls involved in an organization's information risk management processes [and] define how to implement, monitor, maintain, and continually improve […] divisions of responsibility, availability, access control, security, auditing, and corrective and preventive measures" (Microsoft, 2020).

CCSDS states that these controls "should be satisfied for space systems where third-party services are used" (CCSDS, 2019, p. 16). ISO 27001 addresses "*information security in supplier relationships* to ensure protection of the organization's assets that is accessible by suppliers" (CCSDS, 2019, p. 16) through

– A.15.1.1 Information security policy for supplier relationships;
– A.15.1.2 Addressing security within supplier agreements; and
– A.15.1.3 the IT SC which "should be satisfied for all facilities and systems affiliated with a mission" (CCSDS, 2019, p. 16).

Furthermore, ISO 27001 suggests "*supplier service delivery management* to maintain an agreed level of information security and service delivery in line with supplier agreements" (CCSDS, 2019, p. 16) through

– A.15.2.1 Monitoring and review of supplier services; and
– A.15.2.2 Managing changes to supplier services [which] should be satisfied for all systems (CCSDS, 2019, p. 16).

### NATIONAL AEROSPACE STANDARD NAS9933

The *AIA National Aerospace Standard NAS9933 on Critical Security Controls for Effective Capability in Cyber Defense* consists of 20 control families published by the Centre for Internet Security (CIS) and two additional families. Each family consists of *Critical Security Controls* (CSC) "to provide companies with a methodology to evaluate their systems and processes" (AIA, 2018). This includes several sub-controls, being categorized into five levels of capability "instead of a one-size-fits all checklist for compliance" (AIA, 2018). This standard aims at fulfilling two primary goals:

"[1] To provide industry partners an indication of a company's cybersecurity profile (beyond compliance-based controls), as a way to measure a company's cybersecurity risk [and 2] To align the fragmented and conflicting requirements […] contracting process imposes on industry. Rather than different DOD organizations using different tools to assess a company's security across different contracts, this standard is designed to apply common and universal elements of cybersecurity across each enterprise" (AIA, 2018).

AIA states that it aims at

"true risk- and threat-based cybersecurity [to] enable reciprocity, so that a company's level of security is accepted by all prime contractors, systems integrators, and DOD […] to establish the cybersecurity baseline in the aerospace and defence industry" (AIA, 2018).

### NIST SPECIAL PUBLICATIONS 800-37 AND 800-53

The most straightforward set of standards to introduce across NATO MS are NIST standards, as these are already widely adopted and trusted (Yingst, 2020). NIST standards recommend a risk management approach, lining out what can be done to prevent acquisition of compromised software and hardware (Lister, 2020). As stated by NIST, organizations should be aware "of the expanded attack surface that results from an interconnected and globally complex SC […] progress remains to be made on proactive incident detection, response, and recovery" (NIST, 2020a, p. 18). NIST recommends that

> "Contractual terms and conditions include insurance, access requirements, and background checks; Suppliers are contractually obligated to disclose component vulnerabilities, data loss, and security incidents; […] Contractual terms and conditions include a specific section on information security requirements as well as consequences if there is a failure to comply with the requirements for security, quality, and integrity" (NIST, 2020a, p. 12).

With regards to supplier management, and in alignment with 6.1.3, NIST recommends three key focuses (NIST, 2020a, pp. 11-12):

1. *Determining criticality of suppliers* through business impact if the supplier fails or is compromised, the supplier's operational stability, and classification level of data needed. Each supplier should be subsequently *score-graded*: if suppliers require logical and physical access to the agency's network, or products contribute to long-term strategic aims, they elevate in criticality and thus should become subject to ongoing security monitoring.

2. *Establishing cybersecurity requirements within contracts,* including a standardized set of cybersecurity conditions and requirements in contract negotiations to ensure a minimum security level, incorporated security considerations, and aligned relationship expectations early in the process of acquisition.

3. *Conducting annual organizational surveys of critical suppliers*, to ensure data, people, policies, and cybersecurity practices are up to date to support and ensure contracts. This includes surveying the conduct of supplier cybersecurity training and physical, network and IT security measures, and security assessment of organizational data exposed to external networks.

NIST *Special Publication (SP) 800-37* updates the NIST *"Risk Management Framework for Information Systems and Organizations: A System Life Cycle Approach for Security and Privacy"* (NIST, 2018). NIST SP 800-37 describes in further detail vulnerabilities within an organization's SC (NIST, 2018, p. 42). It incorporates SC risk management processes, hence security and privacy risks, and acknowledges that "adversaries are using the supply chain as an attack vector and effective means of penetrating our systems, compromising the integrity of system elements, and gaining access to critical assets" (NIST, 2018, p. v). Hence, NIST SP 800-37 suggests to specifically introduce SC risk assessment concepts "to address untrustworthy suppliers, insertion of counterfeits, tampering, unauthorized production, theft, insertion of malicious code, and poor manufacturing and development practices" (NIST, 2018, p. vi), further referring to ISO. It encourages to build up trust relationships and good communicating with both external and internal stakeholders, determining appropriate risk mitigating and management plans, document mitigating actions, and monitor the performance of such plans (NIST, 2018, p. 21). Thereby, it suggests that monitoring should include "regularly reviewing supplier foreign ownership, control, or influence (FOCI), monitoring inventory forecasts, or requiring on-going audits of suppliers" (NIST, 2018, p. 35).

NIST *Special Publication (SP) 800-53* on *Recommended Security Controls for Federal Information Systems and Organizations* aims to "reduce the extent of malicious code propagation […] including those failures *induced by SC attacks* […] employing different information technologies [and] using different suppliers" (NIST, 2013, p. 362). Additionally, it promotes *cyber resiliency*[82] measures to shorten and prevent SC attacks and reduce the level of adversarial impact.

### NIST SP 800-171

In 2018, *NIST SP 800-171* on *Protecting Controlled Unclassified Information in Nonfederal Systems* was issued, directing DOD prime and sub-contractors to comply with 110 new security requirements. Thereby, NIST states that

> "defence contractors must implement the recommended requirements contained in NIST SP 800-171 to demonstrate their provision of adequate security to protect the covered defence information included in their defence contracts […] if a manufacturer is part of a DOD, General Services Administration (GSA), NASA or other federal or state agencies' supply chain" (NIST, 2019).

Whilst many large contractors in the defence are already maintaining a sufficiently high level of cybersecurity against most foreign sabotage or intellectual espionage attacks, many smaller- and medium-size companies are still alarmingly unprepared. Such subcontractor companies typically handle so-called *controlled unclassified information*[83] (CUI) throughout manufacturing and systems, thus being foreign hacking prime targets. NIST requirements include, beneath others, "training necessary for […] SC security within the context of organizational information security programs" (Ross, Dempsey, et al., 2020, p. 94), reporting SC events, and cyber incident response actions. This involves "notifying relevant external organizations, for example, external mission/business partners, Supply Chain partners, external service providers, and peer or supporting organizations" (Ross, Dempsey, et al., 2020, p. 125).

### NIST 800-172 (DRAFT)

At the time of this thesis, NIST SP 800-171 is being reviewed and *NIST SP 800-172* on *Enhanced Security Requirements for Protecting Controlled Unclassified Information: A Supplement to NIST Special Publication 800-171* being created. Thereby, industrial sector companies were invited to provide online comments to the draft (NIST, 2020). NIST 800-172 points especially to "the protection of Controlled Unclassified Information […] to successfully conduct its essential missions and functions" (NIST, 2020). It underlines that "many federal contractors, for example, routinely process, store, and transmit sensitive federal information in […] *communications, satellite, and weapons systems*" (NIST, 2020, p. 13). Furthermore, NIST promotes the development of a CUI Registry, promoting that such a registry

> "identifies approved CUI categories, provides general descriptions for each, identifies the basis for controls, and sets out procedures for the use of CUI, including but not limited to

---

[82] *Cyber resiliency:* the ability to "anticipate, withstand, recover from, and adapt to adverse conditions, stresses, attacks, or compromises on systems that use or are enabled by cyber resources [to] can continue to operate even in a degraded or debilitated state, carrying out mission-essential functions" (NIST, 2019, p. 16).
[83] *Controlled Unclassified Information (CUI):* a category "of unclassified information within the US Federal government that requires safeguarding or dissemination controls pursuant to and consistent with applicable law, regulations, and government-wide policies but is not classified under Executive Order 13526" (US National Archives, 2016)

> marking, safeguarding, transporting, disseminating, reusing, and disposing of the information" (NIST, 2020, p. 13).

This concerns "*acquisition or procurement responsibilities (e.g., contracting officers)* […] as the entity responding to and complying with the security requirements set forth in contracts or agreements" (NIST, 2020, p. 15). Regarding the SC, it suggests to

> "employ automated discovery and management tools to maintain an up-to-date, complete, accurate, and readily available inventory of system components [including] *manufacturer, supplier information,* component type, date of receipt, cost, model, serial number, and physical location**"** (NIST, 2020, p. 27).

NIST requests to "*assess, respond to, and monitor SC risks* associated with organizational systems and system components" (NIST, 2020, p. 34), stating that

> "managing SC risk is a complex, multifaceted undertaking that requires a coordinated effort across an organization to build trust relationships and communicate with both internal and external stakeholders. SC risk management (SCRM) activities involve identifying and assessing risks, determining appropriate mitigating actions, developing SCRM plans to document selected mitigating actions, and monitoring performance against plans" (NIST, 2020, p. 34).

NIST 800-172 is a "very practical, collaborative effort [however] you need to make sure that what is going on in the US is also being looked at globally, ensuring that there is collaboration across borders" (Bander et al., 2020).

### *US SPD-5*

The recently released US *Memorandum on Space Policy Directive-5—Cybersecurity Principles for Space Systems* (SPD5)**,** issued on 4 September 2020, is

> "billed as the first comprehensive government policy related to cybersecurity for satellites and related systems, and outlines a set of best practices, but not firm requirements, that agencies and companies should follow to protect space systems from hacking and other cyberthreats" (Foust, 2020).

This includes principles and best practices regarding authentication, encryption of C2 links from and to satellites, jamming or spoofing of communications, protection of GS, and "intrusion detection systems for all aspects of space system architectures" (Foust, 2020). Furthermore, SPD5 points towards the Space ISAC for collaboration. It particularly considers SC issues, stating that it „is necessary for developers, manufacturers, owners, and operators of space systems to design, build, operate, and manage them so that they are resilient to cyber incidents" (White House, 2020). Also, SPD5 stresses that "most space vehicles in orbit cannot currently be physically accessed. For this reason, integrating cybersecurity into all phases of development and ensuring full life-cycle cybersecurity are critical for space systems" (White House, 2020). Finally, SPD5 suggests that "space system owners and operators should collaborate to promote the development of best practices [and] share threat, warning, and incident information within the space industry" (White House, 2020), and emphasises on the need for cybersecurity plans which

> "ensure the ability to verify the integrity, confidentiality, and availability of critical functions and the missions, services, and data they enable and provide [including] *Management of SC risks* that affect cybersecurity of space systems through tracking manufactured products;

requiring sourcing from trusted suppliers; identifying counterfeit, fraudulent, and malicious equipment; and assessing other available risk mitigation measures" (White House, 2020).

Miller points to SPD5 as being instrumental to "collaborate with international partners and raise the security posture for the entire space sector, to be able to understand what the best regulations are that need to be put in place" (Miller, 2020).

## *6.3 Summary*

Two major collaboration initiatives should be followed to allow NATO to timely recognise cyberthreats to mission-critical data early on along the SASC: a) raising *awareness* and b) pushing forward the *creation of an adaptive framework of regulation*. In doing so, NATO would be able to fully leverage the potentials of its new Space Policy and its Declaration of Space as an Operational Domain. Thus, it would function as a formal and informal MS consultation and conversation forum, making full use of its broad collective MS portfolio, as suggested by NATO SG Stoltenberg. Hence, *hypothesis two (H2) is confirmed*. However, it has become apparent that a safer supply of NATO mission-critical data depends on more factors than only on the establishment of a strict international legal and operational framework, or transparent development, design, management, and ownership of space assets. In conclusion of this study, the following, final chapter will summarize those findings.

# 7 CONCLUSION

With the launch of the first artificial satellite in 1957, space became vital for enabling national and international security. However, it was only in 2019 that NATO issued its first Space Policy and recognized space as an operational domain. As NATO does not possess its own on-orbit space segments, this was followed in 2020 by a MOU between NATO, France, Italy, the UK, and the US, being those NATO MS with the strongest knowledge of space and cyber. The MOU ensures fifteen more years of critical MS space capacity to the Alliance, and may well counteract the lack of security by design throughout commercial space assets included in NATO's SASC. Numerous cyber vulnerabilities may impact control over such space assets and thus reliability on mission-critical information. Consequences would be dire and could include widespread disruptions or permanent shut-down of space assets, DoS, or attacks on national CI. This threat is aggravated by NATO's mostly unregulated cyber SASC, recent national trends to launch COTS space technology, and therefore augmented complexity of space asset ownership and liability. Hence, this thesis aimed at identifying such cybersecurity gaps along NATO's SASC in order to help readers understand if and how such gaps threaten the integrity and security of NATO's missions and hence support organizational resilience against cyberattacks, so that NATO may better assure its missions against adversaries in cyberspace. By conducting qualitative, empirical research, and following a thorough academic literature research in *Chapter 2,* this thesis was guided by a twofold research question:

> a) *What are current cybersecurity gaps along NATO's global SASC; and*
>
> b) *How can NATO and its allied MS gain greater control over such gaps to safeguard the supply of NATO mission-critical information?*

To answer these questions, *Chapter 3* identified prominent use cases of space assets for NATO, as well as relevant NATO-internal contracts, division of responsibilities and according mechanisms. It became clear that space assets are key enablers for NATO's most advanced technological systems, missions, and operations. Key services are *SATCOM,* providing crucial telecommunications and capacities such as broadband internet, mobile services; *ISR,* allowing to provide joint force planners with multi-spectral information on subsurface, surface, and air conditions; *PNT,* enabling precise navigation and timing; and *deterrence,* facing a rapid arms race in and through space. As stated above, NATO does not own on-orbit space segments, yet currently operates twenty GS and uses national MILSATCOM services. SATCOM remains under responsibility of NCIA, whilst ISR is mostly handled by JISR. Additionally, JAPCC, CCDCOE as well as the soon to be established Space COE provide awareness and counselling. SpSC coordinates operational space functions. However, this wide spread of responsibilities over NATO's internal and external system obscures liability in case of an attack.

*Chapter 4* investigated crucial space asset cybersecurity gaps. By taking out a single space asset segment, thus either GS, space segment or data link, the whole space asset system is likely to become inoperable. Actors behind attacks are classified in either *nation-state actors*, *professional or amateur hackers*, *organised criminals*, or *insiders*. Their attribution and localization become extremely difficult due to both cyberspace's and outer space's ambiguous and completely borderless nature. Nonetheless, attribution remains an important piece of deterrence to ensure that adversaries realize that there will be consequences for actions taken in the cyber domain, hence adding to NATO's resilience posture. As further concluded in this chapter, main cyber-related risks in space asset segments include:

1. *Legacy and/or unsegmented soft- and hardware being installed in GS and space segment's systems*, allowing for eased interception of data;

2. *The increased use of COTS technology for SmallSats and insecure management of encryption,* easing tampering of data during its transfer;

3. *A lack of expertise on the nexus between space and cybersecurity and governmental reluctance to report cyber incidents,* hence hindering awareness; and

4. *The absence of international instruments to regulate conduct in space,* adding to geographical limits of NATO Treaty Article 6 preventing the triggering of the Principle of Individual or Collective Defence, and diminishing NATO's capacity to deter.

*Chapter 5* applied and expanded these findings to NATO's space asset cybersecurity. The rapidly increasing number of globally widespread suppliers involved in proliferating the various components of space assets multiplies to an uncontrollable number of possible cyberthreats and vulnerabilities. This might crucially impact NATO as well as its staff on the ground, including possible casualties. Whilst outsourcing to external public and private companies is needed to stay competitive and up to date with innovative technology to execute NATO's operations to the best possible extent, this increased obscurity of SCs eases the implementation of third-party vulnerabilities in space assets' soft- and hardware. This is aggravated by

1. *NATO's dependency on dual-use space assets*, leaving open responsibility in case of a SC cyberattack;

2. *Market pressure increasing NATO's need to stay competitive,* trading off cybersecurity concerns through public and private SC partners;

3. *Classification restrictions and formatting misalignment between NATO, MS, and private companies*, limiting the sharing of essential threat intelligence and intrusion prevention;

4. *Lacking SASC risk management training and awareness*, diminishing performance during crises such as the ongoing COVID pandemic; and

5. *Significant understaffing and lack of clarity in coordinating space-related tasks* between NATO entities, aggravated by a lack of resources and/or willingness to focus on cybersecurity enhancement.

Therefore, this thesis has shown that NATO missions and operations are indeed threatened by the integrity of mission-critical data being relatively easy to compromise. Any sophisticated, SC-enabled cyberattack on NATO-utilized SATCOM, ISR, or PNT space assets and services may jeopardize NATO's missions, undermine its capability to deter, and cause severe consequences for MS national security. These findings confirm *hypothesis one* stating that

*(H1) Cybersecurity gaps along NATO's SASC are likely to originate from a) outsourcing of space capabilities to private companies, b) the use of outdated software and c) an almost complete lack of international laws and regulations.*

However, following preceding findings, H1 needs to be extended as follows:

> **1**
>
> *Current cybersecurity gaps along NATO's cyber SASC are caused by*
>
> *a) Cyber vulnerabilities, including legacy and unsegmented space asset systems, use of COTS technology, and lacking expertise and SASC regulation; and*
>
> *b) Inadequate handling of NATO's cyber SASC, such as misaligned classification and formatting, lacking risk management training, or significant operational, space-related understaffing.*

The same is true for *hypothesis two,* stating that

*(H2) A safer supply of NATO mission-critical information is likely to be ensured through the establishment of a strict international legal and operational framework, establishing transparent development, design, management, and ownership of space assets.*

As shown throughout the preceding chapter, additional measures are needed to timely recognise SASC cyberthreats to mission-critical data. Thus, H2 is extended as follows:

> **2**
>
> *NATO and its allied MS can gain greater control over such cyber SASC gaps by*
>
> *a) Rising awareness through increased intersectoral collaboration and strategic level engagement, mission-specific risk assessments and training, and spurring R&D; and*
>
> *b) Providing a platform for regulation making, whilst centralizing responsibilities, streamlining security, classification, and formatting, and creating a standardized contract security framework.*

Concluded, required policy changes and specific recommendations as highlighted throughout *Chapter 6* include:

Table vi: Conclusion: Policy Recommendations

| POLICY RECOMMENDATIONS |
|---|

| | |
|---|---|
| ***AWARENESS*** | ***Initiate Intersectoral Collaboration:***<br>• Aligning differing security and ethical standards between countries;<br>• Organizing at least one annual space asset CSCRM symposium by executives from both sides to exchange on current challenges and estimated organizational impact;<br>• Establishing further PPPs along the SASC, for example with:<br>    ○ *CCSDS and ISO;*<br>    ○ *Space ISAC;*<br>    ○ *US AIA;*<br>    ○ *CERTs, CSIRTs and ENISA; and/or*<br>    ○ *OSA*;<br>• Utilizing the upcoming *CCDCOE Tallinn Manual 3.0* multi-stakeholder project, as well as the *2020 EU Cybersecurity Strategy for the Digital Decade* to address gaps.<br><br>***Increase NATO Strategic Level Engagement:***<br>• Creating mission-specific guidance, agreements, and security considerations;<br>• Establishing recurrent executive-level SC leadership councils;<br>• Timely increase of resources and staffing.<br><br>***Spur Exchange of Academic Expertise:***<br>• Strategically investing in space asset Research and Development (R&D) and the development of viable alternatives, such as through NATO STO's Information Systems Technology Panel;<br>• Realizing the establishment of a NATO Space Technology Centre under guidance of NCIA;<br>• Considering attack vectors originating from new and upcoming technology, such as AI being increasingly implemented in space assets;<br>• Inviting outside organisations and cybersecurity experts to provide expertise and differing perspectives on the topic, such as through cooperation between NCIA and Europol EPE.<br><br>***Conduct Mission-Specific Cyber SASC Risk Assessments:***<br>• Ensuring knowledge on specific procurement and SC environment around each mission, thus sourcing the materials to the level of integrity required;<br>• Moving from detective to preventive third-party cybersecurity management capabilities;<br>• Establishing a formal programme through manifesting policies, governance, procedures, tools, and processes on data sent to and received by any supplier;<br>• Assessing NATO-procured ICT components, such as hardware, software, and services;<br>• Analysing suppliers, suppliers' sources, and the larger SC ecosystem;<br>• Dedicating specified NATO personnel to effectively assessing and communicating supplier risks to NATO's executive leadership. |
| | ***Train Cyber SASC Risk Management throughout NATO Exercises***:<br>• Including cyber SASC risk management at least in every midsize to large NATO exercise to support a forward-leaning, collective defence approach;<br>• Facilitating space asset CSCRM training for relevant stakeholder organization-wide as well as for key suppliers;<br>• Facilitating individual end user training for each redeployment;<br>• Engaging white hat hackers to penetration test for possible vulnerabilities throughout trainings;<br>• Continuous phishing simulation exercises, mapping upcoming cyber SC threats and priorities, or tabletop exercises to determine key stakeholders and their responsibilities;<br>• Sharing training and test reports to both the military and intelligence community and the commercial sector. |
| ***REGULATION*** | ***Streamline Security and Classification Levels:***<br>• Implementing formatting standardization for eased data integration and processing, e.g., agreeing on specific SATCOM frequencies, software, timing formatting;<br>• Streamlining classification levels;<br>• Streamlining strict clearances of staff and a good operational understanding of interacting with commercial representatives and users.<br><br>***Enhance Procurement Contract Requirements:***<br>• Creating a standardized public-private security framework for NATO-specific SASC cybersecurity:<br>   o *Streamlining working language, external and internal assessments, incident reporting and communications across all partners and entities, thus creating a single-entry point for suppliers;*<br>   o *Adopting a standards-oriented, streamlined supplier risk approach to CSCRM processes;*<br>   o *Incorporating strengthened insurances, such as for data privacy, into negotiations with suppliers;*<br>   o *Better monitoring progress and operations via customer and supplier assessments to determine CSCRM maturity;*<br>   o *Ensuring that contractors comply with military protection standards.*<br>• Including mandatory reports of all space asset cyber breaches, and guidance on prioritization and determination of critical components and responsibilities;<br>• Substituting Security by Obscurity by Security by Design and Default;<br>• Integrating or partially adopting industry standards and procedures into supplier contracts:<br>   o *NAS9933;*<br>   o *DOD CMMC;*<br>   o *ISO 27001;*<br>   o *NIST SP 800-37;*<br>   o *NIST SP 800-53;*<br>   o *NIST SP 800-171;*<br>   o *NIST 800-172; and*<br>   o *US SPD-5.* |

By completely or partly adopting these policy recommendations, NATO would strive to fully leverage the potential of its *2019 Space Policy* and its *Declaration of Space as an Operational Domain*. Additionally, this would allow NATO to function as a formal and informal MS consultation forum on SASC cybersecurity. This way, it would make full use of its broad collective MS portfolio, best practices and lessons learned, as suggested by NATO SG Stoltenberg.

These findings are in line with the assumptions of *neoliberal institutionalist theory*. As has been shown, NATO and national actors seek to maximize their gain from space assets to maintain relative power and the ability to deter. This behaviour is enforced by commercial competition and innovation. Throughout this thesis, this was proven by findings which align with neoliberal institutionalist theory, thus highlighting that

a) *Lacking cooperation between NATO as an international organization and its MS, as well as commercial SC partners, hinders to find problem solutions and create internationally binding regulations;*

b) *Commercial competitiveness spurred by the search for economic efficiency, competitive advantage, and cheaper labour trades off security concerns.*

Following neoliberal institutionalism, and aligned with above presented policy recommendations, NATO's role should be to guide and advise MS in resolving global political and economic issues, to reduce individual cost and uncertainty. Therefore, NATO should consider a shift from regarding cyber SASC risk management from being 'nice-to-have' to being mandated and critical. Doing so will ease safeguarding of delivery and integrity of NATO mission-critical data, better integrate suppliers and manufacturers along the cyber SASC, and increase transparency, responsibility, and liability. Furthermore, this will ease NATO to stay competitive, deterring and up to date with innovative technology, enhancing outcomes of NATO missions and operations to the best possible extent.

To this end, and with the awareness that certain MS such as the US are already taking concrete steps to better protect their cyber SASC, this thesis provided complementary insights and analysis regarding the challenges at stake. Whilst the limited extent and semi-structured nature of interviews may generate results which cannot be generalized across the whole of NATO system and SC, they do provide detailed insight into and understanding of interviewees' professional perception, experience, and in-depth expertise. Research results clearly illustrated the need for centralized regulations and streamlined handling of this SC aligned with industrial, well proven standards. However, those results have also raised the question of how these considerations should be implemented in an innovative, sustainable, and technical detail. In conclusion, this study aims to constitute a significant enrichment to existing academic literature, as well as a valuable starting point for future research encompassing gaps at the military SC nexus between cybersecurity and space and the future threat landscape. This includes AI being increasingly employed in space assets, which will NATO need to worry about both cyber doctrine and how it interacts with space, and AI doctrine and how it intersects with cyber and space. It is time to focus on SASC cybersecurity and safeguard NATO, MS, and industry from hostile interference.

# ANNEX: COMPLETE INTERVIEW GUIDELINE

The following is a complete list of questions as posed throughout the interviews providing the research base for this thesis. As becomes apparent, *five major fields* determined the interviews: *a) General Introduction, b) NATO and military use of space, c) space asset cybersecurity, d) the nexus between space asset cybersecurity and their supply chain,* and *e) policy recommendations.* Given that the interviews were conducted in a *semi-structured* manner, specific questions, either based on the preceding literature research or key words that prompted additional questions, were posed when appropriate and with regard to the interviewee's position, rank, company sector, general direction of answers, and field-specific knowledge.

---

## A) GENERAL

- Welcome! Allowed to record? Use excerpts later in work? Level of secrecy (e.g., Chatham House Rule)?
- Background, experience, tasks/ responsibilities, organizational structure, background of organization, why field chosen?

## B) USE OF SPACE

- What most used/ what for?
- Role military/ government?
- Forms of use of commercial space assets for military purposes (e.g., deterrence or critical infrastructures)?

## C) NATO/ MILITARY SPACE ASSET CYBERSECURITY

- What is the significance of NATO declaring space/cyberspace as an operational domain? What will follow now? (consequences)? New alliances, therefore? Where does funding come from? How much share is taken off the NATO budget?
- Overall significance of space capabilities? Relevance to NATO operations? How are operational tasks divided?
- Which NATO standards used? MILSATCOM criteria as defined by NATO? Code of Conduct for outer space activities? Most recent standards, guidelines and best practices used by the private and governmental sector? Tech protectionism/ EU Digital Services Act?
- Where are NATO and other significant space ports/ ground stations located? What is leaving there and for what reason?
- Most (im-) pressing new developments in space assets? What are the most promising space systems/ applications? Digital trends?
- Key players? France: new military space strategy? Any change because of the UK Brexit? Role CIRTS/ CERTS? Which countries are the main point of contact? To whom most allied?
- Most recent relevant cyberthreat actors and their technical intrusion tools and motives?

- Most recent efforts in streamlining in the public and private sectors to secure the space industry? Influence of the commercial sector on governmental procurement and securitization efforts of space asset components?
- Current overall impression of space assets cybersecurity? Experience with the military using satellites? What are cyber downfalls?
- To what extent risk and threats assessments, how conducted?
- What are usual uncertainties/ unforeseen events? What are core questions/ considerations for you? What is the aim of attackers? Types of satellite hacking? How to block jamming/ spoofing?
- Which systems are used for what tasks? Where is data stored? Who has access to it? Which software is used? Configuration management hardware/ software? What technology is used?
- COTS security strategies? How documented?
- Role Critical Infrastructures?
- Role internationalization/ distances?
- What types of satellites are most appropriate/ what used for? Usage/ advantages/ disadvantages of CubeSats? Which orbit used most and what for? Satcom-as-a-service (SAAS)?
- Verification mechanisms? Role of quantum satellites? Hack proof? Use of cryptography? Processes for ensuring organizational trust.
- How to keep autonomy?
- Influence COVID?

## D) SPACE ASSET CYBER SUPPLY CHAIN

- Advantages and downfalls of outsourcing the development of space assets to the commercial sector (including the use of COTS/ CubeSats) and key actors (suppliers, customers)?
- Key players? Who are your main suppliers? Which countries are the main point of contact? To whom most allied? Level of complexity in SC? What are each party's roles and responsibilities for secure product acquisition? Key companies: components/ receiver manufacturers, system integrators, service providers?
- Who are the main international collaboration partners? (e.g., NATO to ESA/ US satellite agencies/ other institutional space agencies/ private sector)? Role of ISACs?
- Legislation/ regulation? STANAGs?
- Hot to shield supply chain from/ deter (cyber) intrusions along its military and commercial supply chains? Standards or good practices related to SC integrity? Frameworks, guidelines, and best practices used by the private and governmental sector to do so?
- How up/ downstream defined? How would you describe the type of collaboration between countries to supply satellite assets? Who are 1st, 2nd, 3rd tier suppliers/ customers?
- Difference between SC management and logistics vs procurement?
- In how far risk and threats assessments? How did SC risk management and threat landscape change over time? How will supply chain management criteria be developed and enforced?
- How are supply chain risks factored into day-to-day product acquisitions? What are the controls necessary to eliminate all reasonable supply chain risks?
- Current overall impression of the SASC? What are usual uncertainties/ unforeseen events? Role internationalization/ distances? How would you categorize the risks you are facing? "Island hopping"?
- Does NATO act as a wholesaler to the different MS?

- How are products tracked? Benefits/ downfalls? How are products tagged (Barcode/ RFID)? Which frequency? Who controls? Tracing problems?
- Where is data stored? Who has access to it? How documented?
- COTS strategies? Security configuration management hardware/ software?
- Processes for ensuring organizational trust? Most/ less trusted suppliers/ customers? Why? Processes to develop supply chain assurance?
- How are security checks executed? Measurement measures/ benchmarks?
- Typical assets/ supplier resources?
- What contract types are used? (e.g. transaction authority agreements, space act agreements (US), simplified R&D contracts)?
- How are responsibilities divided between different governmental sectors? Who makes regulations? With whom in contact? Programmes/ procurement rules/ policies? Signing of security clauses?
- Supply chains well traced by primers/ governments alike?
- Clearly distributed roles and responsibilities?
- Mergers to streamline the supply chain? Single IT system? Sharing of supply chains?
- Metrics related to the supply chain?
- Technical level gaps? Risk analysis gaps? Standardization scheme gaps? What kind of information should (not) be shared?
- Models used to develop standardization? NATO codification system?
- Solutions: Software Bill of Materials (SBOM)? Blockchain? Streamline due diligence to focus on critical risks? Establish internal triggers to monitor for change? Create controls and incentives to monitor for change?

## E) POLICY RECOMMENDATIONS

- Most pressing renovation and securitization needs for governments and military (cooperation)?
- Further questions? Documents to disclose? Further research suggestions?
- Further organisational matters; Thank you for taking the time!